\documentclass[traditabstract]{aa} 
\usepackage{graphicx,amsmath,mathtools}
\usepackage[varg]{txfonts}
\usepackage{natbib}
\usepackage{siunitx}
\usepackage{float}

\usepackage{wrapfig}
\usepackage{lscape}
\usepackage{rotating}
\usepackage{epstopdf}
\usepackage{longtable}
\usepackage{gensymb}
\usepackage{footnote}
\usepackage{caption}
\usepackage{subfig}

\makesavenoteenv{tabular}
\makesavenoteenv{table}

\usepackage{threeparttable}
\usepackage{multirow}

\bibpunct{(}{)}{;}{a}{}{,} 

\def\gtrsim{\mathrel{\hbox{\rlap{\hbox{\lower4pt\hbox{$\sim$}}}\hbox{$>$}}}}
\def\lesssim{\mathrel{\hbox{\rlap{\hbox{\lower4pt\hbox{$\sim$}}}\hbox{$<$}}}}
\newcommand{\hi}{H{\sc i}}
\newcommand{\hinospace}{H{\sc i}}

\newcommand{\hdue}{H$_2$}

\newcommand{\kms}{km s$^{-1}$}
\newcommand{\Kkms}{K km s$^{-1}$}
\newcommand{\msun}{M$_{\odot}$$\,$}

\newcommand{\arsec}{$^{\prime\prime}\!$ }
\newcommand{\arsecnospace}{$^{\prime\prime}$}
\newcommand{\micron}{$\mu$m}

\newcommand{\D}{{$\mathcal D$}}
\newcommand{\hers}{{\it Herschel }}

\begin{document}

\title{Star-forming dwarf galaxies in the Virgo cluster: the link between molecular gas, atomic gas, and dust\thanks{Based
   on observations carried out with the IRAM 30m Telescope. IRAM is supported by INSU/CNRS (France), MPG (Germany) and IGN (Spain).}}

   \subtitle{}

   \author{M. Grossi\inst{1}
          \and E. Corbelli\inst{2}
          \and L. Bizzocchi\inst{3}
          \and C. Giovanardi\inst{2}
          \and D. Bomans\inst{4}
          \and B. Coelho\inst{1}
          \and I. De Looze\inst{5,6}
          \and T. S. Gon\c{c}alves\inst{1}
          \and L. K. Hunt\inst{2}
          \and E. Leonardo\inst{7,8}
          \and S. Madden\inst{9}
          \and K. Men\'endez-Delmestre\inst{1}
          \and C. Pappalardo\inst{7,8}
          \and L. Riguccini\inst{1}
}
\institute{
Observat\'orio do Valongo, Universidade Federal do Rio de Janeiro, Ladeira Pedro Ant\^onio 43, Rio de Janeiro, Brazil\\
\email{grossi@astro.ufrj.br}
\and
INAF-Osservatorio Astrofisico di Arcetri, Largo Enrico Fermi 5, 50125 Firenze, Italy
\and
Center for Astrochemical Studies, Max-Planck-Institut f{\"u}r extraterrestrische Physik (MPE),
Giessenbachstra{\ss}e 1, 85748 Garching, Germany
\and
Astronomical Institute of the Ruhr-University Bochum (AIRUB), Universit\"{a}tstr. 150, D-44801 Bochum, Germany
\and
Sterrenkundig Observatorium, Universiteit Gent, Krijgslaan 281, B-9000 Gent, Belgium
\and
Department of Physics and Astronomy, University College London, Gower Street, London WC1E 6BT, UK
\and
Instituto de Astrof\'isica e Ci\^encias do Espa\c{c}o, Universidade de Lisboa, OAL, Tapada da Ajuda,
PT1349-018 Lisboa, Portugal
\and
Departamento de F\'isica, Faculdade de Ci\^encias, Universidade
de Lisboa, Campo Grande, PT1749-016 Lisbon, Portugal
\and
Laboratoire AIM, CEA/DSM - CNRS - Universit\'e Paris Diderot, IRFU/Service d'Astrophysique, CEA Saclay, 91191 Gif-sur-Yvette, France
}

   \date{}

\abstract{We present $^{12}$CO(1-0) and $^{12}$CO(2-1) observations of a sample of 20
star-forming dwarfs selected from the \hers Virgo Cluster Survey,
with oxygen abundances ranging from 12 + log(O/H) $\sim$8.1 to 8.8.
CO emission is observed in ten galaxies and marginally detected in another one.
CO fluxes correlate with the FIR 250 $\mu$m emission, and the dwarfs follow the same linear relation that holds for more massive spiral galaxies extended to a wider dynamical range.
We compare different methods to estimate \hdue\ molecular
masses, namely a metallicity-dependent CO-to-\hdue\ conversion factor
and one dependent on $H$-band luminosity.
The molecular-to-stellar mass ratio
remains nearly constant at stellar masses
$\lesssim 10^9$ \msun, contrary to the atomic hydrogen fraction, M$_{HI}/$M$_*$, which increases inversely with M$_*$. The flattening of
the M$_{H_2}/$M$_*$ ratio  at low stellar masses
does not seem to be related to the effects of the cluster environment because it occurs for both \hi-deficient and \hi-normal dwarfs.
The molecular-to-atomic ratio is more tightly correlated with stellar surface density than metallicity, confirming that the interstellar gas pressure plays
a key role in determining the balance between the two gaseous components of the interstellar medium.
Virgo dwarfs follow the same linear trend between molecular gas mass and star formation rate as more massive spirals,
but gas depletion timescales, $\tau_{dep}$, are not constant and range between 100 Myr and 6 Gyr.
The interaction with the Virgo cluster environment is removing the atomic gas and dust components of the dwarfs,
but the molecular gas appears to be less affected
at the current stage of evolution within the cluster.
However, the correlation between \hi\ deficiency and the molecular gas depletion time suggests that the lack of gas replenishment from the outer regions of the disc is
lowering the star formation activity.
}


\maketitle

\section{Introduction}
\label{sec:intro}

Star-forming dwarf (SFD) galaxies are rich in atomic hydrogen (\hinospace), but despite their star-formation
activity, the detection of molecular gas in these systems is challenging. The lack of low-energy
transitions in the molecular hydrogen molecule (\hdue), which
is due to the absence of a permanent dipole moment, implies that
the \hdue\ content has to be inferred from the emission of carbon monoxide \citep[CO;][]{1991ARA&A..29..581Y}.
While \hdue\ is self-shielded from the UV radiation,
dust is primarily responsible for preventing CO photodissociation \citep{2010ApJ...716.1191W},
therefore only dense CO clumps can survive in dust-poor, low-metallicity systems \citep{1994A&A...292..371L,2003A&A...406..817I,2009ApJ...702..352L}.
It is unclear to which extent
a lack of CO emission means a correspondingly low \hdue\ content
because a significant \hdue\ mass may lie outside the CO region, in the outer parts of the molecular cloud where carbon is in the atomic phase
\citep[][]{1995ApJ...454..293P,1997ApJ...483..200M}.
Indeed, carbon monoxide is often undetected in very metal-poor low-mass galaxies with a threshold abundance usually observed at
 Z $\simeq$ 0.2 Z$_{\odot}$ \citep{1998AJ....116.2746T,2005ApJ...625..763L,2012AJ....143..138S}, and
the CO detection rate is usually higher in more massive Magellanic-type dwarf galaxies
than in irregulars and blue compact dwarfs \citep[BCDs;][]{2004A&A...414..141A}.

\begin{figure*}
  \centering
\includegraphics[bb=10 -30 560 290,width=6.cm]{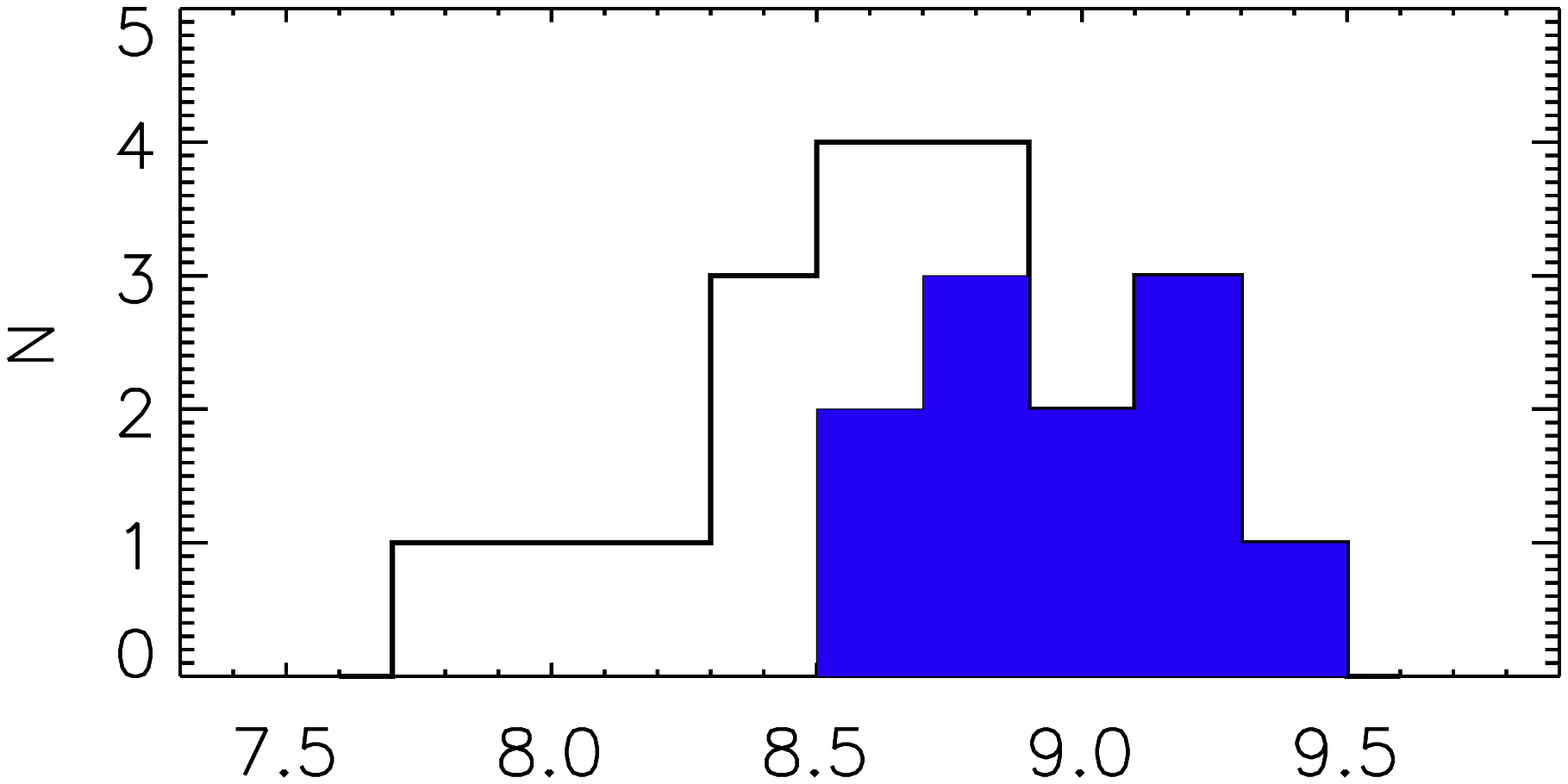}
\includegraphics[bb=10 -30 560 290,width=6.cm]{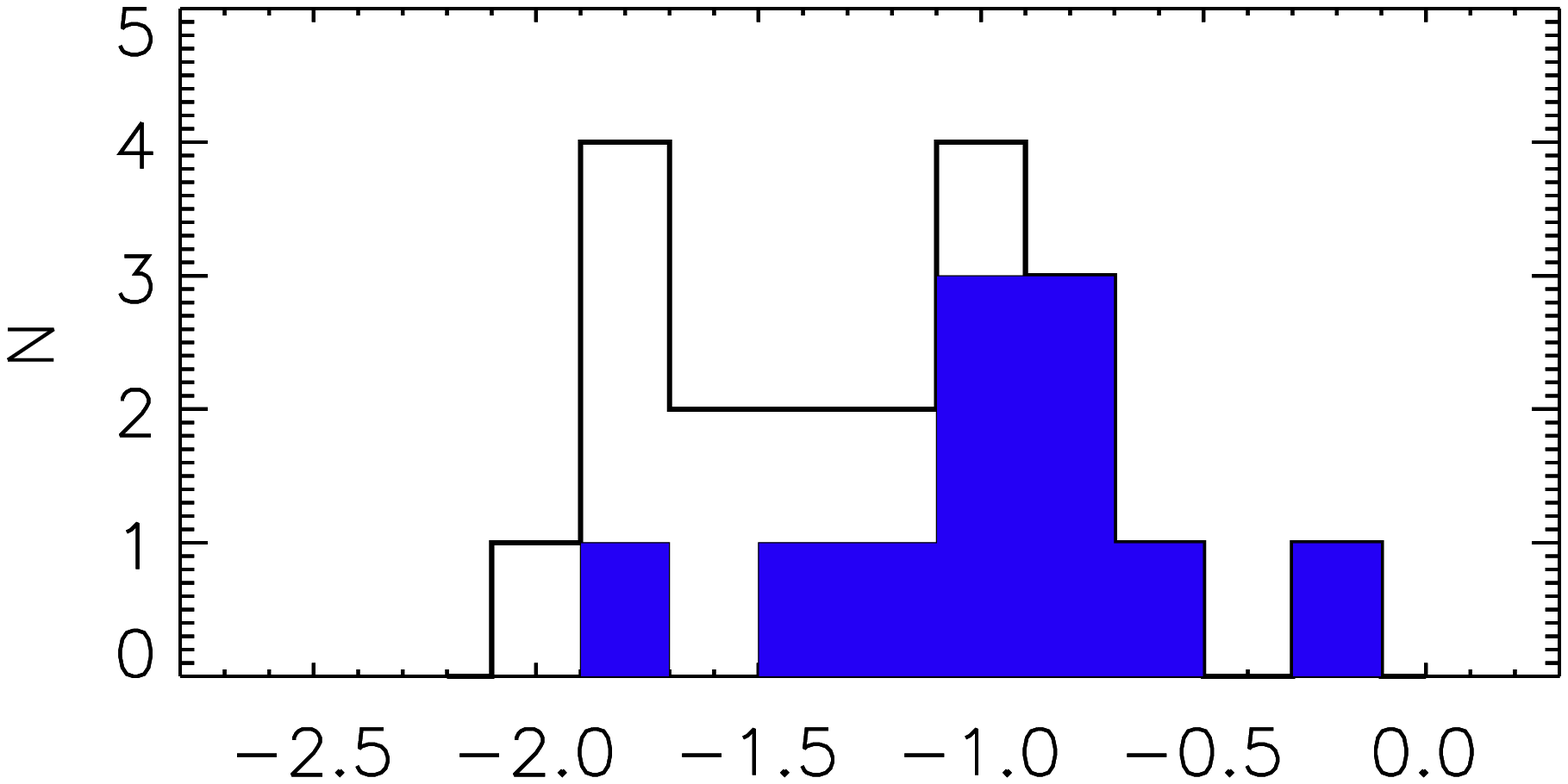}
\includegraphics[bb=10 -30 560 290,width=6.cm]{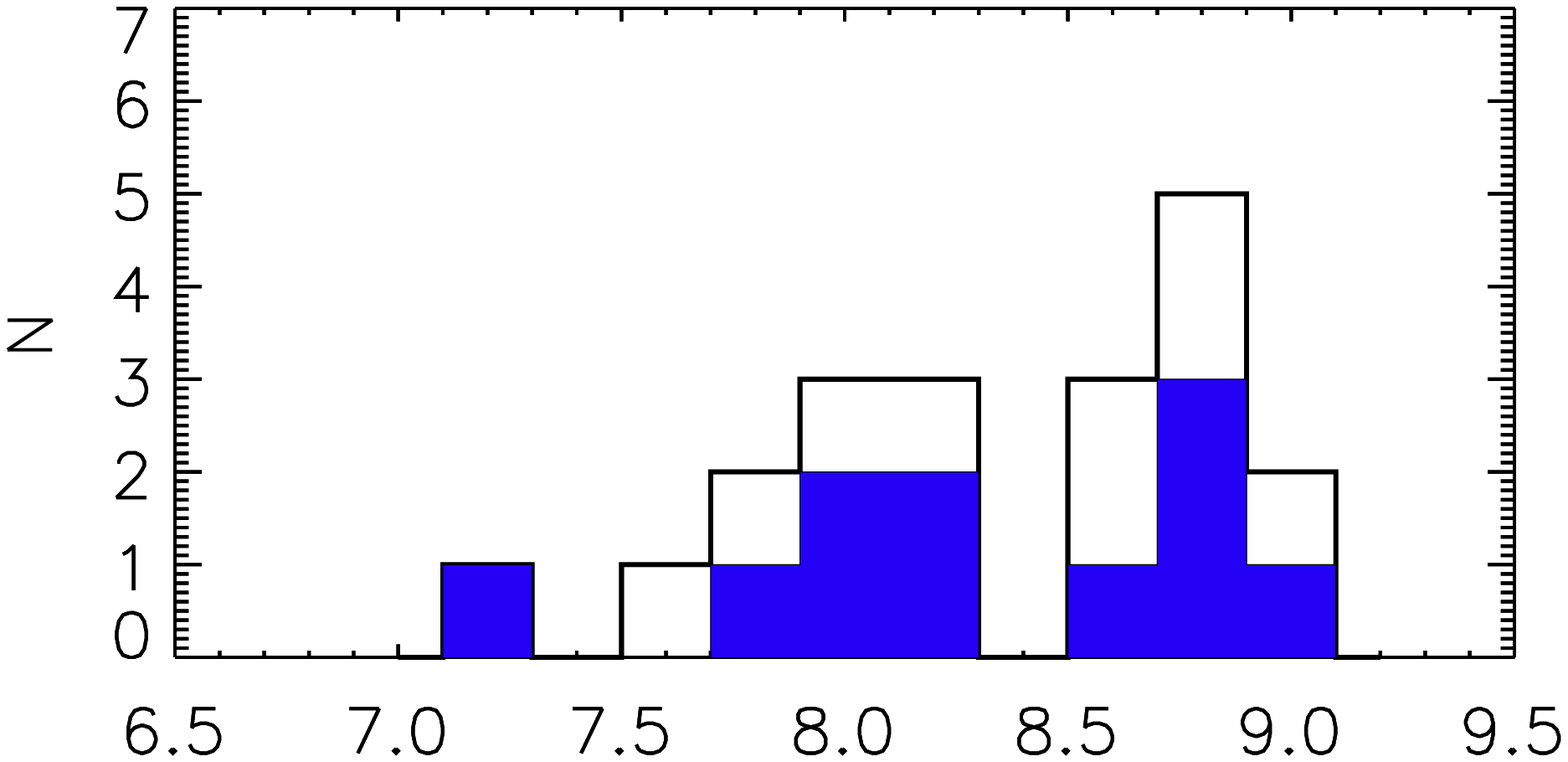}
\includegraphics[bb=10 -30 560 290,width=6.cm]{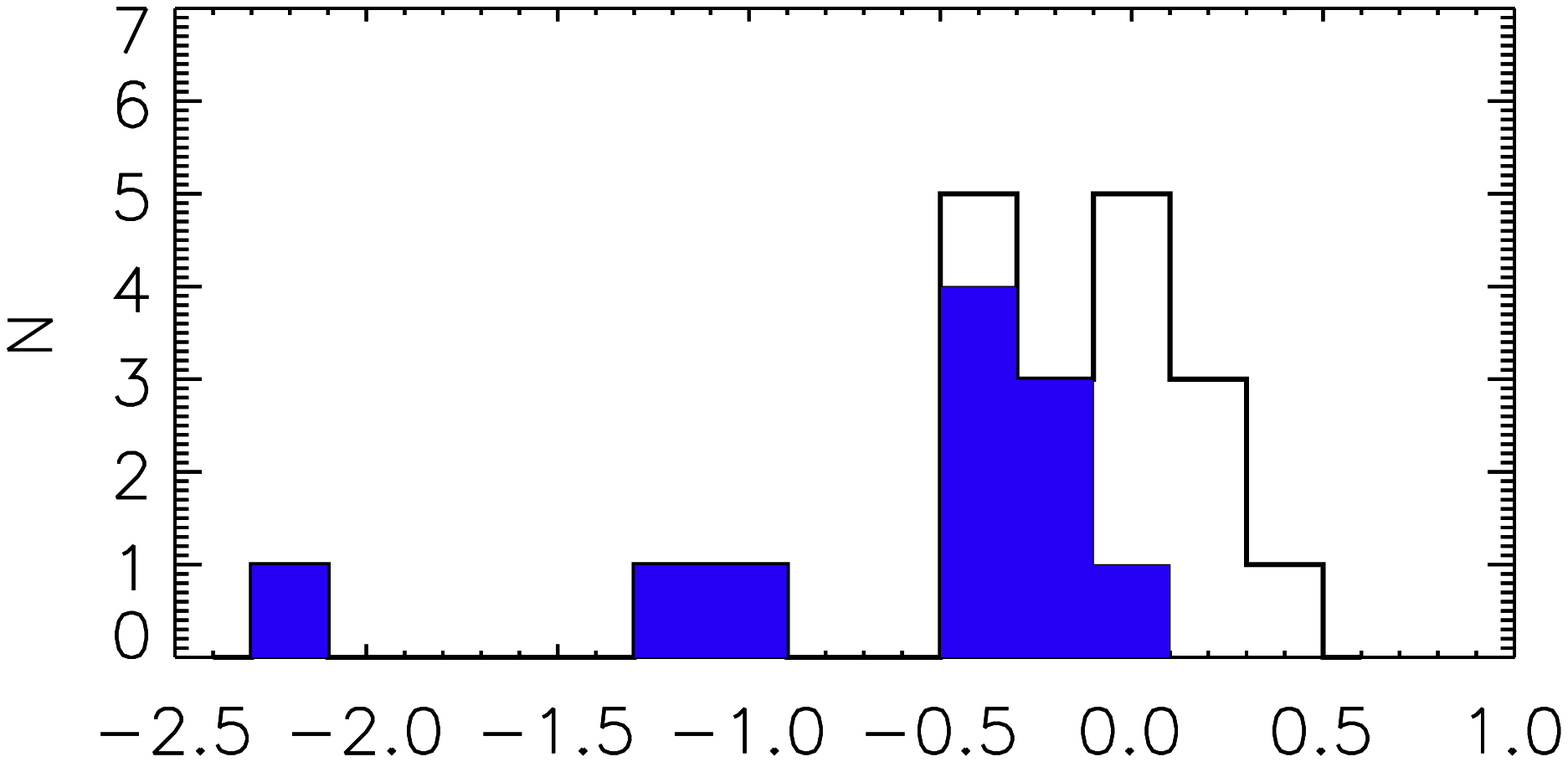}
\includegraphics[bb=10 -30 560 290,width=6.cm]{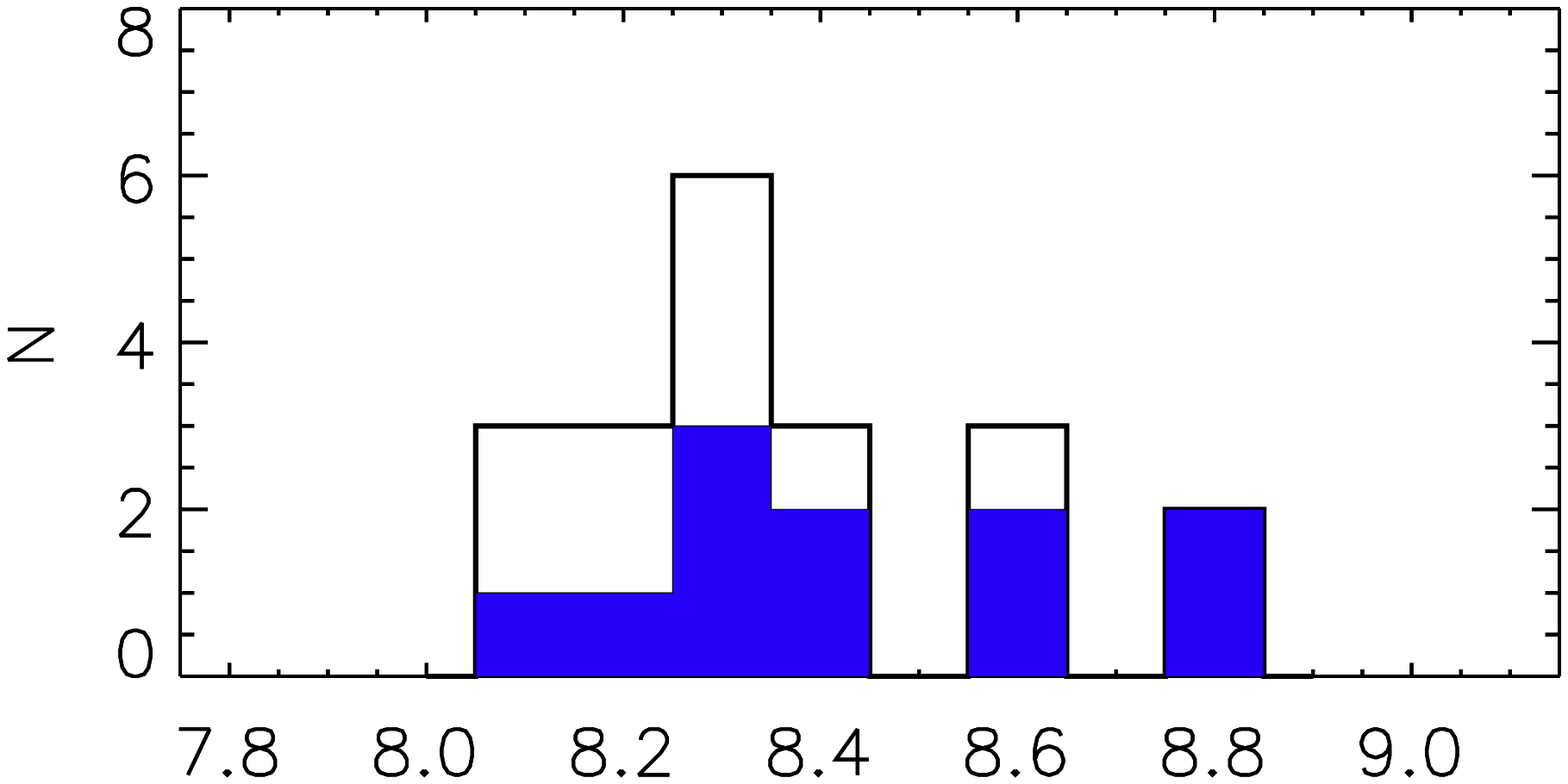}
\includegraphics[bb=10 -30 560 290,width=6.cm]{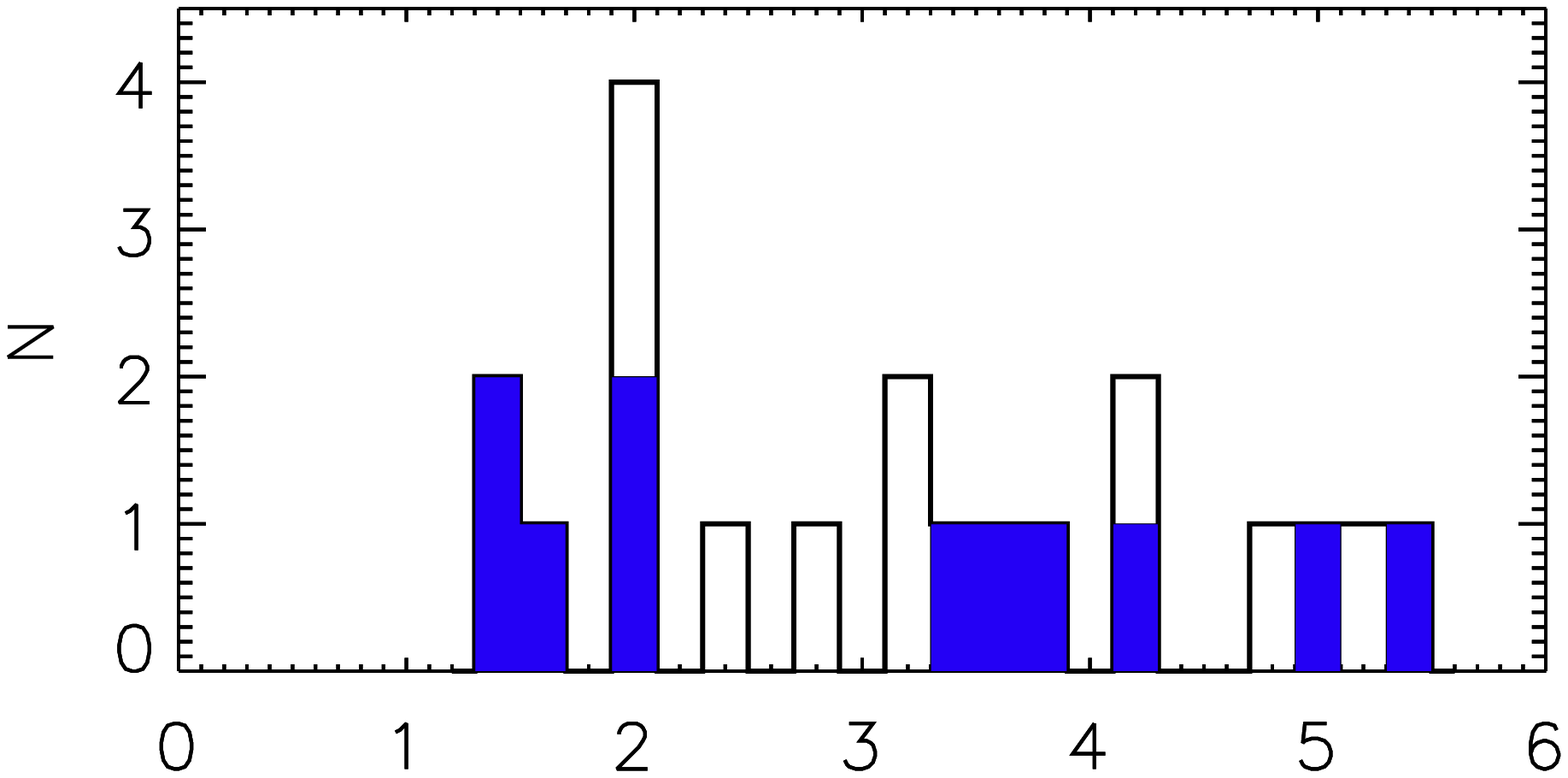}
\caption{Main properties of the Virgo SFDs observed in this work: stellar mass, SFR, \hi\ mass, \hi\-to-stellar mass ratio,
metallicity, and projected distance from either M87 or M49. Filled histograms show the CO-detected galaxies.}
   \label{fig:sample_properties}%
\end{figure*}

The conversion between CO intensity and \hdue\ abundance has been studied for more than two decades \citep[for a review, see][]{2013ARA&A..51..207B}.
The existence of a standard CO-to-H$_2$ mass conversion factor, $X_{CO}$, and its dependence on the physical conditions of the interstellar medium (ISM)
has been highly debated \citep{1997A&A...328..471I,2011ApJ...737...12L}:
metallicity, ionising stellar radiation field strength and the density of the gas
are among the main parameters that can affect the value of the $X_{CO}$ factor in galaxies
\citep{1995ApJ...448L..97W,1997A&A...328..471I,2002A&A...384...33B,2008ApJ...680..246T,2011ApJ...737...12L,2011MNRAS.415.3253S,2012AJ....143..138S,2013ApJ...777....5S}

Because large \hdue\ envelopes may be missed in the census
of the ISM gas components, cold dust (T $<$ 30 K) has recently been used as an indirect
tracer of the total gas content \citep[e.g.][]{2007A&A...471..103B,2011ApJ...737...12L,2015ApJ...799...96G}. Dust is observed to be well mixed with gas
\citep[e.g.][]{1978ApJ...224..132B,1996A&A...312..256B} and can be mapped by its emission at far-infrared (FIR) wavelengths.
\citet{2012A&A...542A..32C} showed that there is a linear relation between cold dust emission and CO brightness
in spiral galaxies down to $S_{CO}$ = 200 K \kms (or M$_{H_2} = 2 \times 10^8$ \msun, {assuming a galactic conversion factor}). However,
no metallicity dependence was investigated in their sample because of the lack of systems with abundances below the galactic value.

In this paper
we present and analyse $^{12}$CO(1-0) and $^{12}$CO(2-1) observations of SFDs in the Virgo cluster with the Institut de Radio Astronomie Millim\'etrique (IRAM) 30m telescope. These galaxies were detected at FIR/submillimetre (submm) wavelengths with {\em Herschel} \citep{2015A&A...574A.126G}, as part of
the \hers\ Virgo Cluster survey \citep[HeViCS;][]{2012MNRAS.419.3505D,2013MNRAS.428.1880A}.
Oxygen abundances span a range of
solar and sub-solar values with 12 + log(O/H) between 8.1 and 8.8 dex\footnote{Throughout this work we assume a solar oxygen
abundance of 12 + log(O/H) = 8.69 \citep{2009ARA&A..47..481A}.}.

Such a sample of dwarf galaxies is ideal to extend the dynamical range of the cold dust-CO emission correlation
established for Virgo spiral galaxies down to lower stellar masses
and sub-solar metallicities.
Moreover, using the CO lines to infer the molecular hydrogen masses,
we aim to shed light on the ISM of dwarf galaxies and on the relation between dust, atomic and molecular gas components, and on possible environmental dependencies.

HeViCS observations of higher mass galaxies combined with \hi\ and CO data have shown that dust removal in Virgo
\hinospace-deficient spiral
galaxies is lower than the amount of stripped neutral hydrogen, and that the molecular component,
well confined into the deep potential well, is affected even less by the dense environment
\citep{2010A&A...518L..49C,2011MNRAS.415.1797C,2012A&A...542A..32C}.
\citet{2014A&A...564A..67B} showed that galaxies in the core of the
Virgo cluster are modestly deficient in molecular gas,
in agreement with previous results based on a smaller sample of galaxies \citep{2006PASP..118..517B,2009ApJ...697.1811F}.
Highly \hi-deficient Virgo SFD galaxies are mostly characterised by reduced star formation activity and
lower dust fractions \citep{2015A&A...574A.126G}, therefore
we here wish to investigate how the
molecular component of dwarfs can be affected by the cluster environment.

The paper is organised as follows.
In Sect. \ref{sec:obs}
we describe IRAM observations and data reduction.
The analysis of the CO observations is discussed in Sect. \ref{sec:COanalysis}. The properties of the detected and non-detected
galaxies are compared in Sect. \ref{subsec:Conodet}. In Sect. \ref{sec:FIR-CO} we analyse the correlation
between FIR flux densities and integrated CO-line intensities.
We estimate the molecular masses in Sect. \ref{sec:h2masses}.
The molecular hydrogen properties of the Virgo SFDs are presented and discussed in Sect. \ref{sec:scaling_relations} and
include scaling relations of the \hdue\ component (Sect. \ref{subsec:h2frac});
the molecular-to-atomic gas mass ratios (Sect. \ref{subsec:h2tohi}), dust-to-gas ratios (Sect. \ref{sec:d2g}),
the molecular, atomic, and total gas mass star-formation laws (Sect. \ref{sec:sfr-law}), and
the environmental effects on the molecular gas (Sect. \ref{subsec:sfe}).
Finally, in Sect. \ref{sec:conclusions} we summarise our results and conclusions.

\begin{table*}
\centering
\caption{Main properties of the Virgo SFDs.
}
\begin{tabular}{lcclccSccrr}
\hline \hline
    ID   &  log ($M_{\star}$)    & log ($M_{HI})$  &  log ($M_d$)              &   log($SFR$)            &  12 + log(O/H)  &    \si{$\: \: Def_{HI}$}    & $\Delta V_{HI}$  &  D      & $a_{25}$ & $b_{25}$  \\ 
         &  [M$_{\odot}$]        & [M$_{\odot}$]   & $\: \: \:$[M$_{\odot}$]   & [M$_{\odot}$ yr$^{-1}$] &                 &                             & [km s$^{-1}$]    & [Mpc]   &  (\arsec)    & (\arsec)  \\
 \hline \hline
  VCC10 & 8.95 $\pm$ 0.04 & 8.74 $\pm$ 0.01 & 6.13$_{-0.06}^{+0.06}$           & -1.05 $\pm$ 0.08 & 8.56 $\pm$ 0.10 &            0.10 &  186   &  32.0  &  61.8  &   13.2   \\
  VCC87 & 8.39 $\pm$ 0.04 & 8.51 $\pm$ 0.01 & 5.91$_{-0.06}^{+0.07}$           & -1.62 $\pm$ 0.07 & 8.25 $\pm$ 0.10 &            0.17 &   99   &  17.0  &  87.0  &   43.2   \\
 VCC135 & 9.44 $\pm$ 0.04 & 7.19 $\pm$ 0.08 & 6.19$_{-0.06}^{+0.06}$           & -1.03 $\pm$ 0.08 & 8.65 $\pm$ 0.10 &            1.73 & 123\tablefootmark{(a)}    &  32.0  &  69.6  &   34.2   \\
 VCC144 & 8.81 $\pm$ 0.05 & 8.76 $\pm$ 0.01 & 5.70$_{-0.06}^{+0.06}$           & -0.27 $\pm$ 0.05 & 8.21 $\pm$ 0.10 &           -0.21 &   64   &  32.0  &  37.8  &   19.2   \\
 VCC172 & 8.88 $\pm$ 0.04 & 8.95 $\pm$ 0.01 & 6.04$_{-0.08}^{+0.08}$           & -1.45 $\pm$ 0.09 & 8.58 $\pm$ 0.10 &            0.01 &  126   &  32.0  &  75.6  &   33.6   \\
 VCC213 & 8.89 $\pm$ 0.04 & 7.84 $\pm$ 0.03 & 5.84$_{-0.06}^{+0.06}$           & -1.20 $\pm$ 0.06 & 8.77 $\pm$ 0.12 &            0.57 &  122   &  17.0  &  55.8  &   42.6   \\
 VCC324 & 8.72 $\pm$ 0.04 & 8.23 $\pm$ 0.01 & 5.50$_{-0.06}^{+0.06}$           & -0.75 $\pm$ 0.07 & 8.14 $\pm$ 0.10 &            0.40 & 45\tablefootmark{(a)}    &  17.0  &  81.0  &   69.0   \\
 VCC334 & 8.04 $\pm$ 0.04 & 7.95 $\pm$ 0.01 & 4.94$_{-0.07}^{+0.07}$           & -1.77 $\pm$ 0.17 & 8.22 $\pm$ 0.10 &            0.17 &   42   &  17.0  &  33.6  &   30.6   \\
 VCC340 & 9.11 $\pm$ 0.04 & 8.89 $\pm$ 0.01 & 6.05$_{-0.06}^{+0.06}$           & -0.84 $\pm$ 0.07 & 8.26 $\pm$ 0.10 &           -0.01 &   56   &  32.0  &  66.0  &   25.8   \\
 VCC562 & 7.76 $\pm$ 0.04 & 7.74 $\pm$ 0.03 & 5.00$_{-0.07}^{+0.07}$           & -1.74 $\pm$ 0.17 & 8.10 $\pm$ 0.10 &            0.44 &   42    &  17.0  &  37.8  &   29.4   \\
 VCC693 & 8.33 $\pm$ 0.04 & 8.27 $\pm$ 0.01 & 5.55$_{-0.06}^{+0.07}$           & -1.93 $\pm$ 0.22 & 8.43 $\pm$ 0.10 &            0.27 &   99   &  17.0  &  69.6  &   60.0   \\
 VCC699 & 9.19 $\pm$ 0.04 & 8.94 $\pm$ 0.01 & 6.26$_{-0.06}^{+0.06}$           & -0.63 $\pm$ 0.06 & 8.30 $\pm$ 0.10 &            0.08 &   85   &  23.0  & 117.0  &   82.8   \\
 VCC737 & 8.35 $\pm$ 0.04 & 8.66 $\pm$ 0.01 & 5.73$_{-0.07}^{+0.07}$           & -1.85 $\pm$ 0.17 & 8.28 $\pm$ 0.10 &           -0.17 &  164    &  17.0  &  64.2  &   21.0   \\
 VCC841 & 8.12 $\pm$ 0.04 & 7.68 $\pm$ 0.03 & 5.20$_{-0.07}^{+0.08}$           & -1.62 $\pm$ 0.07 & 8.33 $\pm$ 0.10 &            0.68 &   39   &  17.0  &  50.4  &   17.4   \\
VCC1437 & 8.52 $\pm$ 0.04 & 8.03 $\pm$ 0.02 & 5.23$_{-0.06}^{+0.06}$           & -1.78 $\pm$ 0.17 & 8.38 $\pm$ 0.10 &            0.11 &   57   &  17.0  &  35.4  &   27.0   \\
VCC1575 & 9.25 $\pm$ 0.04 & 7.97 $\pm$ 0.02 & 6.24$_{-0.06}^{+0.06}$           & -0.90 $\pm$ 0.10 & 8.76 $\pm$ 0.10 &            0.89 &   88   &  17.0  & 120.0  &   84.6   \\
VCC1686 & 9.07 $\pm$ 0.04 & 8.68 $\pm$ 0.01 & 6.44$_{-0.06}^{+0.06}$           & -0.89 $\pm$ 0.07 & 8.33 $\pm$ 0.15\tablefootmark{(b)} &    0.38 &  105   &  17.0  & 167.4  &  102.6   \\
VCC1699 & 8.57 $\pm$ 0.04 & 8.77 $\pm$ 0.01 & 5.46$_{-0.06}^{+0.07}$           & -1.12 $\pm$ 0.08 & 8.07 $\pm$ 0.12 &           -0.06 &   86   &  17.0  &  93.0  &   49.8   \\
VCC1725 & 8.59 $\pm$ 0.04 & 8.21 $\pm$ 0.01 & 5.78$_{-0.06}^{+0.07}$           & -1.36 $\pm$ 0.07 & 8.25 $\pm$ 0.10 &            0.50 &   76   &  17.0  &  93.0  &   58.2   \\
VCC1791 & 8.52 $\pm$ 0.04 & 8.72 $\pm$ 0.01 & 5.71$_{-0.06}^{+0.07}$           & -1.08 $\pm$ 0.05 & 8.16 $\pm$ 0.10 &           -0.11 &   93   &  17.0  &  77.4  &   38.4   \\
\hline \hline
\end{tabular}
\tablefoot{
\tablefoottext{a} \citep{2005ApJS..160..149S}
\tablefoottext{b} Metallicity derived from the mass-metallicity relation of \citet{2013A&A...550A.115H}.
}
\label{tab:01}
\end{table*}

\section{Sample selection, observations, and data reduction}
\label{sec:obs}

\subsection{Sample selection}

The sample of SFD galaxies observed at the IRAM 30m telescope was defined according to the following criteria: i) morphological classification
as blue compact dwarfs (BCD) or
Magellanic spirals and irregulars (Sm, Im), according to the Virgo Cluster Catalog \citep[VCC;][]{1985AJ.....90.1681B} and GOLDMine
\citep{2003A&A...400..451G,2014arXiv1401.8123G} and ii) detectable
FIR emission in four {\em Herschel} bands (100, 160, 250, and 350 $\mu$m) above 5$\sigma$. 
These criteria produced a selection of 23 galaxies.
We added the galaxy VCC172 to the sample because of its proximity to VCC144 both in projected distance ($\sim$ 1$\degree$)
and velocity ($\Delta V \sim 150$ \kms), although its detection at 350 $\mu$m is slightly below 5$\sigma$.
Only 20 targets were observed because of bad weather conditions during the observing
runs\footnote{Of the four non-observed galaxies, VCC1554 was detected at 3
mm by \citet{2014A&A...564A..65B} as part of the CO follow-up observations of the $Herschel$ Reference Survey
\citep[][]{2010PASP..122..261B}.}.
Table \ref{tab:01} displays the main properties of the observed sample as derived in
\citet{2015A&A...574A.126G}. In Fig. \ref{fig:sample_properties} we show
the range of stellar and \hi\ masses, star formation rates (SFRs), metallicities,
\hi-to-stellar mass ratios, and the projected distance from either M87 or M49, the elliptical galaxies at the core
of the two main substructures in Virgo, cluster A and cluster B \citep{1985AJ.....90.1681B}.
Filled histograms correspond to the detected galaxies in at least one CO line.
We briefly summarise below the methods followed to estimate these parameters, but we
refer to \citet{2015A&A...574A.126G} for further details.

Stellar masses were calculated following the approach of \citet{2013MNRAS.433.2946W}, which is based on 3.4 $\mu$m photometry  with the
WISE telescope \citep{2010AJ....140.1868W}.
Atomic hydrogen (\hi) masses were derived from the Arecibo Legacy Fast ALFA (ALFALFA) blind
\hi\ survey \citep{2005AJ....130.2598G}, using the $\alpha$.40 catalogue release \citep{2011AJ....142..170H}.
The \hi\ deficiency parameter, $Def_{HI}$\footnote{The \hi\ deficiency is defined as the logarithmic difference between the \hi\ mass of a reference sample of isolated galaxies
for a given morphological type and the observed \hi\ mass, $Def_{HI}$ = log M$_{HI}^{ref}$ - log $M_{HI}^{obs}$} \citep{1984AJ.....89..758H}, was estimated following
\citet{2013A&A...553A..89G}.
Star-formation rates were calculated from H$\alpha$
photometry obtained from the GOLDMine data base \citep{2003A&A...400..451G,2014arXiv1401.8123G}.
Mid-infrared photometry at 22 $\mu$m from WISE archive images was measured to take into account the contribution of dust-obscured
star formation using the relation of \citet{2014MNRAS.438...97W}.
For those galaxies without a WISE detection,
we calculated the SFR from the H$\alpha$ fluxes alone, using \citet{1998ARA&A..36..189K} for a \citet{2001MNRAS.322..231K} IMF,
after correcting the H$\alpha$ fluxes for internal extinction from the Balmer decrement measured in the Sloan Digital Sky Survey (SDSS) spectra \citep[][]{2000AJ....120.1579Y}. 
Oxygen abundances were derived from the SDSS based with the method of
\citet{2013A&A...550A.115H}. The method combines five metallicity calibrations and converts them into a base metallicity,
that is, the O3N2 = [O{\sc iii}]$\lambda$5007/[N{\sc ii}]$\lambda$6584 index of \citet{2004MNRAS.348L..59P}. However,
because the [O{\sc ii}]$\lambda$3727 line is out of the measured wavelength range of the SDSS at the distance of Virgo,
the only applicable calibrations for our sample were those based on the N2 = [N{\sc ii}]$\lambda$6584/H$\alpha$ and O3N2 indices
of \citet{2004MNRAS.348L..59P}.
The oxygen abundance of VCC1686 was derived
from the mass metallicity relation of \citet{2013A&A...550A.115H} because of the lack of SDSS spectroscopical observations.
Dust masses were calculated by fitting a modified black-body (MBB) function to the FIR spectral energy distribution (SED) between 100 and 350 $\mu$m
assuming a fixed emissivity index $\beta = 1.5$ and $\kappa_0$ = 3.4 cm$^2$ g$^{-1}$ at $\lambda$ = 250 $\mu$m, following the prescription
of \citet{2013A&A...552A..89B}. The emissivity index $\beta = 1.5$ was found to best fit the FIR SEDs minimising the fraction
of residuals in the four Herschel bands among a range of five adopted values
(1.0,1.2,1.5,1.8,2.0; see Grossi et al. 2015).
Isophotal major ($a_{25}$) and minor diameters ($b_{25}$) at the 25th $B$-magnitude
arcsec$^{-2}$ and distances to the galaxies were extracted from the GOLDMine data base.

\begin{figure*}
  \centering 
\includegraphics[bb=180 1510 3380 4020,width=14cm]{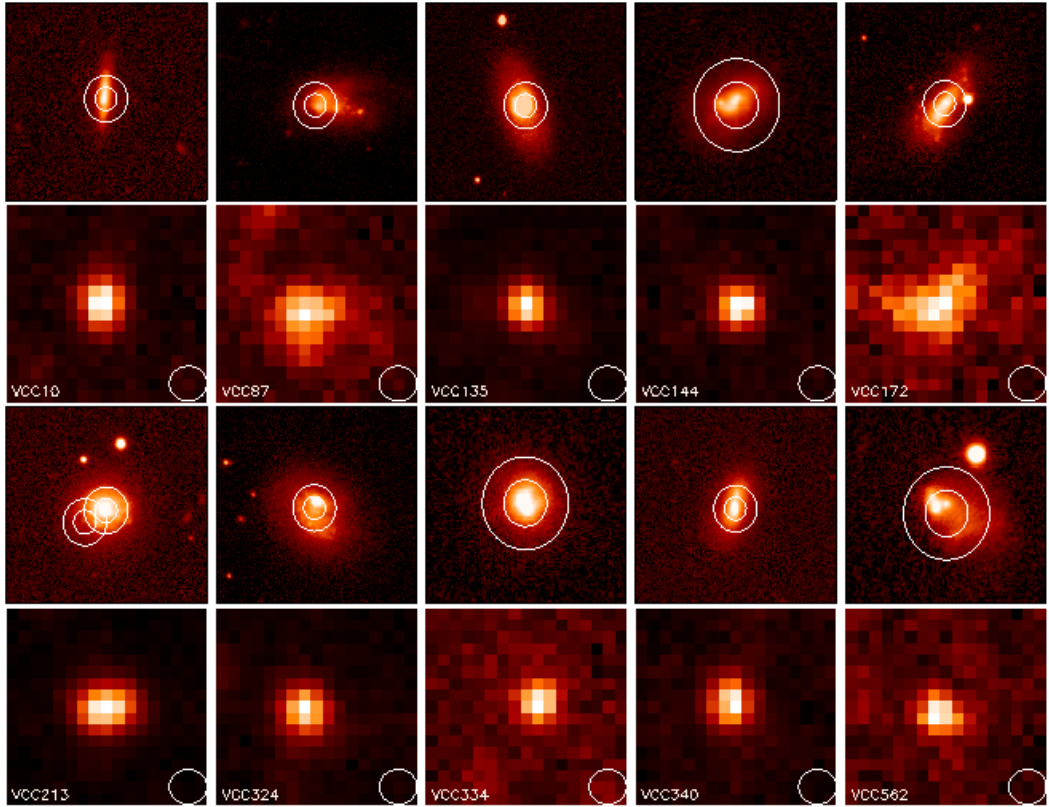}
\caption{Sample of Virgo SFDs observed at IRAM: the 30m telescope pointings are overlaid on the SDSS $g$ images (first and third rows).
The larger and smaller circles indicate the beam sizes at 115 GHz and 230 GHz, respectively.
{\em Herschel}/SPIRE 250 $\mu$m images
are also displayed for each target (second and fourth rows) with the 250 $\mu$m beam size shown in the
bottom right corner.}
   \label{fig:sample_images_a}%
\end{figure*}

\subsection{IRAM observations}

CO(1-0) and CO(2-1) observations with the IRAM 30m telescope (Pico Veleta, Spain) were obtained over the course of three observing runs:
2011 December 9-10, 2012 May 6-7, and 2012 29 June - 2 July.
The telescope full width half maximum (FWHM) is 21\farcs3 and 10\farcs7 at 115.27 GHz  and 230.54 GHz, respectively. This
corresponds to $\sim$1.8 kpc and $\sim$0.9 kpc at a distance of 17 Mpc.
The 30m telescope beam size at 115 GHz is comparable to the {\em Herschel}/SPIRE resolution at 250 \micron\ (18\arsecnospace).

Most of our targets were observed with
a single pointing because their optical extension and the size of the FIR-emitting region are comparable to
the IRAM beam at 115 GHz
(Fig. \ref{fig:sample_images_a}).
However, to obtain a more complete coverage
of the molecular gas distribution, four galaxies were observed with more than
one pointing with a spacing of half the beam size
at 115 GHz  (VCC213, VCC699, VCC1575, VCC1686; see Fig. \ref{fig:sample_images_a}).
Total integration times per target varied between 10 and 130 minutes (see Table \ref{tab:point_list}).

Observations were performed
in reasonable weather conditions
($\tau \textrm{(230 GHz)} \sim$ 0.07-0.42) using the Eight MIxer
Receiver ({\em EMIR\footnote{http://www.iram.es/IRAMES/mainWiki/EmirforAstronomers
}}) in bands E090 and E230 to simultaneously detect CO(1-0) and CO(2-1) lines.
However, during the third run, we used only the E090 band, hence for seven dwarf galaxies we only have CO(1-0)
line observations.
To increase the redundancy, Fast Fourier Transform Spectrometer (FTS) and WIdeband Line Multiple Autocorrelator (WILMA)
were used in parallel to sample the data.
FTS provides a bandwidth of 8 GHz in each of the two orthogonal linear polarisations
and a channel resolution of 200 kHz.
The bandwidth of WILMA autocorrelator is 1 GHz with a spectral resolution of 2 MHz.
The wobbler-switching mode, with a 120\arsec throw, was preferred given the small sizes of the targets.
Pointing and focus were checked on nearby strong continuum sources and Mars and were
monitored every two hours in stable conditions and every one hour during
sunrise. The pointing accuracy was $\lesssim$ 3\arsec and the system temperatures varied
between $\sim$280 and $\sim$700 K at 115 GHz on the antenna temperatures (T$_A^*$) scale.
To convert $T_A^*$ into main beam temperatures ($T_{mb}$), we
used $T_{mb} = T_a^* \times F_{eff}/B_{eff}$,
with forward efficiencies ($F_{eff}$) of 0.94 and 0.91 and main beam efficiencies ($B_{eff}$)
of 0.78 and 0.58, at 115 and 230 GHz, respectively.
Conversion between main beam temperature in Kelvin and flux in Jansky ($S_{\nu}$) for a point-like source were made using the
relation\footnote{Kramer 1997, IRAM report on Calibration of 30m data}

\begin{equation}
G = \frac{S_{\nu}}{T_{mb}} = 8.16 \times 10^{-7} \left( \frac{\nu}{\textrm{GHz}} \right)^2 \left( \frac{\theta_{beam}}{\textrm{arcsec}} \right)^2
\textrm{[Jy/K]}
\label{eq:gain}
,\end{equation}

\noindent where $\theta_{beam}$ is the FWHM of the beam at the observing frequency $\nu$, corresponding to a
gain factor $G =$ 4.94 Jy/K.

\begin{figure*}
\ContinuedFloat
  \centering
\includegraphics[bb=200 1550 3370 4040,width=14cm]{pic_second10.eps}
\caption{Sample of Virgo SFDs observed at IRAM: continued.}
\end{figure*}

\subsection{Data reduction}

The data were reduced with the GILDAS-CLASS90\footnote{http://www.iram.fr/IRAMFR/GILDAS/} software package \citep{2005sf2a.conf..721P}.
Scans were averaged
after fitting polynomial baselines, and
they were boxcar-smoothed to a resolution of $\delta V$ = 8.1 \kms.
The rms noise level of the final spectra ranges between 3 mK  and 11 mK in both bands (Table \ref{tab:COfluxes}).

Integrated intensities, $I_{CO}$, were calculated in two ways: i) we fitted a Gauss function (single-peak profiles) or a linear combination of the first two Hermite functions \citep[double-peak profiles;][]{2007AJ....133.2087S}
to the spectra, and
ii) we summed the spectra over the observed velocity range.
Both methods give consistent values, which we display in Table \ref{tab:COfluxes}.
We assumed that a galaxy was detected in one beam when the peak emission was above 3$\sigma$ or when the integrated emission over the defined velocity window was above 4$\sigma$.
Of a total of 20 galaxies, we detected the CO(1-0) line in
nine sources and the CO(2-1) line in ten targets.
VCC10 and VCC144 were only detected at 230 GHz, while VCC1725 was marginally detected only at 115 GHz.
Upper limits for the undetected galaxies were derived using

\begin{equation}
I_{CO}^{up} = 4 \sigma \sqrt{\Delta V_{up} \delta V}
\label{eq:upper_limits}
,\end{equation}

\noindent where $\delta V$ is the spectral resolution and $\Delta V_{up}$ is the assumed line width of the galaxy. When no CO lines were detected, $\Delta V_{up}$ corresponds to the \hi\
line FWHM given in Table \ref{tab:01}, and when only one line was detected, we used the measured FWHM to calculate the upper limit to
the intensity of the other
undetected line (e.g. VCC10, VCC144, VCC1575(e), VCC1686(c), VCC1725).
The spectra of all the observed galaxies are displayed in Figs. \ref{fig:CO_spec} and \ref{fig:CO21_spec}.

In the rest of this work we present the results derived from the sum of the spectra over the observed velocity width.
The errors on the integrated intensities
are determined according to

\begin{equation}
\Delta I_{CO} = \sigma \sqrt{\Delta V_{CO} \delta V_{CO}}
,\end{equation}

\noindent where $\sigma$ is the rms noise of the spectrum, $\Delta V_{CO}$ the CO
line width, and $\delta V_{CO}$ the spectral resolution.

\small
\begin{table*}
\centering
\caption{CO line intensities, systemic velocities, line widths, and noise level at 115 and 230 GHz.}
\begin{tabular}{lccccccccc}
\hline \hline
ID & \tablefootmark{(a)}$I_{CO,w}^{(1-0)}$ &  \tablefootmark{(b)}$I_{CO,g}^{(1-0)}$ &  \tablefootmark{(a)}$I_{CO,w}^{(2-1)}$    &  \tablefootmark{(b)}$I_{CO,g}^{(2-1)}$        &   \tablefootmark{(c)}$V_{sys,g}$         &  $\Delta V^{(1-0)}_g$  &  $\Delta V^{(2-1)}_g$  &  \tablefootmark{(d)}$\sigma^{(1-0)\,}$ & \tablefootmark{(d)}$\sigma^{(2-1)\,}$ \\
   &  [\Kkms]           & [\Kkms]             &   [\Kkms]              &    [\Kkms]                 & [km/s]               &  [km/s]               &   [km/s]           & [mK]  & [mK]                           \\
\hline \hline
VCC10        &       $<$ 0.41         &    $<$ 0.41            &     0.99$\pm$0.13    &   1.11$\pm$0.19   & $\:\:\:$--$\:\:\:$;~1938                  &     --           &  59     &   4.7   &  5.0   \\
VCC87        &       $<$ 0.42         &    $<$ 0.42            &          --          &   --              &  -157                                     &     --           &  --     &   3.7   &  --    \\
VCC135       &       1.71$\pm$0.30    &    1.84$\pm$0.34       &     5.25$\pm$0.40    &   5.37$\pm$0.47   &  2409;~2406                               &     112          & 112     &   9.4   &  10.8  \\
VCC144       &       $<$ 0.39         &    $<$ 0.39            &     1.11$\pm$0.12    &   1.13$\pm$0.14   & $\:\:\:$--$\:\:\:$;~2026                  &     --           &  69     &   4.1   &  4.0   \\
VCC172       &       $<$ 1.06         &    $<$ 1.06            &     $<$ 1.26         &   $<$ 1.26        &  2175                                     &     --           &  --     &   8.3   &  9.9   \\
VCC213(a)    &       3.57$\pm$0.25    &    3.44$\pm$0.26       &     3.08$\pm$0.33    &   3.66$\pm$0.31   &  -164;~-167                               &     80           &  69     &  10.3   & 11.3   \\
VCC213(b)    &       2.45$\pm$0.19    &    2.64$\pm$0.23       &     3.64$\pm$0.21    &   3.54$\pm$0.25   &  -131;~-125                               &     44           &  29     &   6.0   &  9.4   \\
VCC324       &       0.51$\pm$0.11    &    0.47$\pm$0.16       &     0.74$\pm$0.09    &   0.69$\pm$0.08   &  1540;~1530                               &     42           &  26     &   2.8\tablefootmark{(e)}   &  3.3   \\
VCC334       &       $<$ 0.22         &    $<$ 0.22            &         --           &    --             &  -240                                     &     --           &  --     &   3.0   &  --    \\
VCC340       &       0.58$\pm$0.13    &    0.62$\pm$0.16       &     0.90$\pm$0.11    &   0.81$\pm$0.08   &  1519;~1513                               &     46           &  32     &   4.0   &  3.5   \\
VCC562       &       $<$ 0.80         &    $<$ 0.80            &         --           &    --             &    44                                     &     --           &  --     &  10.9   &  --    \\
VCC693       &       $<$ 0.58         &    $<$ 0.58            &     $<$ 0.55         &   $<$ 0.55        &  2048                                     &     --           &  --     &   5.1   &  4.9   \\
VCC699(a)    &       0.66$\pm$0.13    &    0.80$\pm$0.13       &     0.65$\pm$0.09    &   0.63$\pm$0.09   &   720;~711                                &     69           &  29     &   4.4   &  4.0   \\
VCC699(b)    &       0.65$\pm$0.13    &    0.64$\pm$0.23       &     0.60$\pm$0.11    &   0.55$\pm$0.15   &   743;~734                                &     48           &  17     &   6.2   &  7.5   \\
VCC737       &       $<$ 0.39         &    $<$ 0.39            &         --           &    --             &  1725                                     &     --           &  --     &   2.7   &  --    \\
VCC841       &       $<$ 0.31         &    $<$ 0.31            &         --           &    --             &   491                                     &     --           &  --     &   4.4   &  --    \\
VCC1437      &       1.25$\pm$0.13    &    1.29$\pm$0.17       &     2.62$\pm$0.12    &  2.74$\pm$0.11    &  1157;~1160                               &     56           &  51     &   4.8   &   3.8  \\
VCC1575(a)   &       3.21$\pm$0.22    &    3.33$\pm$0.23       &     5.54$\pm$0.26    &  5.26$\pm$0.27    &   577;~579                                &     39           &  32     &   7.7   &  10.6  \\
VCC1575(b)   &       4.48$\pm$0.23    &    4.51$\pm$0.27       &     6.95$\pm$0.30    &  6.83$\pm$0.34    &   583;~583                                &     48           &  43     &   8.5   &  11.4  \\
VCC1575(c)   &       4.83$\pm$0.25    &    5.11$\pm$0.25       &     3.93$\pm$0.21    &  3.80$\pm$0.22    &   608;~610                                &     50           &  32     &   6.4   &   9.5  \\
VCC1575(d)   &       4.41$\pm$0.19    &    4.38$\pm$0.22       &     6.38$\pm$0.23    &  6.66$\pm$0.24    &   621;~616                                &     42           &  42     &   7.6   &   9.6  \\
VCC1575(e)   &       1.20$\pm$0.20    &    0.91$\pm$0.16       &     $<$ 0.50         &  $<$ 0.50         &   574;$\:\:$--$\:\:\:$                    &     20           &  --     &   7.6   &   9.7  \\
VCC1686(a)   &       0.74$\pm$0.12    &    0.90$\pm$0.13       &     0.81$\pm$0.11    &  0.83$\pm$0.14    &  1123;~1121                               &     37           &  28     &   5.8   &   6.4  \\
VCC1686(b)   &       $<$ 0.90         &    $<$ 0.90            &     $<$ 0.96         &  $<$ 0.96         &  $\:\:\:$--$\:\:\:$;$\:\:\:$--$\:\:\:$    &     --           &  --     &   9.5   &  10.0  \\
VCC1686(c)   &       1.20$\pm$0.30    &    1.31$\pm$0.41       &     $<$ 1.06         &  $<$ 1.06         &  1130;$\:\:\:$--$\:\:\:$                  &     70           &  --     &   6.3\tablefootmark{(e)}   &  11.1 \\
VCC1699      &       $<$ 0.30         &    $<$ 0.30            &         --           &    --             &  1632                                     &     --           &  --     &   2.8   &  --    \\
VCC1725      &      0.41 $\pm$ 0.12   &    0.46$\pm$0.13       &     $<$ 0.34         &  $<$ 0.34         &  1049;$\:\:\:$--$\:\:\:$                  &     28           &  --     &   4.5   &   5.6  \\
VCC1791      &       $<$ 0.46         &    $<$ 0.46            &         --           &    --             &  2075                                     &     --           &  --     &   4.2   &  --    \\
\hline \hline
\end{tabular}
\tablefoot{
\tablefoottext{a}{Measured summing of the spectrum over the observed velocity width.}
\tablefoottext{b}{Measured by fitting a Gaussian function or a linear combination of the first two symmetric Hermite functions.}
\tablefoottext{c}{Systemic velocity derived from the fit of the CO(1-0) and CO(2-0) lines. For non-detections we display only the systemic velocities determined from the SDSS optical spectra.}
\tablefoottext{d}{Noise level at a spectral resolution of 8.1 \kms.}
\tablefoottext{e}{Noise level measured at a spectral resolution of 16.2 \kms.}
}
\label{tab:COfluxes}
\end{table*}
\normalsize

\section{Analysis of CO observations}
\label{sec:COanalysis}

The majority of the galaxies in the sample were observed with
a single pointing in their central position.
Consequently,  we may expect two types of situations
when estimating the total CO brightness of the galaxy: 1) the telescope beam is smaller than the size of the CO-emitting
area and corrections for  incomplete coverage should be applied,
or 2) the CO emission originates from a region smaller than the telescope beam, and we therefore need to determine its size and the corresponding correction factor after assuming the source geometry
(usually a uniform disc or Gaussian.)
We  discuss the strategy to account for these effects below.

\subsection{Aperture correction for extended emission}
\label{sec:ap_cor_extended}

\subsubsection{Single-pointing observations}

To determine the aperture correction for an emitting region more extended than the beam,
we followed two approaches: i) an empirical  method   calibrated on
nearby mapped galaxies \citep{2011MNRAS.415...32S}, and ii) an analytic method that assumes an exponential radial profile for the
distribution of the molecular gas \citep{2011A&A...534A.102L}.

The first
method was proposed by \citet{2011MNRAS.415...32S}, who simulated the effects of a single-beam observation on a sample of galaxies
with IRAM 30m telescope CO maps \citep{2007PASJ...59..117K}, calculated the ratio between the total flux and the flux observed by
a $\sim$22\arsec\ beam, and obtained an empirical calibration. With this technique,
the aperture correction, $f_{ap}$, defined
as the ratio between the total (extrapolated)
CO flux ($S_{CO}^{tot}$) and the flux in the central pointing
($S_{CO}^{central}$)

\begin{equation}
f_{ap} = S_{CO}^{tot}/S_{CO}^{central}
,\end{equation}

\noindent is given by the following relation:

\begin{equation}
S_{CO}^{tot} = \frac{S_{CO}^{central}}{1.094 - 0.008 D_{25} + 2 \times 10^{-5} D_{25}^2}
,\end{equation}

\noindent where $D_{25}$ is the diameter at the 25th $B$-magnitude isophote in arcseconds,
$S_{CO}^{central}$ is the measured flux that is related to the integrated line intensity by $S_{CO}^{central} = G \times I_{CO}^{central}$,
and $G$ is the gain of the telescope.

The second method, introduced by \citet{2011A&A...534A.102L}, provides a more analytic approach
to derive the aperture correction.
The starting assumption is that the radial surface distribution of the CO emission $S_{CO}(r)$
follows an exponential law with a scale length
$r_e$, which is related to the radius at the 25th $B$-magnitude isophote, $r_{25}$, according
to $r_e = 0.2 r_{25}$ \citep{2008AJ....136.2782L,2011A&A...534A.102L}.
The total CO flux, $S_{CO}^{tot}$, is calculated by spatially integrating the CO exponential profile \citep[see Eq. 4 in ][]{2011A&A...534A.102L}.
$S_{CO}^{central}$ is derived by convolving the CO exponential distribution with a Gaussian beam
\citep[see Eq. 5 in ][]{2011A&A...534A.102L}. The aperture correction, $f_{ap}$,
in this case depends on
the ratio of the scale length and the beam size, $r_e$/$\theta_{beam}$, as well as
on the galaxy inclination $i$ \citep{2011A&A...534A.102L}.

We calculated the aperture corrections for both methods
and obtained comparable values ranging between 1.2 and 4.4. The largest discrepancies
are found in the most extended objects of the sample (VCC1575, VCC1686, VCC1699, VCC1725: see Table \ref{tab:apcorr}).
Comparing the extrapolation of a single-beam
observations to complete CO maps, \citet{2014A&A...564A..66B} concluded that the analytic prescription of Lisenfeld et al. is more appropriate than the empirical relation of Saintonge et al. (2011). Moreover, the COLD GASS sample consisted of massive ($> 10^{10}$ \msun) and metal-rich galaxies, fairly different from our sample of low-mass dwarfs with intermediate metallicities. Therefore in the rest of this work we adopt
the method of \citet{2011A&A...534A.102L} to correct for extended emission, although
this choice does not significantly affect the estimate of the final \hdue\ masses. \footnote{We note that the single
CO pointing is offset from the centre of the galaxy in three objects where we chose to target the peaks of the FIR emission.  Two galaxies were not detected (VCC1791 and VCC1699), and one (VCC1725)
is a tentative detection. Applying these techniques, which are calibrated for a central pointing, means that we are obtaining more conservative upper limits for VCC1791 and VCC1699, and that we are slightly overestimating the total CO intensity of VCC1725.}

\subsubsection{Multiple-pointing observations}

For galaxies with more than one pointed observation, hereafter extended sources, we  computed the total
CO flux by fitting the radial surface distribution of the emission.  We assumed that the emitting region fills the beam and that
the brightness decreases exponentially with radius \citep{2009AJ....137.4670L}.
At each observed position we convolved the modelled brightness distribution with the telescope beam and integrated over the
beam extent to derive the predicted line intensity.
We compared the modelled to the observed flux and determined the exponential scale length that best fitted the data using a $\chi^2$ test.
We followed this procedure for both the $J$=1-0 and $J$=2-1 lines.
A flat brightness
distribution provided the best fit to the $J$=2-1 emission of VCC213 and to the $J$=1-0 emission of VCC699. In these cases
we only integrated  out to 33~arcsec (equivalent to 1.5 times the FWHM of the 115 GHz beam) to determine the total CO flux.
We compared these results to the method described in the previous subsection and obtained similar results within the uncertainties.

\subsection{Correction for a source size smaller than the beam}
\label{sec:ap_cor_beam_dilution}

The underlying assumption in the calculation of the aperture corrections derived in Sect. \ref{sec:ap_cor_extended} is that
the emission fills the beam.
If the angular size of a source, $\Omega_s$, is smaller than the beam size, $\Omega_{beam}$,  the main beam
temperature, T$_{mb}$, will be lower than the line brightness temperature, T$_B$, by a factor $\Omega_s/\Omega_{beam}$. In this case the observed integrated
line intensity, $I_{CO} = \int T_{mb}dv$, measures the beam diluted brightness temperature, while the intrinsic
CO line intensity is the source brightness
temperature integrated over the velocity width.
Here we describe a method to estimate the size of the emitting region
based on the comparison between
the observed main beam temperature line ratio, $I_{2-1}$/$I_{1-0}$, and the intrinsic brightness temperature ratio
$R_{21} \equiv$ T$_B^{2-1}$/T$_B^{1-0}$. We assume that the source geometry is described by either a uniform disc or a gaussian,
and we determine the corresponding corrections factors.

\subsubsection{Estimate of source sizes}
\label{sec:sizes}

For a uniformly bright disc of radius $R_u$ and angular size $\Omega_{u}$ =$\pi R_u^2$,
the main beam temperature to brightness temperature ratio is given by \citep{1961ApJ...133..322H,2011A&A...528A.116C}

\begin{equation}
{T_{mb}\over  T_B} = 1-e^{-( R_u/0.6 \theta_{beam})^2}= 1-e^{-\Omega_{u}/\Omega_{beam}},
\label{eq:unif_source}
\end{equation}

\noindent  where $\Omega_{beam}$=1.13 $\theta_{beam}^2$ is the effective solid angle
for a Gaussian beam of FWHM $\theta_{beam}$.
For a Gaussian source of angular size $\Omega_g = 1.13 \, \theta_g^2$ (and FWHM $\theta_g$),
this ratio can be written as \citep{1993ApJ...417..305B}

\begin{equation}
{T_{mb}\over  T_B} = {\Omega_g \over \Omega_g + \Omega_{beam}}.
\label{eq:gauss_source}
\end{equation}

We can then use the observed main beam temperature line  ratio, I$_{2-1}$/I$_{1-0}$, to estimate the size of the emitting region.
For a uniform and a Gaussian brightness distribution we have

\begin{equation}
{I_{2-1}\over I_{1-0}}= R_{21} { 1- e^{-\Omega_u/\Omega_{2-1}} \over 1- e^{-\Omega_u/\Omega_{1-0}}}
\label{eq:line_rat_unif}
\end{equation}

\begin{equation}
{I_{2-1}\over I_{1-0}}= R_{21} { \Omega_g+\Omega_{1-0} \over  \Omega_g + \Omega_{2-1} },
\label{eq:line_rat_gauss}
\end{equation}

\noindent where we assumed $R_{21}$= 0.8, the intrinsic line ratio  typically observed in nearby resolved galaxies \citep{2009AJ....137.4670L}.
Because $\Omega_{beam}^{1-0}/\Omega_{beam}^{2-1}=4$ for the IRAM 30m-telescope,  ${I_{2-1}\over I_{1-0}}$
can be up to a factor 4 higher than the brightness temperature line ratio (if $\Omega_u$, $\Omega_g \ll \Omega_{beam}$), while
for an emitting region larger than the beam the two ratios will be equal (${I_{2-1}\over I_{1-0}} \simeq R_{21}$)

When both lines were detected, we derived a range of size estimates from Eqs. \ref{eq:line_rat_unif} and \ref{eq:line_rat_gauss}, considering
that the observed integrated intensities
of the two lines vary between
$I_{CO}-\delta I_{CO}$ and $I_{CO}+\delta I_{CO}$,
where $\delta I_{CO}$ is the measured uncertainty.
Then we took the mean value of the upper and lower estimates of both cases.
When only CO(2-1) was detected (as in VCC10 and VCC144), we only set upper limits to the source size.
In this case we considered the whole range of sizes compatible with the observations (from point-like to the maximum value defined by the upper limit)
and took the mean of the derived estimates.
The estimated sizes for a uniform and a Gaussian brightness distribution ($2R_u$, $\theta_g$) are displayed in Table \ref{tab:apcorr}.

\subsubsection{Correction factors}
\label{size_factors}

When the sizes were known, we determined the corresponding correction factors to the flux, $f_s^{u,g}$,
which account for the non-negligible extension of the source compared to
the beam FWHM \citep{1961ApJ...133..322H}.
For a uniformly bright disc ($\Omega_u$) and a Gaussian source ($\Omega_g$),
we define using Eqs. \ref{eq:unif_source} and \ref{eq:gauss_source}

\begin{equation}
f_{s}^u =  \frac{\left( \frac{\Omega_u}{\Omega_{beam}} \right)}{\left( 1 - e^{-\left(\Omega_u/\Omega_{beam} \right)}\right)}
\end{equation}

\begin{equation}
f_{s}^g = \left( 1 + \frac{\Omega_g}{\Omega_{beam}} \right)
,\end{equation}

\noindent where $\Omega_{beam}$ corresponds
to the IRAM 30m telescope beam at 115 GHz.
For point-like sources ($\Omega_u$, $\Omega_g \ll \Omega_{beam}$) the correction
is negligible ($f_{s}^u = f_s^g \rightarrow 1$), while
the highest values
are obtained for sizes comparable to the beam. The correction factors $f_{s}^u$, $f_s^g$ are given in Table \ref{tab:apcorr}.

\subsection{Final CO fluxes}
\label{sec:final_CO}

After assessing the different correction factors
that take into account the finite source size ($S^{tot}_{s} = f_{s}^{u,g} G I_{CO}^{beam}$) and that the emitting region can be more 
extended than the beam ($S^{tot}_{Ext} = f_{ap} G I_{CO}^{beam}$),
the total CO fluxes were computed as the average between the minimum and maximum
flux obtained for all possible sizes. The difference between the mean and extreme values defines the one-sigma uncertainty. For VCC135 we considered only the correction
for source size smaller than the beam because of its small extent.
When no lines were detected,
we converted the upper limit to the main beam temperature, estimated with Eq. \ref{eq:upper_limits}, to a flux upper limit by multiplying it by the gain
and by the aperture correction for extended regions
derived in Sect. \ref{sec:ap_cor_extended} (see Table \ref{tab:apcorr}).

\begin{figure*}
  \centering
\includegraphics[bb=50 -10 570 550,width=6cm]{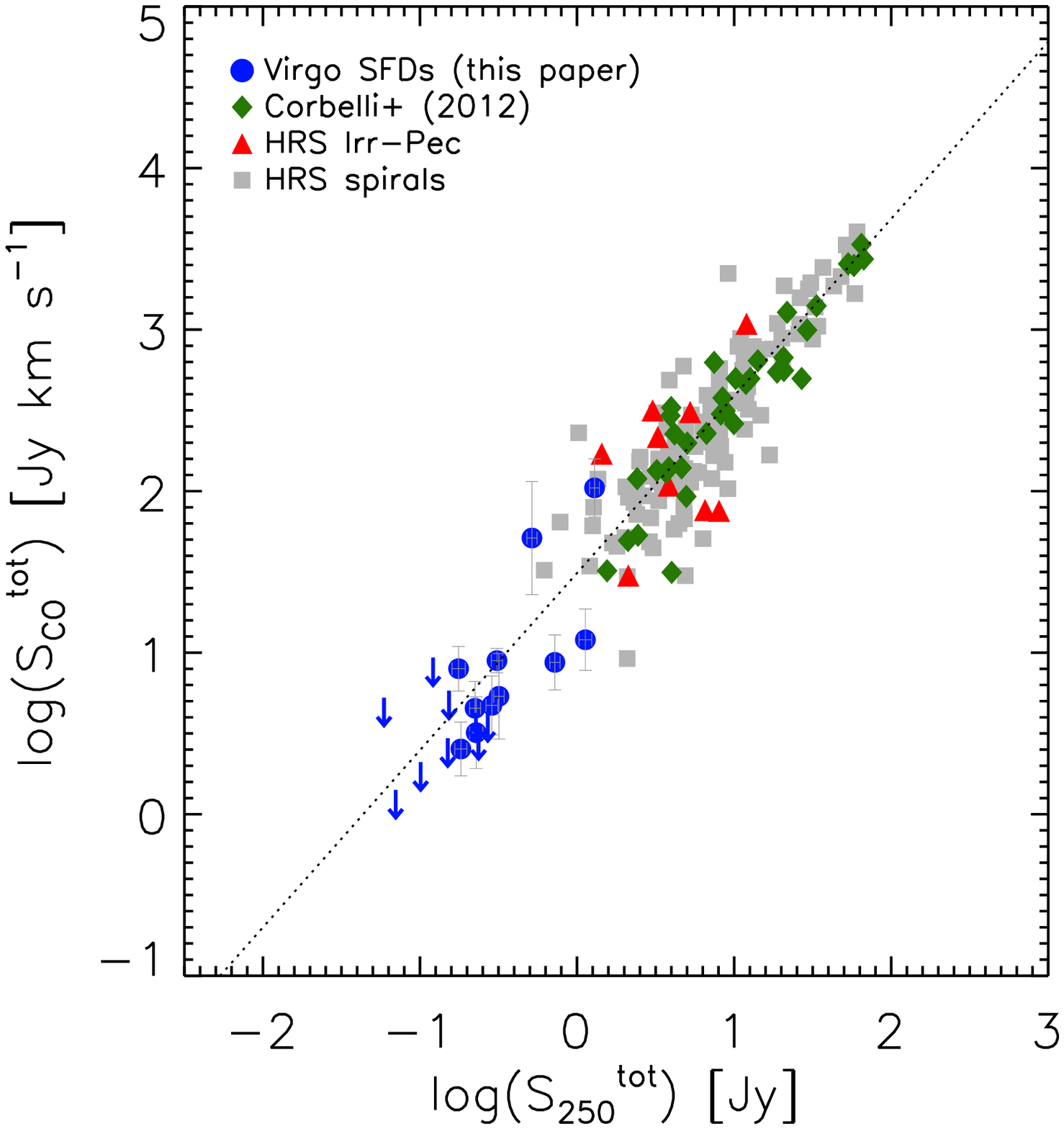}
\includegraphics[bb=50 -10 570 550,width=6cm]{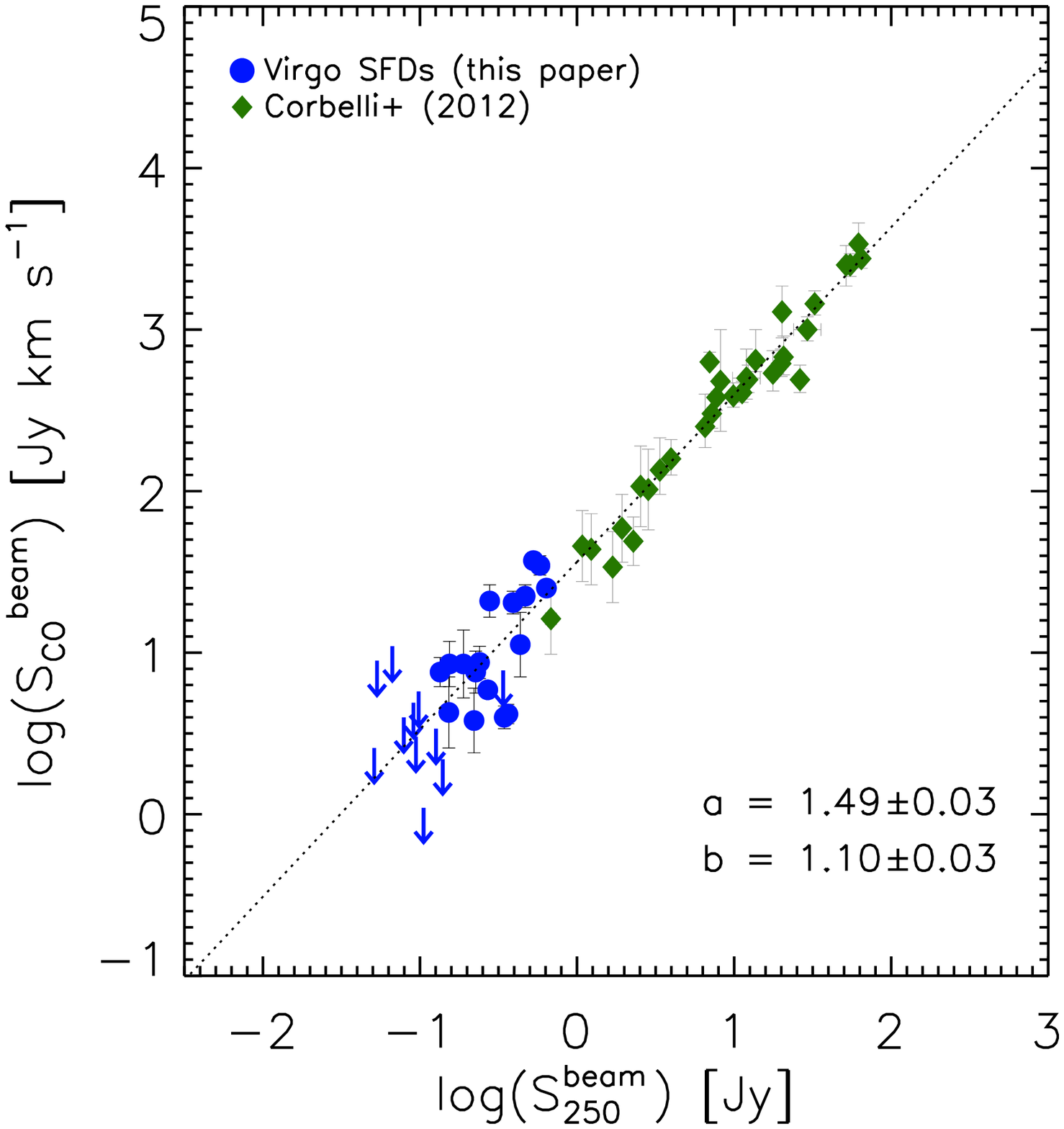}
\includegraphics[bb=50 -10 570 550,width=6cm]{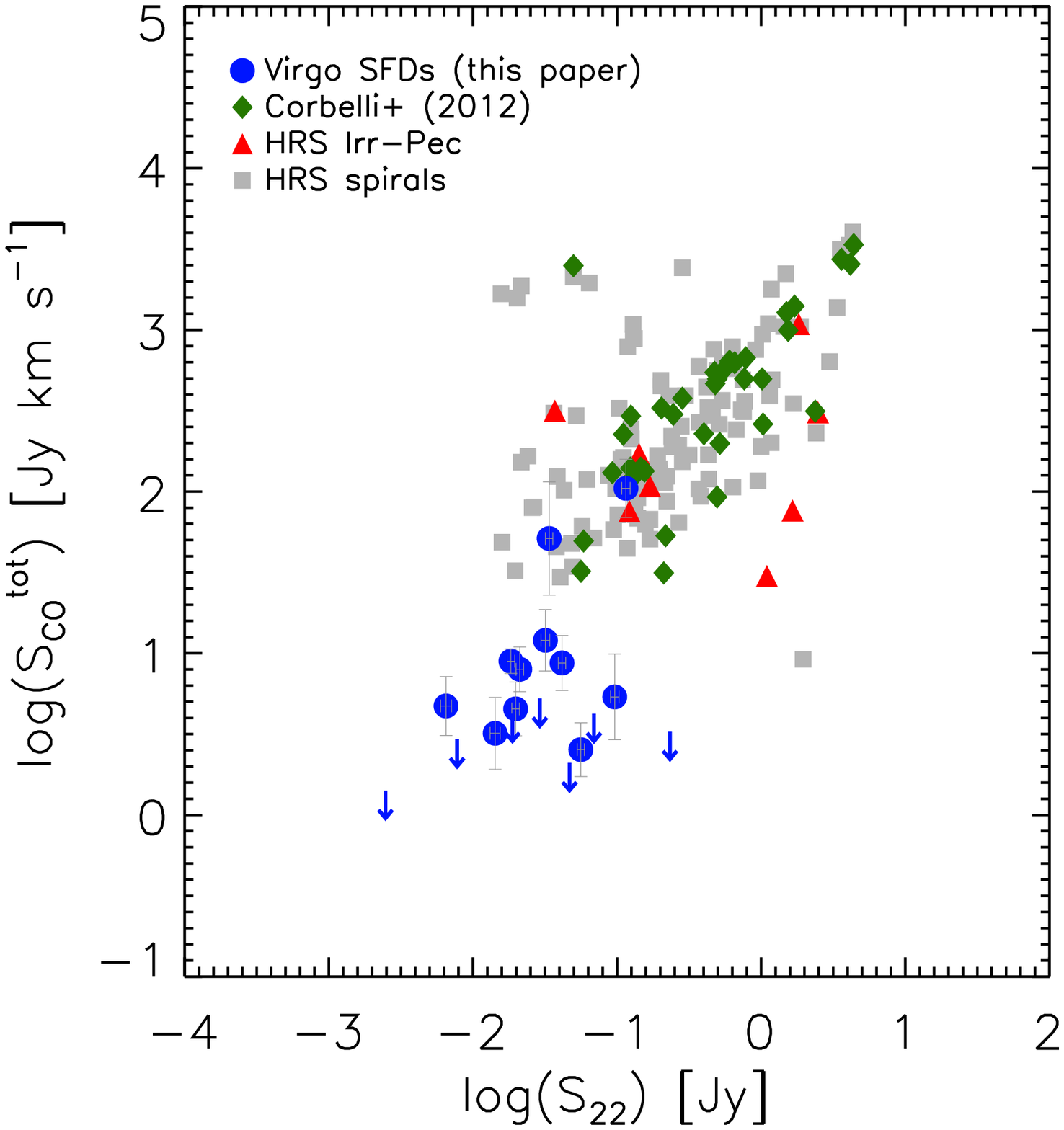}
\caption{{\em Left:} Total CO(1-0) fluxes against total $Herschel$/SPIRE flux densities at 250 $\mu$m.
Virgo SFD dwarfs (blue dots) are compared
to Virgo spirals (green diamonds) from Corbelli et al. (2012), and spiral and irregular or peculiar galaxies from the
HRS survey (grey squares and red triangles, respectively). The dotted line shows the best fit to the three samples.
{\em Centre:} CO(1-0) fluxes within
the $\sim$22$\arcsec$ IRAM 30m telescope beam against 250 $\mu$m flux densities measured within the same area. Here only Virgo SFD
and spiral galaxies are shown. {\em Right:} Total CO(1-0) fluxes against total WISE flux densities
at 22 $\mu$m for Virgo and HRS samples. Symbols are the same as in the left panel.}
   \label{fig:CO_FIR}%
\end{figure*}

\section{CO detections versus non-detections}
\label{subsec:Conodet}

Figure \ref{fig:sample_properties} compares
the global properties
of galaxies with a detected CO emission with those of non-detections.
The main parameter that appears to separate detections from non-detections is stellar mass: more massive
dwarf galaxies  are more likely to be detected in CO \citep{2004A&A...414..141A,2005A&A...438..855I,2005ApJ...625..763L}.
The detected galaxies have stellar masses in between the Large and Small Magellanic Cloud, with log($M_{*}/$\msun) =  9.3
\citep{2002AJ....124.2639V} and 8.5 \citep{2004AJ....127.1531H}, respectively.
Our dwarfs lie along the main-sequence relation of star-forming galaxies \citep{2004MNRAS.351.1151B,2015ApJS..219....8C},
therefore, the detections also have overall the
highest SFRs.
The majority of the detections also correspond to galaxies with the lowest \hi\ fractions (M$_{HI}$/M$_*$), which is
usually found to be
anti-correlated with stellar mass \citep{2010A&A...519A..31L,2012ApJ...756..113H,2013A&A...553A..89G}.
The higher CO detection rate at higher M$_*$ can be easily understood. With increasing stellar mass, all global quantities related
to the inner disc increase, including the CO line intensity; moreover, a higher stellar density favours the formation of molecules \citep{1993ApJ...411..170E}.
This is a consequence of the higher potential well in the inner disc that is provided by the stellar component, which favours higher gas densities
and higher \hdue\ formation rates, as we discuss in Sect. \ref{subsec:h2tohi}.
Non-detections are more \hi-rich than detections.
Lower stellar surface densities lead to
a more diffuse, low-pressure distribution of atomic gas, which
is a less suitable environment for
converting atomic into molecular gas \citep{2005ApJ...625..763L}.

The detection rate
increases in the higher metallicity bins,
but we caution about
the uncertainties in deriving the metal abundances from the SDSS spectra that lack the [{O \sc ii}]$\lambda$3727
line (see Grossi et al. 2015 for details).
The galaxy with the lowest metallicity detected in our survey is VCC324 with 12 + log(O/H) = 8.14 $\pm$ 0.1.

\citet{2015A&A...574A.126G} discussed that the whole sample of FIR-detected SFDs is preferentially located in the less dense regions of the cluster.
In the sub-sample that we observed at IRAM, only one dwarf, VCC1686, is within two degrees of M87.
The others are distributed in the outskirts of subcluster B (centred on M49), in the
low-velocity cloud (LVC), a subgroup of galaxies at V $\lesssim$ 0 \kms superposed on the
M87 region that is thought to be infalling towards the cluster
core from behind \citep{1989ApJ...339..812H}, and in the Virgo southern extension. Six\footnote{VCC10, VCC135,
VCC144, VCC172, VCC340, VCC699} out of 20 galaxies are at larger distances, and they
are located in the background structures called W$^{\prime}$, W, and M clouds to the south-west and to
the north-west of the cluster core, respectively \citep{1984ApJ...282...19F,1987AJ.....94..251B}.
These structures are outside the Virgo virial radius and represent an intermediate density environment.
It is not possible to relate the lack of CO emission in the non-detected
galaxies to their location within the cluster, since they show a projected spatial distribution similar to the CO-detected dwarfs.

\section{FIR-CO correlation}
\label{sec:FIR-CO}

The combined analysis of HeViCS data and CO(1-0) data for a sample of spiral galaxies in Virgo has
shown that a linear relation holds between cold dust emission and CO brightness down to F$_{250} = 2.5$ Jy and S$_{CO} = $
100 Jy \kms \citep{2012A&A...542A..32C}. The main sample examined in this work comprised spiral galaxies with metallicities
above solar. Here we wish to assess whether the correlation holds at lower stellar masses and subsolar metallicities. 

In Fig. \ref{fig:CO_FIR} (left panel) we plot the observed total CO fluxes ($S_{CO}^{tot}$) against {\em Herschel}/SPIRE flux densities
at 250 $\mu$m ($S_{250}^{tot}$) for our sample.
For those galaxies with only a  CO(2-1) detection, we estimated the CO(1-0) flux assuming a (2-1)/(1-0) ratio of 0.8
\citep{2009AJ....137.4670L}, after taking into
account corrections for an emitting region smaller or larger than the beam (see Sect. \ref{sec:COanalysis}).
Flux densities at 250 $\mu$m were taken from Grossi et al. (2015).
In the figure we compare our galaxies to two samples from the literature: i) the Virgo spirals of Corbelli et al. (2012) and
ii) the {\em Herschel} Reference Survey
\citep[HRS;][]{2010PASP..122..261B}, a volume-limited (15 Mpc $\leq D \leq$ 25 Mpc), $K$-band-selected set of 323 galaxies,
of which 260 with morphology later than Sa and 63 with morphological types S0 and earlier. In this and the following figures we use only the
HRS late-type galaxies as a comparison sample, which we divided for simplicity into two morphological bins: spirals (from Sa to Sd) and
irregulars-peculiars (later than Sd), which are expected to have more similar properties to the Virgo SFDs.
HRS CO fluxes were measured in \citet{2014A&A...564A..65B} and 250 $\mu$m flux densities in
\citet{2012A&A...543A.161C}\footnote{FIR flux densities were corrected for the revised SPIRE beam areas and calibration similarly
to what was done with the HeViCS data in Grossi et al. (2015).}. The same quantities for Virgo spirals were taken from Corbelli et al. (2012)$^9$.

CO fluxes correlate strongly with FIR emission even at the lower stellar masses probed by our sample of dwarfs.
The correlation
extends over three orders of magnitude in both $S_{CO}^{tot}$ and $S_{250}^{tot}$ (at least for galaxies with subsolar
metallicities), and it
is close to linear, as found by Corbelli et al. (2012).
The best-fit relation for the three data sets is

\begin{equation}
\log S_{CO}^{tot} = 1.49(\pm0.03) +  1.10(\pm0.03) \log S_{250}^{tot},\end{equation}

\noindent and it is plotted in the left panel of Fig. \ref{fig:CO_FIR}\footnote{We also fitted the data on linear scales and found a best-fit line
$\log$ $S_{CO}^{tot}$ = -37.4 ($\pm$14.6) + 44.4 ($\pm$1.8) $\log$ ($S_{250}^{tot}$).}.
In the middle panel of the figure we show the CO(1-0) flux in single-pointing
observations (corrected for a finite source size) and S$_{250}^{beam}$
measured in circular apertures of 18$\farcs$6 radius\footnote{The circular aperture chosen for the 250 $\mu$m photometry corresponds to 88$\%$ of the effective area of the 30 m
telescope beam at 115 GHz.},
centred at the same positions as the IRAM pointings
(see Table \ref{tab:point_list}).
The best-fit relation in this case has the same intercept and slope as the one determined for the total FIR and CO fluxes
(central panel of Fig. \ref{fig:CO_FIR}). The weighted mean of the scatter of the Virgo SFD data about the best fit
\citep[see Eq. (6) in][]{2010MNRAS.409.1330W} is$\text{ about
three}$  times lower when the single pointings are considered because in this case the
uncertainties related to the global content of CO are not taken into account.

\citet{2011AJ....142...37S} showed that spiral galaxies from the HERACLES survey \citep{2009AJ....137.4670L} exhibit a strong
correlation between CO and 24 $\mu$m emission because of the tight link between molecular gas and star formation  that is
traced by mid-infrared (MIR)
radiation at 24 $\mu$m.
As a comparison, in the right panel of Fig. \ref{fig:CO_FIR} we display $S_{CO}$ against the flux density at 22 $\mu$m, $S_{22}$, obtained from the
Wide-field Infrared Survey Explorer \citep[WISE;][]{2010AJ....140.1868W} for all samples.
We used 22 $\mu$m emission instead of 24 $\mu$m because WISE fluxes are available for the three sets of galaxies. WISE flux densities were taken from \citet{2014A&A...570A..69B} for the HRS and Virgo spiral
galaxies and from \citet{2015A&A...574A.126G} for the Virgo SFDs. The figure shows a trend between $S_{CO}^{tot}$ and $S_{22}$,
but the scatter is larger than for the $S_{CO}^{tot}-S_{250}$ correlation both at high and low stellar masses.
Therefore CO emission is consistent with being linearly proportional to FIR emission down to the low stellar masses probed by
our sample of dwarf galaxies; the link between molecular gas and cold dust appears to be stronger than that with warm dust traced
by MIR emission.

\begin{figure}
  \centering
\includegraphics[bb=40 -10 510 285,width=9.5cm]{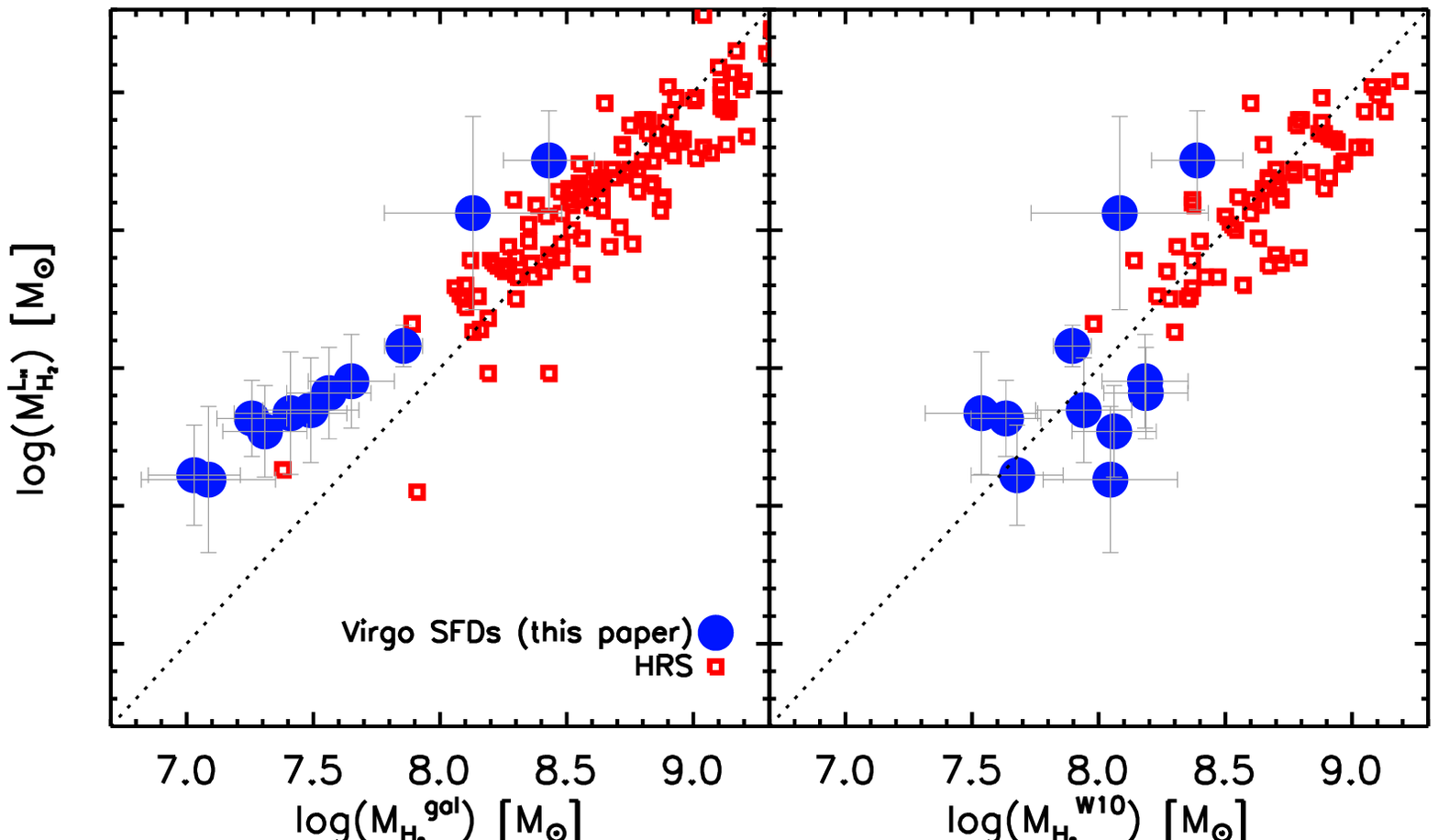}
\caption{Comparison between H$_2$ masses derived with different methods. The $H$-luminosity-dependent estimate \citep{2002A&A...384...33B} is plotted
against the values obtained with a Milky Way
standard conversion factor (left) and
 a metallicity-dependent $X$ factor \citep[][right panel]{2010ApJ...716.1191W}.
As a comparison we also show galaxies from the HRS survey (red squares).
The dotted lines indicate the one-to-one relation.}
   \label{fig:H2masses}%
\end{figure}

\section{Estimates of molecular hydrogen masses}
\label{sec:h2masses}

The conversion between CO intensities and \hdue\ column density is
usually performed using a CO-to-\hdue\ conversion factor, $X_{CO}$,
which is defined as
\begin{equation}
X_{CO} = \frac{N(\textrm{H}_2)} {I_{CO}} \textrm{[}\textrm{cm}^{-2} \textrm{(K km s}^{-1}\textrm{)}^{-1}\textrm{]}.
\end{equation}

In the Milky Way the standard conversion factor is $ X_{CO}^{gal} = 2 \times 10^{20} \: \: \: \textrm{cm}^{-2} \textrm{(K km s)}^{-1}$
\citep[][and references therein]{2001ApJ...547..792D,2006A&A...454..781L,2013ARA&A..51..207B}, equivalent to $\alpha_{CO}^{gal}$ = 3.21 \msun pc$^{-2}$
(K \kms)$^{-1}$ without the correction for helium.

$X_{CO}$ is expected to vary with the physical properties
of the ISM, such as metal content,  ionising radiation field strength, and gas surface density. Metallicity
is the dominant parameter affecting $X_{CO}$ variations, however,
since it is found to increase sharply in systems
with metallicities below 12 + log[O/H] $\approx$ 8.4, or one-half the solar value
\citep{1997A&A...328..471I,1998AJ....116.2746T,2001PASJ...53L..45M,2002A&A...384...33B,2012AJ....143..138S}.
Thus, applying a galactic CO-to-\hdue\ conversion factor can lead
to underestimating the molecular gas content in low-metallicity galaxies.

For this reason, we compare two different methods to estimate \hdue\ masses that take into account the variation of the $X_{CO}$ factor
with the metal content.

 \begin{table}
\caption{Total H$_2$ masses of the Virgo SFD galaxies.}
\centering
\begin{tabular}{lccc}
\hline \hline
ID & $\log M_{H_2}^{gal}$ & $\log M_{H_2}^{L_H}$            & $\log M_{H_2}^{W10}$  \\
   &  [M$_{\odot}$]           & [M$_{\odot}$]                       & [M$_{\odot}$]   \\
\hline \hline
10    &  7.41 $\pm$ 0.22 &    7.84 $\pm$ 0.22        &    7.54 $\pm$ 0.22      \\
87    &      $<$6.98     &     $<$ 7.67              &    $<$7.63              \\
135   &  7.86 $\pm$ 0.08 &    8.08 $\pm$ 0.08        &    7.90 $\pm$ 0.08      \\
144   &  7.31 $\pm$ 0.17 &    7.77 $\pm$ 0.17        &    8.06 $\pm$ 0.17      \\
172   &      $<$7.88     &    $<$8.27                &    $<$7.98              \\
213   &  8.13 $\pm$ 0.35 &    8.56 $\pm$ 0.35        &    8.08 $\pm$ 0.35      \\
324   &  7.09 $\pm$ 0.27 &    7.60 $\pm$ 0.26        &    8.05 $\pm$ 0.27      \\
334   &      $<$6.51     &    $<$7.21                &    $<$7.23              \\
340   &  7.56 $\pm$ 0.17 &    7.91 $\pm$ 0.17        &    8.19 $\pm$ 0.17      \\
562   &      $<$7.08     &    $<$8.15                &    $<$8.17              \\
693   &      $<$7.12     &    $<$7.41                &    $<$7.42              \\
699   &  7.65 $\pm$ 0.17 &    7.95 $\pm$ 0.17        &    8.18 $\pm$ 0.17      \\
737   &      $<$6.82     &    $<$7.45                &    $<$7.40              \\
841   &      $<$6.68     &    $<$7.36                &    $<$7.15              \\
1437  &  7.26 $\pm$ 0.14 &    7.82 $\pm$ 0.14        &    7.63 $\pm$ 0.14      \\
1575  &  8.43 $\pm$ 0.18 &    8.75 $\pm$ 0.18        &    8.39 $\pm$ 0.18      \\
1686  &  7.49 $\pm$ 0.19 &    7.85 $\pm$ 0.19        &    7.94 $\pm$ 0.19      \\
1699  &     $<$6.87      &    $<$7.44                &    $<$8.08              \\
1725  &  7.03 $\pm$ 0.18 &    7.61 $\pm$ 0.18        &    7.68 $\pm$ 0.18      \\
1791  &     $<$6.98      &    $<$7.62                &    $<$7.88              \\
\hline \hline
\end{tabular}
\label{tab:h2masses}
\end{table}

\begin{figure*}
  \centering
\includegraphics[bb=70 -10 590 550,width=8cm]{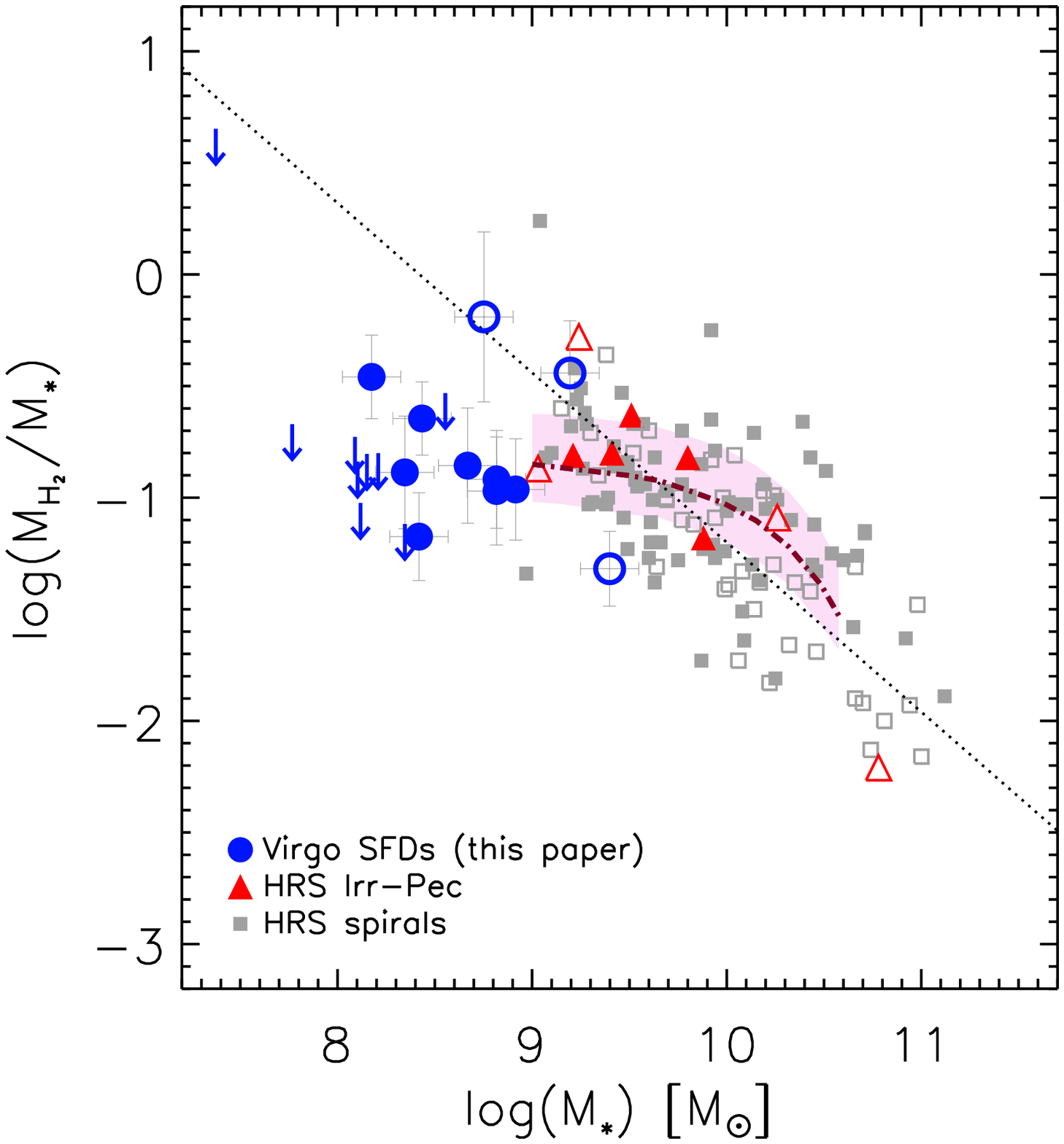}
\includegraphics[bb=50 -10 570 550,width=8cm]{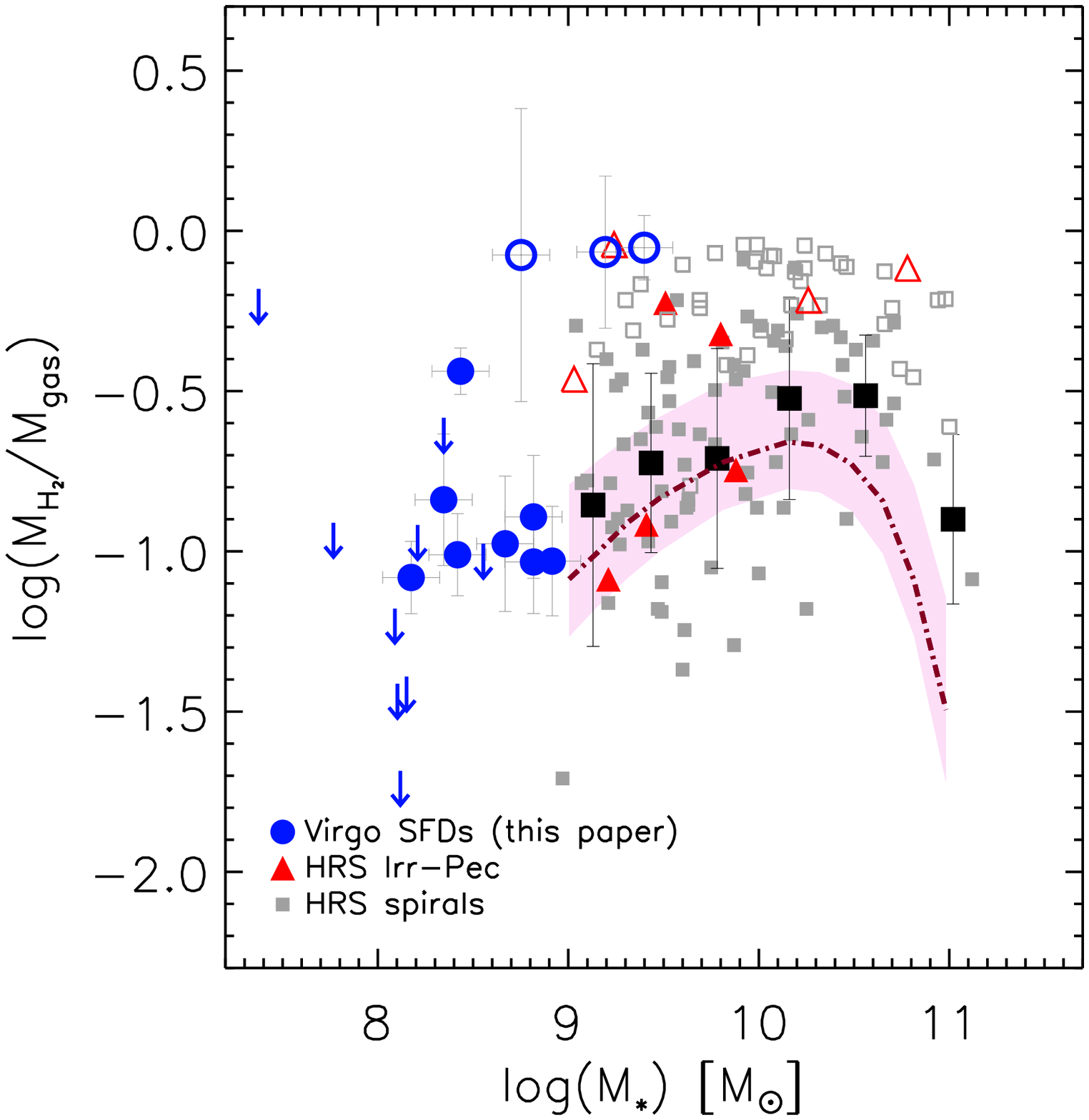}
\caption{{\em Left panel:} Ratio of H$_2$ to stellar mass against stellar mass. Virgo SFDs (blue dots) are compared to spiral (grey squares) and
irregular-peculiar (red triangles) galaxies from the HRS. \hi-deficient galaxies ($Def_{HI} > 0.5$) in both samples are indicated
with empty symbols.
The dotted line shows the scaling relation obtained for the HRS galaxies
with a normal \hi\ content \citep{2014A&A...564A..66B}. {\em Right panel:} Molecular gas mass fraction against stellar mass.
Symbols are the same as in the left panel. Large black squares show the average distribution of the HRS galaxies with a normal \hi\ content.
The dot-dashed lines in both panels indicate
the predictions of the model of \citet{2015MNRAS.450..606L}, and the shaded area corresponds to the uncertainty on the model.}
\label{fig:H2frac}%
\end{figure*}

\begin{enumerate}

\item We use a metallicity-dependent CO-to-\hdue\ conversion factor based on the model of \citet{2010ApJ...716.1191W}, $\alpha_{CO}^{W10}$, which includes the possiblilty that a fraction of \hdue\ mass may lie outside the CO region, where carbon is in the atomic or ionised state:

\begin{equation}
\frac{\alpha_{CO}^{W10}}{\alpha_{CO}^{gal}} =  \exp \left( \frac{0.74 - 0.078 \ln Z^{\prime}}{Z^{\prime}} \right) \times  \exp \left( -0.74 + 0.078 \ln Z^{\prime} \right)
,\end{equation}

where $Z^{\prime} = Z/Z_{\odot} = 10^{(12 + \log (O/H) - 8.7)}$, and $\alpha_{CO}^{MW}$ is the Milky Way conversion factor in units of \msun {pc}$^{-2}$ (K \kms)$^{-1}$ .

 \item We also consider luminosity-dependent $X_{CO}$ \citep{2002A&A...384...33B}, based on the
 luminosity-metallicity relation

\begin{equation}
\log \frac{X_{CO}^{L_{H}}}{\textrm{cm}^{-2} \textrm{(K km s}^{-1}\textrm{)}^{-1}} = -0.38 \: \log L_H + 24.23
,\end{equation}

where $L_H$ is the total galaxy luminosity in the $H$ band. This method was applied by \citet{2014A&A...564A..65B} to estimate molecular gas masses
in the HRS galaxies, which we use in the following sections as a comparison sample to the \hdue\ properties of the Virgo SFDs.

 \end{enumerate}

We calculated the molecular hydrogen mass of the Virgo SFD galaxies from the following relation \citep{2005ARA&A..43..677S}:

\begin{equation}
M_{H_2} = 3.25 \times 10^7 \, {\alpha_{CO}^i} \, S_{CO} \, \nu^{-2} D^2  \, \textrm{[M}_{\odot}\textrm{]}
,\end{equation}

\noindent where $\alpha_{CO}^i$ = [$\alpha_{CO}^{gal}$, $\alpha_{CO}^{L_H}$, $\alpha_{CO}^{W10}$], $S_{CO}$ is the CO(1-0) flux in Jy \kms corrected for the extension of the emitting region
(smaller or larger than the beam;
see Sect. \ref{sec:final_CO}), $\nu$ the frequency of the CO(1-0) line in GHz, and $D$ the distance to the galaxy in Mpc.

In Fig. \ref{fig:H2masses} we compare the H$_2$ masses derived with a $H$-luminosity-dependent $X_{CO}$ to those obtained with a galactic (left panel) and a metallicity-dependent (right panel), $X_{CO}$. 
As expected, when the $H$-luminosity (or metallicity dependence) is taken into account,  we obtain higher molecular masses than
with the assumption of a fixed X$_{CO}$.
The $X$ factor of Boselli et al (2002) is at most four times larger than the galactic value, while
the calibration of \citet{2010ApJ...716.1191W}
gives masses higher by a factor of 9 at most
 (for the lowest metallicity object of the sample, VCC324), compared to the Milky Way conversion.
The $H$-luminosity-dependent and the metallicity-dependent estimates agree reasonably well
(within the uncertainties), and the larger discrepancy between the two methods (a factor of $\sim$ 3) is found for VCC324.

The \hdue\ masses estimated with all the methods discussed in this section are displayed in Table \ref{tab:h2masses}.
In the rest of the paper we adopt the $H$-luminosity-based
estimate because it provides more CO-detected galaxies in the HRS sample.
The choice of this $X$ factor does not remarkably modify the main
results of our work.

\section{Molecular hydrogen properties of Virgo SFDs}
\label{sec:scaling_relations}

\begin{figure*}
  \centering
\includegraphics[bb=60 0 570 550,width=6.05cm,height=6.6cm]{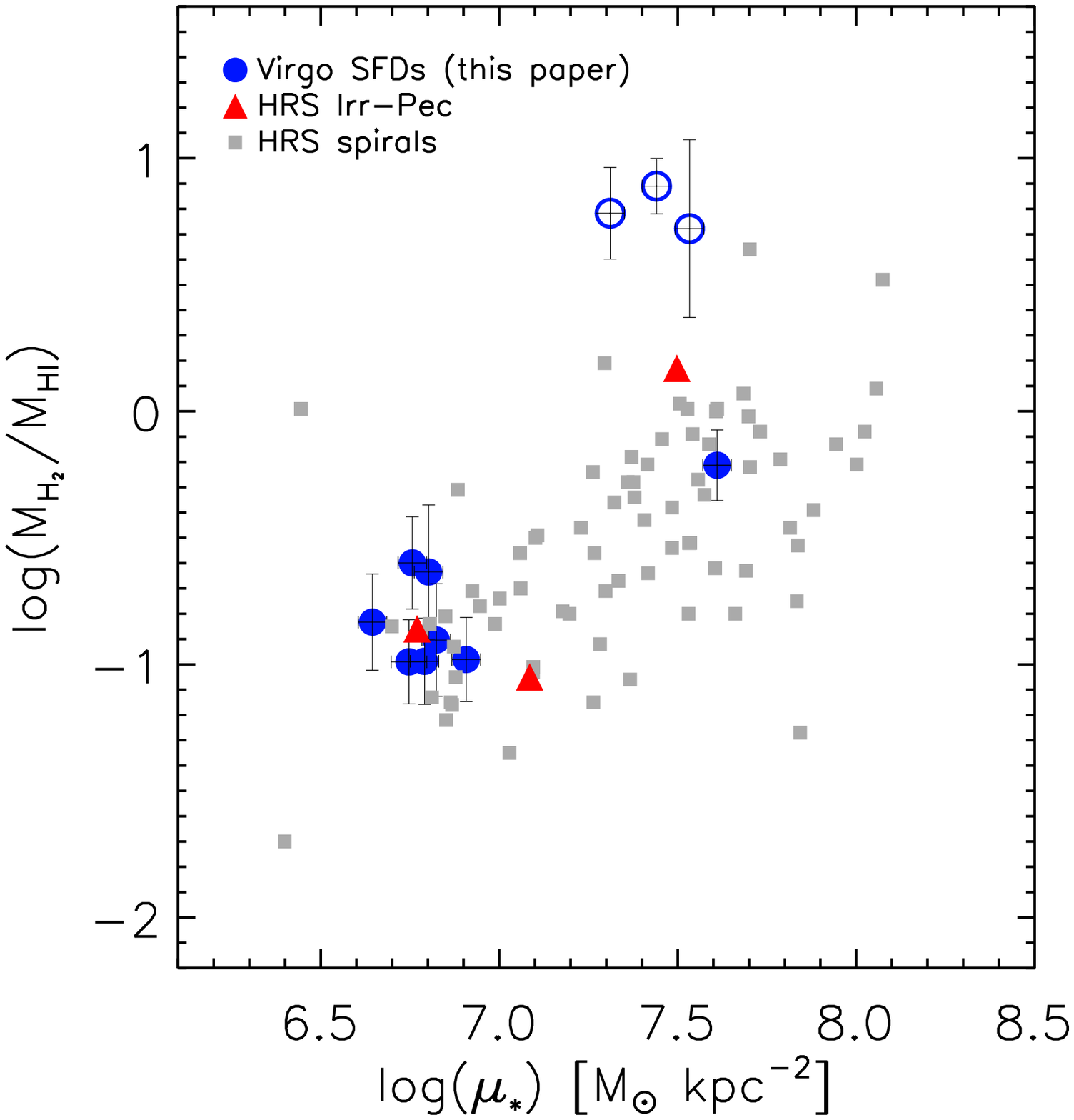}
\includegraphics[bb=70 0 570 550,width=6.05cm,height=6.6cm]{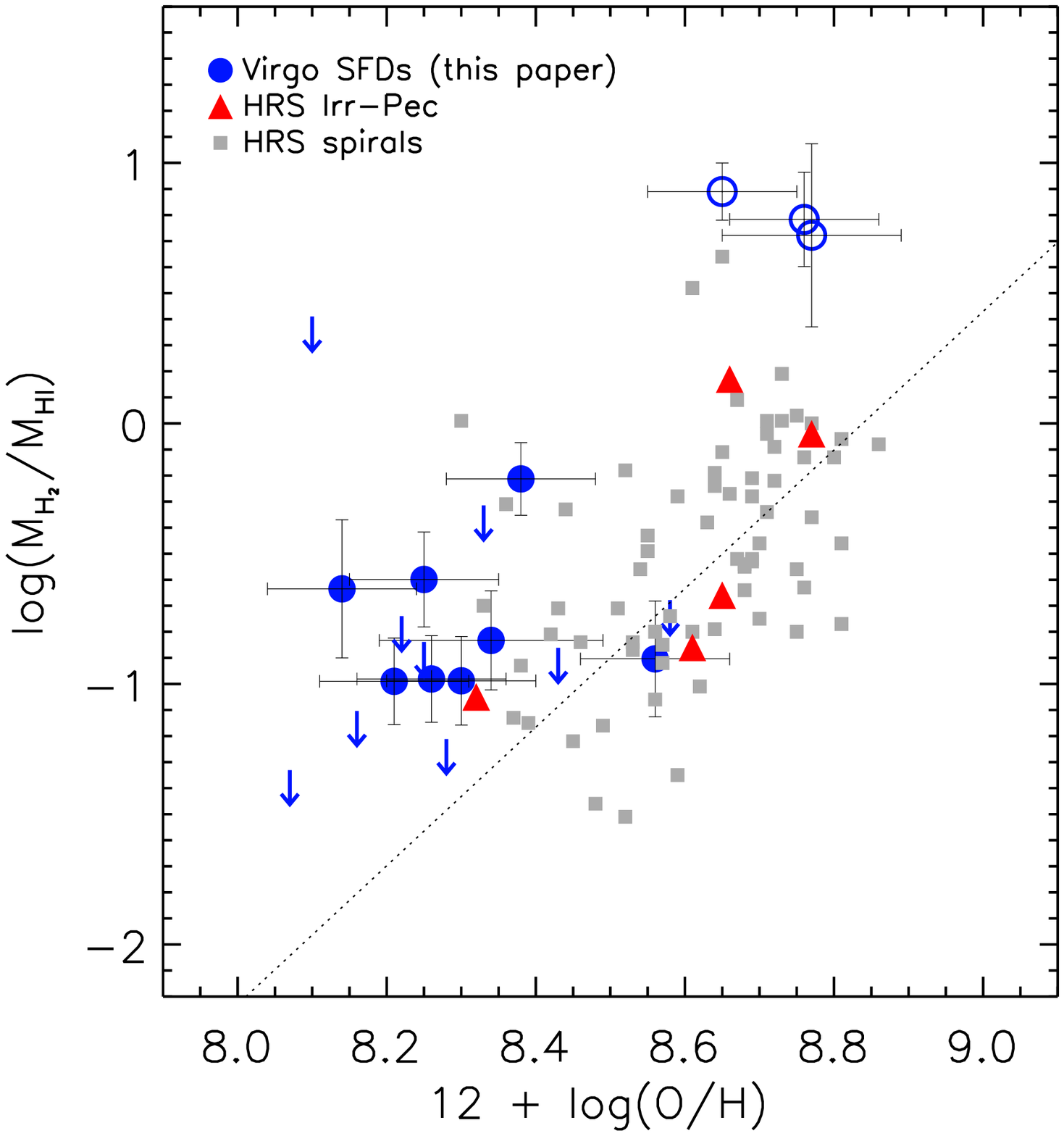}
\includegraphics[bb=70 0 570 550,width=6.05cm,height=6.6cm]{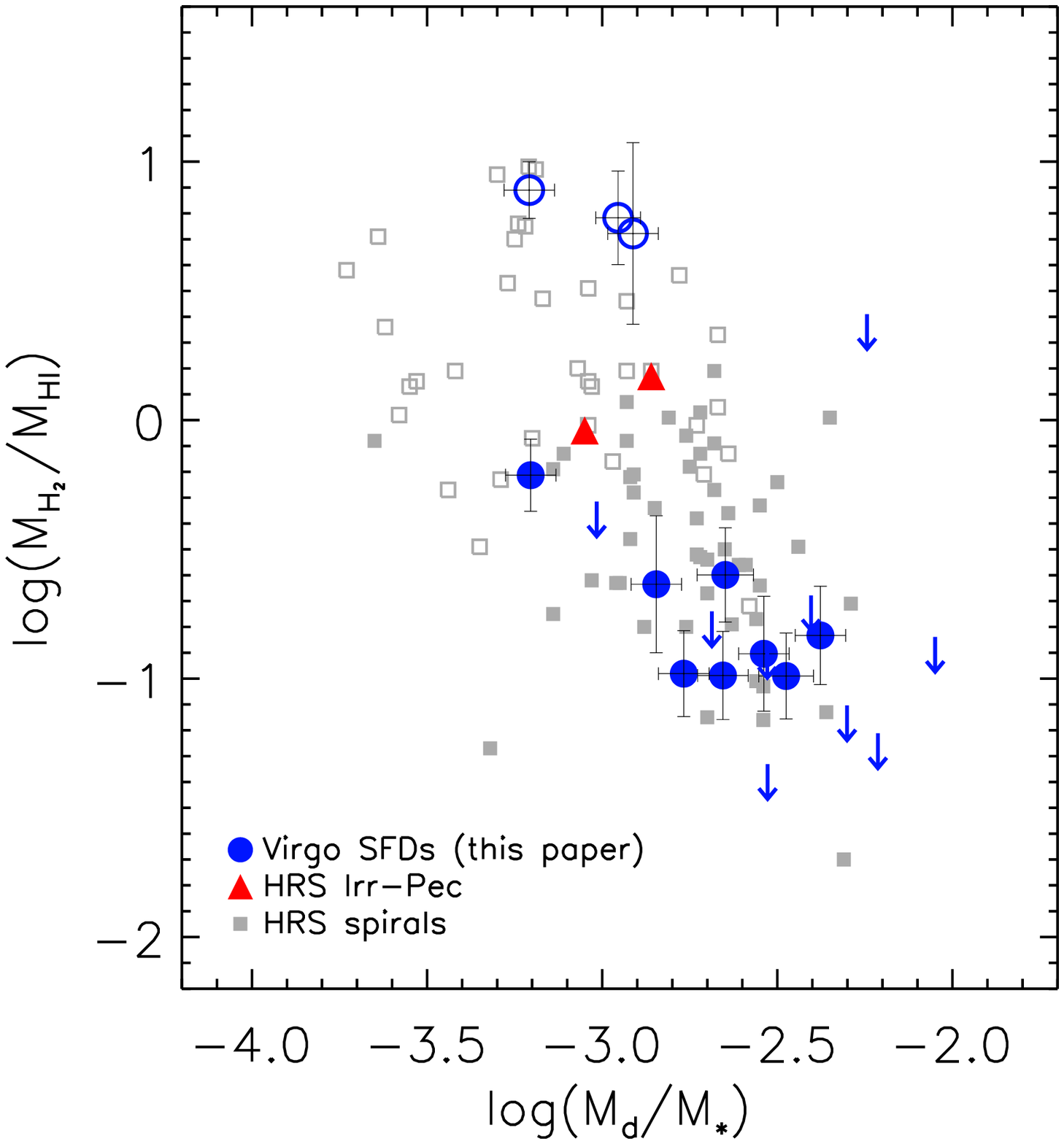}
\caption{H$_2$-to-\hi\ ratio against stellar mass surface density ({\em left panel}), metallicity ({\em central panel}),
and ratio of dust to stellar mass ({\em right panel}).
Virgo SFDs (blue dots) are compared to spiral (grey squares) and irregular-peculiar
(red triangles) galaxies from the HRS. \hi-deficient galaxies in all samples are indicated with empty symbols.
The dotted line in the central panel shows the scaling relation
obtained for the HRS galaxies with a normal \hi\ content \citep{2014A&A...564A..66B}.}
   \label{fig:H2HI_scaling}%
\end{figure*}

\subsection{Molecular-to-stellar mass ratio}
\label{subsec:h2frac}

To investigate the relation between the stellar and the cold gas mass content of the Virgo SFDs, we display
in Fig. \ref{fig:H2frac} (left panel) the ratio of \hdue\  to stellar mass (M$_{H_2}$/M$_*$) versus stellar mass. This scaling relation has been
thoroughly studied by other molecular gas surveys \citep{2011MNRAS.415...32S,2014A&A...564A..66B}.
To inspect a broader dynamical range than that sampled by our set of dwarfs, we compared Virgo SFDs to HRS galaxies.
To make the samples as homogeneous as possible, we recalculated the stellar masses of Virgo SFDs following the method
of \citet{2009MNRAS.400.1181Z} that was adopted by the HRS team \citep{2012A&A...544A.101C}, based on $g$ and $i$ photometry (see Table \ref{tab:mstar_mdust_HRS_like}). 
\hdue\ masses in the figure are derived with a $H$-luminosity dependent $X$ factor.
In the figure we distinguish between \hi-normal ($Def_{HI} \leq 0.5$) and \hi-deficient ($Def_{HI} > 0.5$) galaxies (filled and empty symbols, respectively).
The dotted line indicates the
anti-correlation between M$_{H_2}$/M$_*$ and stellar mass
found in the HRS sample, considering only galaxies with a normal \hi\ content \citep[][]{2014A&A...564A..66B}.
Virgo dwarfs with $Def_{HI} \leq 0.5$ and stellar masses below 10$^9$ \msun\
(filled circles) do not follow the scaling relation determined in the HRS sample.
The trend given by the upper limits also indicates that this deviation extends to the lowest stellar masses sampled by our survey.

The flattening of the molecular-to-stellar mass ratio at low M$_*$ is expected by evolution models of the gaseous content of galaxies.
The dash-dotted line shows the predictions of
the model of \citet{2009ApJ...693..216K} derived in \citet[][]{2015MNRAS.450..606L}, with the shaded area indicating the corresponding
uncertainty. The flattening of the relation can be explained by the role played by
metallicity in \citet{2009ApJ...693..216K}. The model assumes a critical surface density for molecule formation, which is roughly inversely proportional to the gas-phase metallicity.
The lower metallicity in galaxies of lower stellar masses affects the process of molecular hydrogen formation,
but stars can form in \hi-dominated regions, which are very poor in molecules
\citep{2012MNRAS.421....9G,2013MNRAS.436.2747K}. Therefore, according to this model, the flattening of M$_{H_2}/$M$_*$ at
low masses is not the consequence of gas removal due to environmental effects.
Of the most \hi-deficient galaxies, only VCC135 has a low \hdue\ fraction compared to galaxies with a similar stellar mass, while VCC213
and VCC1575 appear to have a similar \hdue\ content as the other dwarfs.

To verify whether \hi\  dominates the total gas budget at low masses, we show in the right panel of Fig.
\ref{fig:H2frac} the scaling relation for the ratio of molecular-to-total gas mass compared to the predictions of
\citet[][ dot-dashed line]{2015MNRAS.450..606L}.
The molecular gas fraction of \hi-normal SFDs ($Def_{HI} \leq 0.5$) ranges between 9\% and 38\% of the total amount of gas (with a mean value of 14\%), and the majority of these dwarfs have M$_{H_2}$/M$_{gas}$ ratios comparable to the predictions of the model
(extrapolated to M$_* < 10^9$ \msun).
The molecular-to-total gas mass ratio is higher in the \hi-deficient galaxies (empty circles and squares), and it
does not seem to vary remarkably with $M_*$ in these systems at low and high stellar masses.
The HRS galaxies with a normal gas content show significant scatter,
although if we consider the mean values of the sample in bins of stellar masses (large filled squares), they appear
to follow the general trend defined by the model.
The main contribution of \hdue\ to the total gas mass occurs at stellar masses around 10$^{10}$
\msun.

\subsection{Ratio of molecular to atomic gas}
\label{subsec:h2tohi}

The molecular and atomic gas masses ratio, $R_{mol} =$ M$_{H_2}$/M$_{HI}$ provides information about the conversion of atomic
into molecular gas and about the balance between \hdue\ formation and destruction.
Models of the \hi\ to \hdue\ transition \citep{1993ApJ...411..170E} showed that in regions with low molecular fractions
$R_{mol} \sim P^{2.2}/j$, where $P$ is the interstellar mid-plane hydrostatic gas pressure
and $j$ is the mean interstellar radiation field.
\citet{2002ApJ...569..157W} and \citet{2004ApJ...612L..29B}
 also claimed that
the interstellar gas pressure
plays the dominant role in determining the balance between \hi\ and \hdue.
However, \citet{2009ApJ...693..216K} predicted that the \hdue-to-\hi\ ratio mainly depends on the total gas column density of a galaxy and only
secondarily on the metallicity.
It is therefore important to address through observations how $R_{mol}$ is related to global galaxy properties to better understand how the conversion of atomic into molecular gas is regulated in galaxies,
especially in low-mass star-forming systems.

In Fig. \ref{fig:H2HI_scaling} we plot $R_{mol}$ against some of the main global properties of our sample: stellar surface
density and metallicity.
While Virgo dwarfs appear to follow the correlation between this ratio and stellar surface density (left panel) defined by the locus
of the HRS galaxies, the dependence of the molecular-to-atomic ratio on
metallicity  is less clear at low stellar masses (central panel). The dwarf galaxies with an apparently normal \hi\ content show
higher $R_{mol}$ ratios than the scaling relation obtained by \citet{2014A&A...564A..66B} for more massive systems.

The gas content of \hi-deficient galaxies (i.e. VCC135, VCC213, and VCC1575; empty circles in Fig. \ref{fig:H2HI_scaling}) is dominated by molecular hydrogen because all have $R_{mol} > 1$.
The survival of molecular gas in the central regions of a
\hi-stripped dwarf irregular stellar disc could be a relevant step
for the formation of nucleated dwarf elliptical galaxies (dEs), given that many
cluster dEs have distinct nuclear blue regions \citep{2006AJ....132.2432L}.
Thus these galaxies could provide hints on the morphological transformation of star-forming dwarf galaxies
into early-type systems in a cluster environment. VCC135 shows a blue core with
strong emission lines overlaid on an extended redder stellar structure with an elliptical symmetry;  VCC213, on the other hand, has a
red core that resembles the nucleus of dwarf elliptical, and star formation is occurring in a ringed or spiral structure around
the core \citep{2014A&A...562A..49M}. Both have global optical colours redder than other BCDs in
Virgo \citep[$g-i > 0.7$;][]{2014A&A...562A..49M}. The redder colours, the structural properties and visual
appearance, and the low \hi\ content provide clear indication of the different evolutionary stage of these galaxies
that is probably due to the influence of the environment.

Lastly, in the right panel of Fig. \ref{fig:H2HI_scaling} we show how the \hdue-to-\hi\ ratio varies with the dust fraction
(M$_d$/M$_*$). Once again, Virgo SFDs are compared to the HRS late-type galaxies.\footnote{Dust masses for the HRS have been calculated in \citet{2014A&A...565A.128C} by fitting models of \citet{2007ApJ...657..810D}  and in \citet{2014MNRAS.440..942C} using an MBB with a fixed emissivity ($\beta=2$ and $\kappa_0 = 3.76$ cm$^2$ g$^{-1}$ at $\lambda =$ 250 $\mu$m).  However, to avoid biases when we compare dust masses of the two samples (HRS and Virgo SFDs), we decided to recalculate M$_d$ for the HRS following the same method
as adopted in \citet{2015A&A...574A.126G}. We fitted the SED between 100 and 350 $\mu$m only (because of the 500 $\mu$m excess detected in most of the Virgo dwarfs) with a single MBB with fixed emissivity ($\beta = 1.5$) and $\kappa_0 = 3.4$ cm$^2$ g$^{-1}$ at $\lambda =$ 250 $\mu$m. The resulting measurements, displayed in the right panel of Fig. \ref{fig:H2HI_scaling} and the bottom left panel of Fig. \ref{fig:tau_HIdef},
are on average lower by a factor 0.08 $\pm$ 0.1 dex  than the dust masses in \citet{2014MNRAS.440..942C} for $\beta = 2$. However, the choice of $\beta$ and $\kappa_0$
to estimate dust masses does not affect the conclusions drawn from these figures because the same trends hold if we use either our method or the approach of \citet{2014MNRAS.440..942C} for both samples.}
The figure shows that the ratio is inversely correlated with $R_{mol}$ and that the same trend holds for both
\hi-deficient and \hi-normal galaxies. This might be related to the fact that \hi-poor galaxies are being stripped of their dust
content, revealing a dust deficiency as shown by previous {\it Herschel} surveys
\citep{2010A&A...518L..49C,2012A&A...540A..52C,2012A&A...542A..32C,2015A&A...574A.126G}. In Sect. \ref{subsec:sfe} we discuss the effects of the environment
on the dust and molecular gas components of our sample of dwarfs and we compare them to the galaxies of the HRS.

\subsection{Dust-to-gas mass ratio}
\label{sec:d2g}

\begin{figure}[h!]
  \centering
\includegraphics[bb=-20 -20 560 770,width=7.6cm]{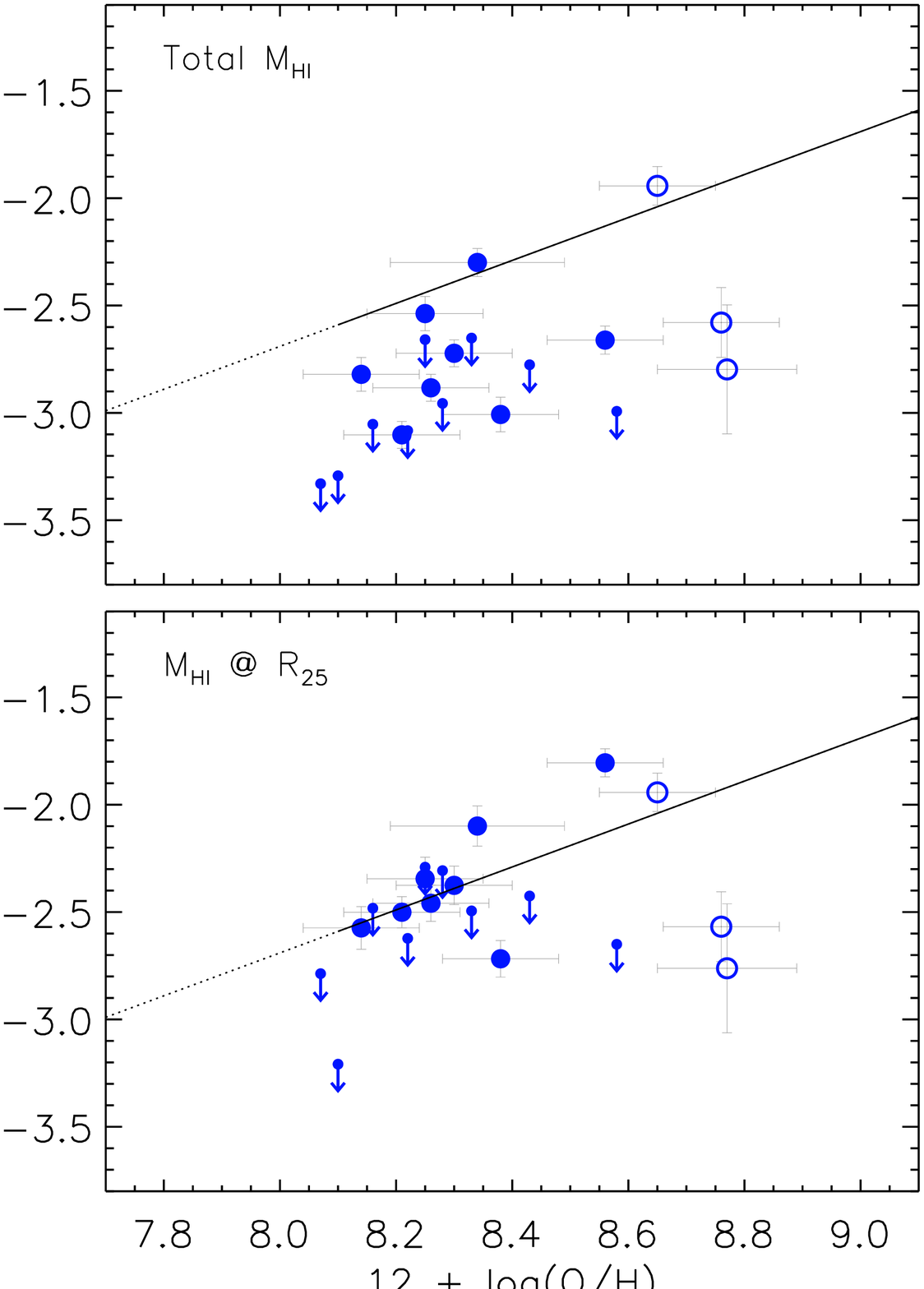}
\caption{{\em Top-panel}: Dust-to-gas mass ratios against metallicity for Virgo SFDs. The solid line indicates
  a linear scaling of the Milky Way dust-to-gas ratio and metallicity which holds for galaxies with 12+ log(O/H) $\gtrsim 8.1$. 
The total gas mass here
is calculated considering the total \hi\ mass from ALFALFA observations. {\em Bottom-panel}: Same as upper panel. The total gas mass here
is calculated considering only the \hi\ mass within $R_{25}$, as explained in Sect. \ref{sec:d2g}.}
   \label{fig:gas2dust}%
\end{figure}

The dust-to-gas mass ratio \D\ (M$_{d}$/M$_{g}$) gives an indication of the enrichment of the gas
by heavy elements produced in stars (C, O, Mg, Si, Fe), the amount of metals that are locked in dust grains, and the net balance between the production and growth of dust grains
and their destruction in the ISM.
If the ratio of dust-to-metals in the ISM does not vary among
galaxies, the relation between \D\ and the oxygen abundance O/H is expected to be linear \citep{2001MNRAS.328..223E}.
Previous studies have shown that indeed there is a nearly linear relation between \D\ and the metallicity at oxygen abundances 12 + log(O/H) $\gtrsim$ 8.1
 \citep{2007ApJ...663..866D,2011ApJ...737...12L,2013ApJ...777....5S}, while the non-linear behavior
becomes notable for metallicities below this value \citep{2014A&A...563A..31R}.

In Fig. \ref{fig:gas2dust} we explore the variation of  \D\ with nebular oxygen abundance for our sample. In the top panel
 the gas-to-dust ratio is derived by combining the molecular and total atomic gas masses.
\D\ correlates with metallicity, but most of the Virgo dwarfs, especially the more gas-rich ones,
have dust-to-gas ratios lower than expected by linearly scaling the Milky Way values (Fig. \ref{fig:gas2dust}, top panel):

\begin{equation}
\mathcal{D} = \frac{M_d}{M_{H \mathsc i} + M_{H_2}} = 0.01 \frac{(O/H)}{(O/H)_{\odot}},
\label{eq:d2g}
\end{equation}

\noindent assuming  a Milky Way dust-to-gas ratio of $\approx$ 0.01 \citep{2011piim.book.....D}
and oxygen abundance 12 + log(O/H) = 8.69 \citep{2009ARA&A..47..481A}\footnote{Nonetheless, we are aware that a wide spread of \D\ values is
observed in the linear regime related to different galactic star formation histories
\citep{2013ApJ...777....5S,2014A&A...563A..31R,2014A&A...562A..76Z}.}.
As we discussed in Sect. \ref{subsec:h2tohi}, our dust mass estimates based on an MMB fit with $\beta = 1.5$ are lower by $\sim$ 0.1 dex than the most commonly adopted emissivity index $\beta = 2$. 
Despite this variation, as a result of the choice of $\beta$,
an average difference of 0.1 dex would still be too low to explain the discrepancy between the linear scaling of the Milky Way and
the measured \D\ of our dwarfs.

A possible explanation could be that the \hi\ masses from the ALFALFA catalogue only provide the global \hi\ content
because our dwarfs are unresolved by the $\sim$ 3\farcm5 Arecibo
beam.
\hi\ envelopes in dwarf galaxies are known to be remarkably more extended than the stellar (and dust) component
\citep{2002A&A...390..829S}. \D\ might therefore be underestimated in some cases because of the different size of the atomic gas disc compared to the extension of the dust
and molecular gas reservoir ($\lesssim R_{25}$),
but also because of dust and gas stripping as galaxies move through the intracluster medium.
The dwarfs with the highest
metallicities in our sample are also the most \hi-deficient, and we have shown in \citet{2015A&A...574A.126G} that such
 galaxies show evidence of dust stripping as well.

\begin{figure*}
  \centering
\includegraphics[bb=60 0 550 550,width=6.05cm,height=6.6cm]{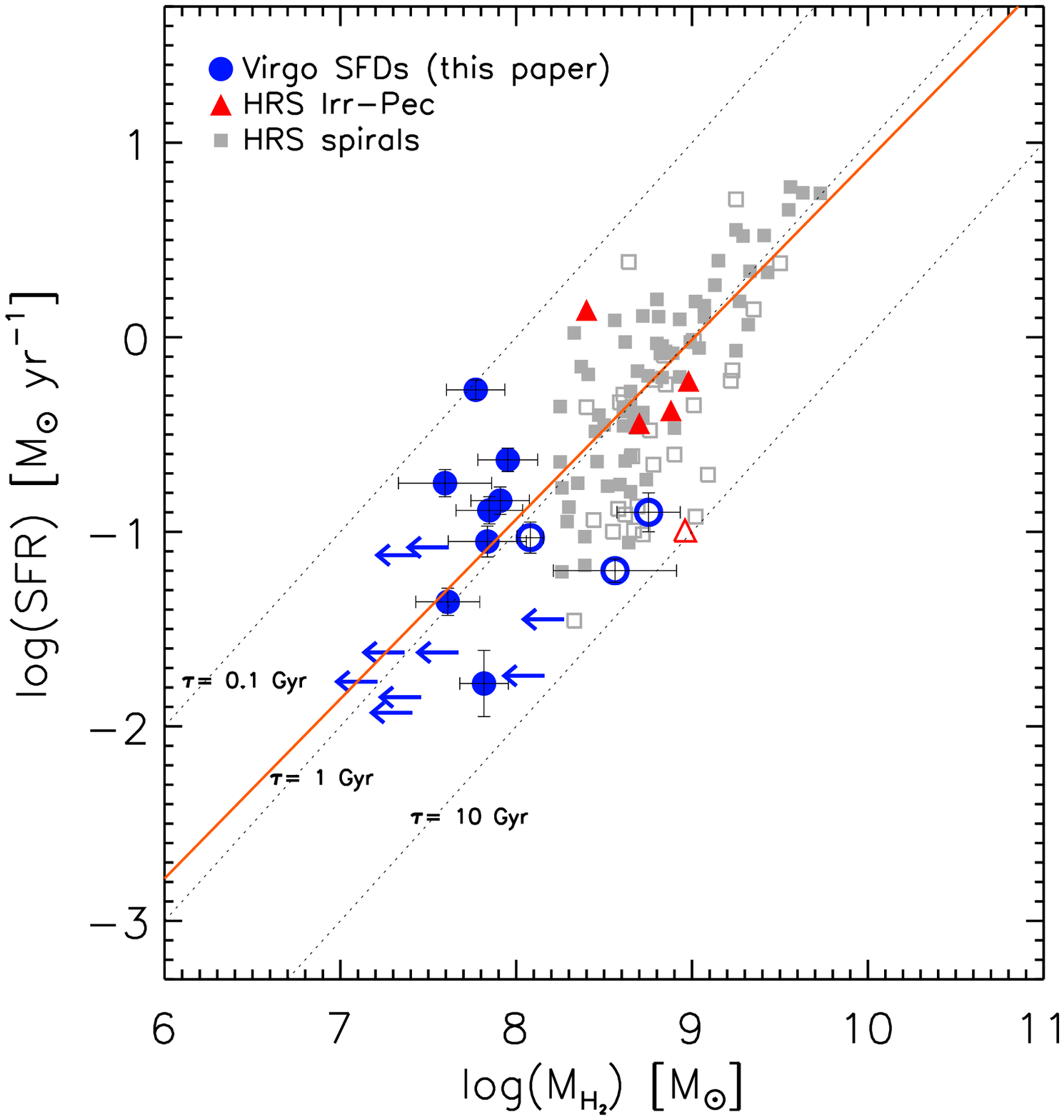}
\includegraphics[bb=80 0 570 550,width=6.05cm,height=6.6cm]{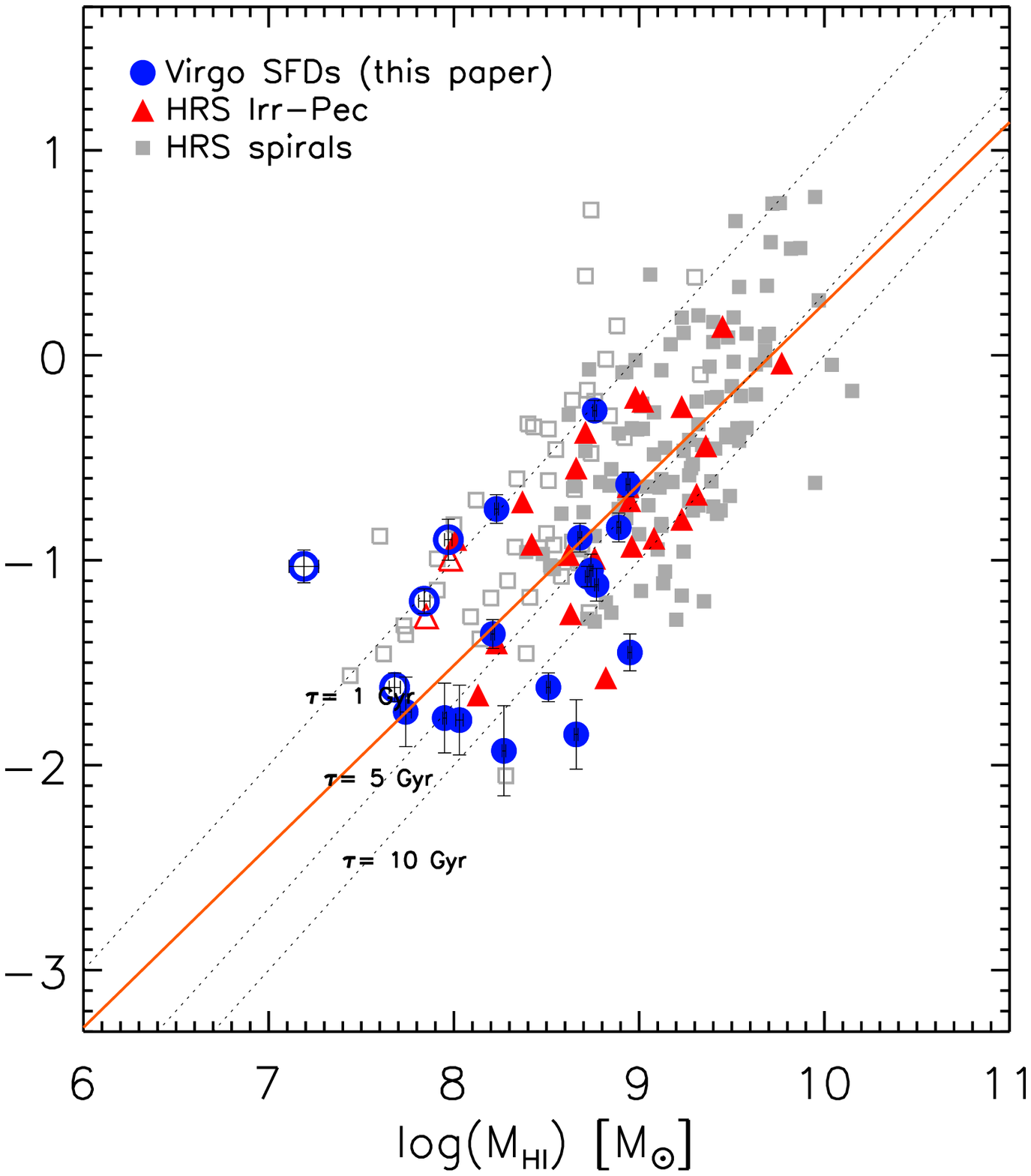}
\includegraphics[bb=80 0 570 550,width=6.05cm,height=6.6cm]{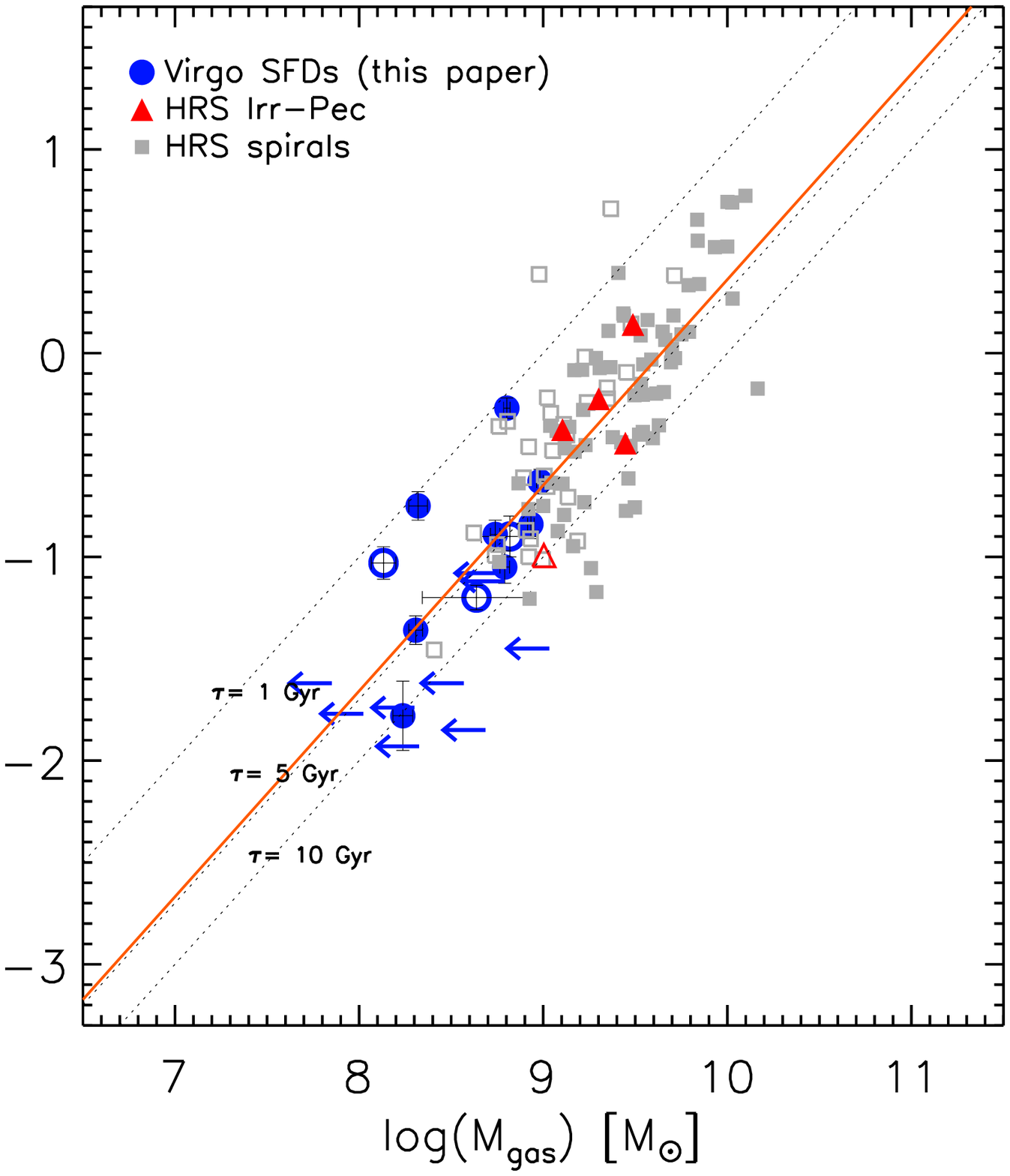}
\caption{{\em Left panel:} Star formation rate against H$_2$ mass. Virgo SFDs (blue dots) are compared to
the HRS spiral (grey squares) and irregular-peculiar (red triangles) galaxies. H{\sc i}-deficient galaxies are indicated with
empty symbols, and blue arrows are Virgo dwarf upper limits.
The solid line shows the best-fit relation.
Dotted lines indicate constant
molecular gas depletion timescales of 0.1, 1, and 10 Gyr.
{\em Central panel:} Star formation rate against \hi\ mass. Symbols are the same as in the left panel.
The solid line shows the best-fit relation.
Dotted lines indicate constant
atomic gas depletion timescales of 1, 5, and 10 Gyr.
{\em Right panel:} Star formation rate against total gas mass (M$_{H_2}$ + M$_{HI}$).
Symbols are the same as in the left panel. The solid line shows the best-fit relation.
Total gas depletion timescales of 1, 5, and 10 Gyr are also overlaid. Coefficients of the best-fit relations are given in Table \ref{tab:sfr_laws}.
}
\label{fig:SFlaw_int}%
\end{figure*}

To fairly estimate \D,\ it is important that we consider only the \hi\ mass within the extension of the dust
disc, which we consider to be equal to $R_{25}$ \citep{2015A&A...576A..33H}.
However, without high-resolution maps of the neutral hydrogen distribution,
we can only derive an approximate estimate of this value.
To this aim we assumed that the \hi\ surface density profile follows an exponential law, $\Sigma_{HI}(r) = \Sigma_{HI}(0) e^{-r/h}$,
where $\Sigma_{HI}(0)$ is the central \hi\ surface density and $h$ the \hi\ disc scale length \citep{2001AJ....122..121V,2002A&A...390..829S}.
From the total \hi\ mass measured with ALFALFA we derived $h$, assuming   $\Sigma_{HI}(0) = $
6 \msun\ pc$^{-2}$, the average central \hi\ surface density observed by \citet{2002A&A...390..829S} for a sample of 73 late-type dwarf galaxies.
This value is also compatible with those obtained from VLA observations of a few of the Virgo dwarfs included in our sample
\citep{2003AJ....126.2774H}.
Then we calculated $\Sigma_{HI}(R_{25})$ and the corresponding \hi\ mass.
The uncertainty on the masses was estimated assuming that $\Sigma_{HI}(0)$ varies between 4 and 8
\msun\ pc$^{-2}$, the 1$\sigma$ dispersion determined in the sample of \citet{2002A&A...390..829S}.

In the lower panel of Fig. \ref{fig:gas2dust}, we show the same plot as in the upper panel, with \hi\ masses calculated within the optical radius
$R_{25}$. The figure shows that when we follow this approach, most of the dwarfs with a normal \hi\ content lie along the linear
relation defined by Eq. \ref{eq:d2g}.
This confirms the importance of calculating
the total gas and dust masses within the same region to prevent biases in the estimate of
\D,\, as has also been stressed by previous works  \citep{2007ApJ...663..866D,2012ApJ...752..112H,2014A&A...563A..31R}.

\subsection{Integrated star formation laws}
\label{sec:sfr-law}

Molecular gas and star formation show an extremely tight correlation across a wide range of galaxy types, both in the local Universe and
at high redshift, through the Kennicutt-Schmidt relation \citep{1959ApJ...129..243S,1989ApJ...344..685K,1998ApJ...498..541K}.
The slope of the power-law relation and the rank correlation depend on
the spatial resolution sampled by the data (Bigiel et al. 2008), the type of galaxy
(star-forming, starburst), the examined gas component (\hdue, \hi, or total gas), and the local environment
\citep[inner or outer disc;][]{2008AJ....136.2846B,2011ApJ...741...12B,2011ApJ...737...12L,2011AJ....142...37S}.

Studying the star formation law in dwarf galaxies allows us to probe
a different physical regime of the ISM compared to spirals because the dominant gas component is neutral hydrogen on a large scale,
and surface gas densities and metal abundances are lower \citep{1997PASP..109..937H,2011AJ....142..121H,2012AJ....143..138S}, similarly to the
outer region of spiral discs \citep{2015ApJ...805..145E}.

We only had integrated measurements available (total \hi, \hdue\ and total SFRs), therefore
we examine in Fig. \ref{fig:SFlaw_int} the integrated star formation laws for the different gas components: molecular, atomic, and total gas.
Again we compared Virgo dwarfs to the HRS sample\footnote{SFRs for the HRS survey have been taken from \citet{2015A&A...579A.102B}
and converted into a Kroupa IMF.
They were derived using the same calibration
(H$\alpha$ + 22 $\mu$m) as was applied to our sample.} --
covering approximately three orders of magnitude in gas masses and SFRs.
A linear fit to the data points (including only galaxies with a normal \hi\ content) is consistent with a linear relation
for the molecular and total components (left and right panels of Fig. \ref{fig:SFlaw_int}, orange lines), with a slope of 0.93 and 1.01,
respectively (Table \ref{tab:sfr_laws}). For the atomic gas component
the slope is slightly lower, 0.88, indicating that overall, molecular gas is better correlated with star formation than \hi\
(see Table \ref{tab:sfr_laws}).

\begin{figure*}
  \centering
\includegraphics[bb=50   0 570 550,width=7.0cm]{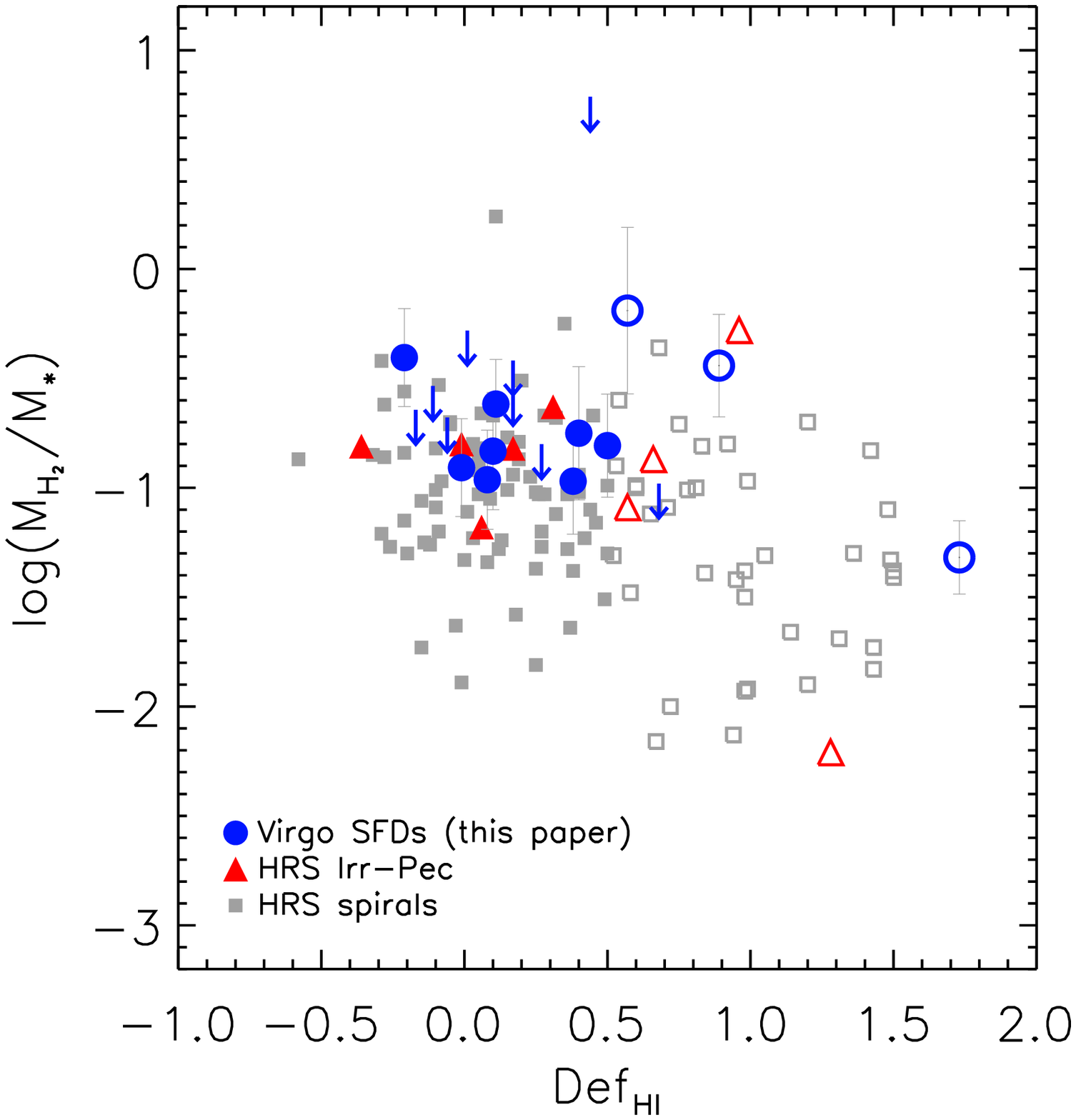}
\includegraphics[bb=50   0 570 550,width=7.0cm]{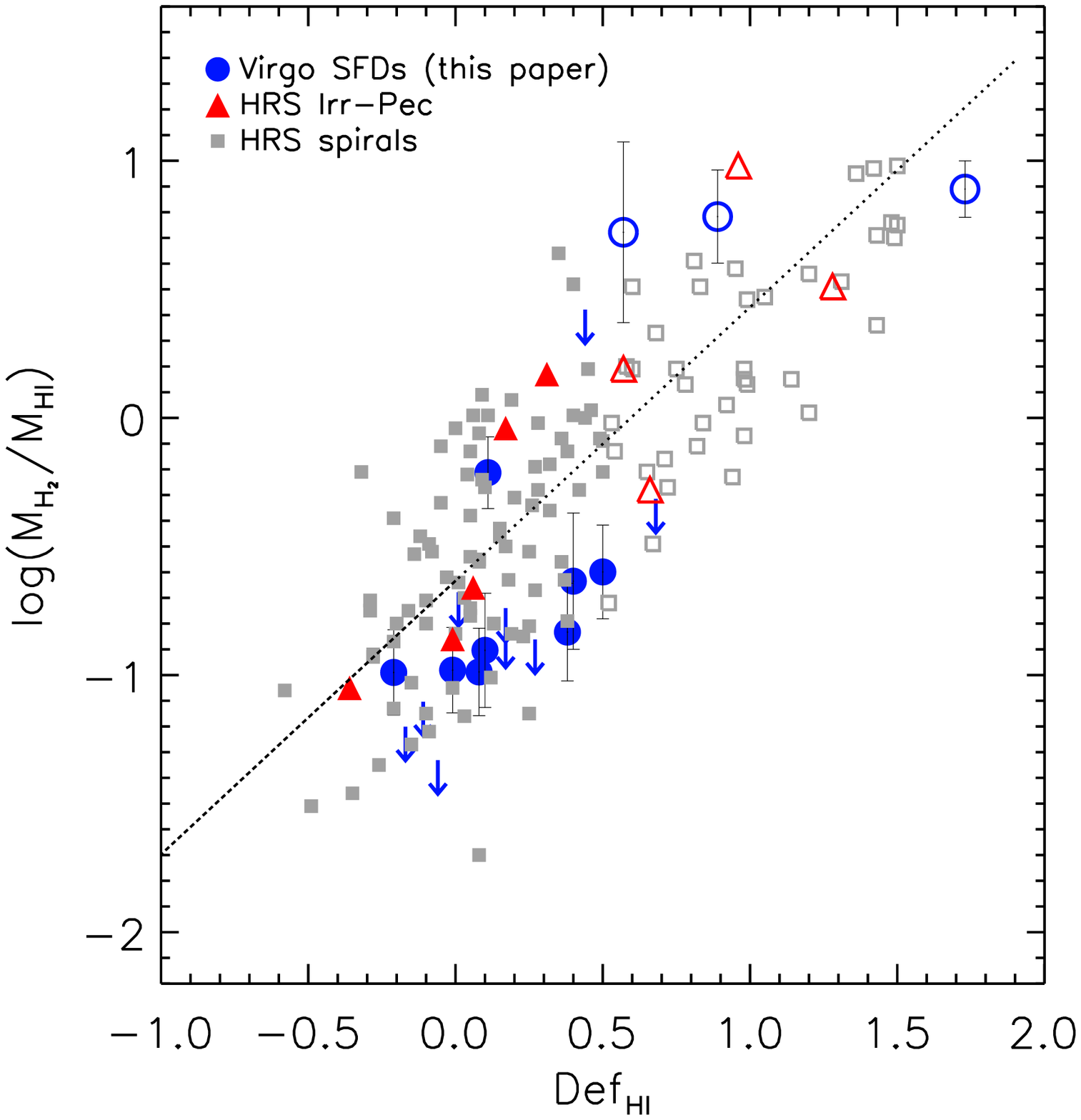}
\includegraphics[bb=50   0 570 550,width=7.0cm]{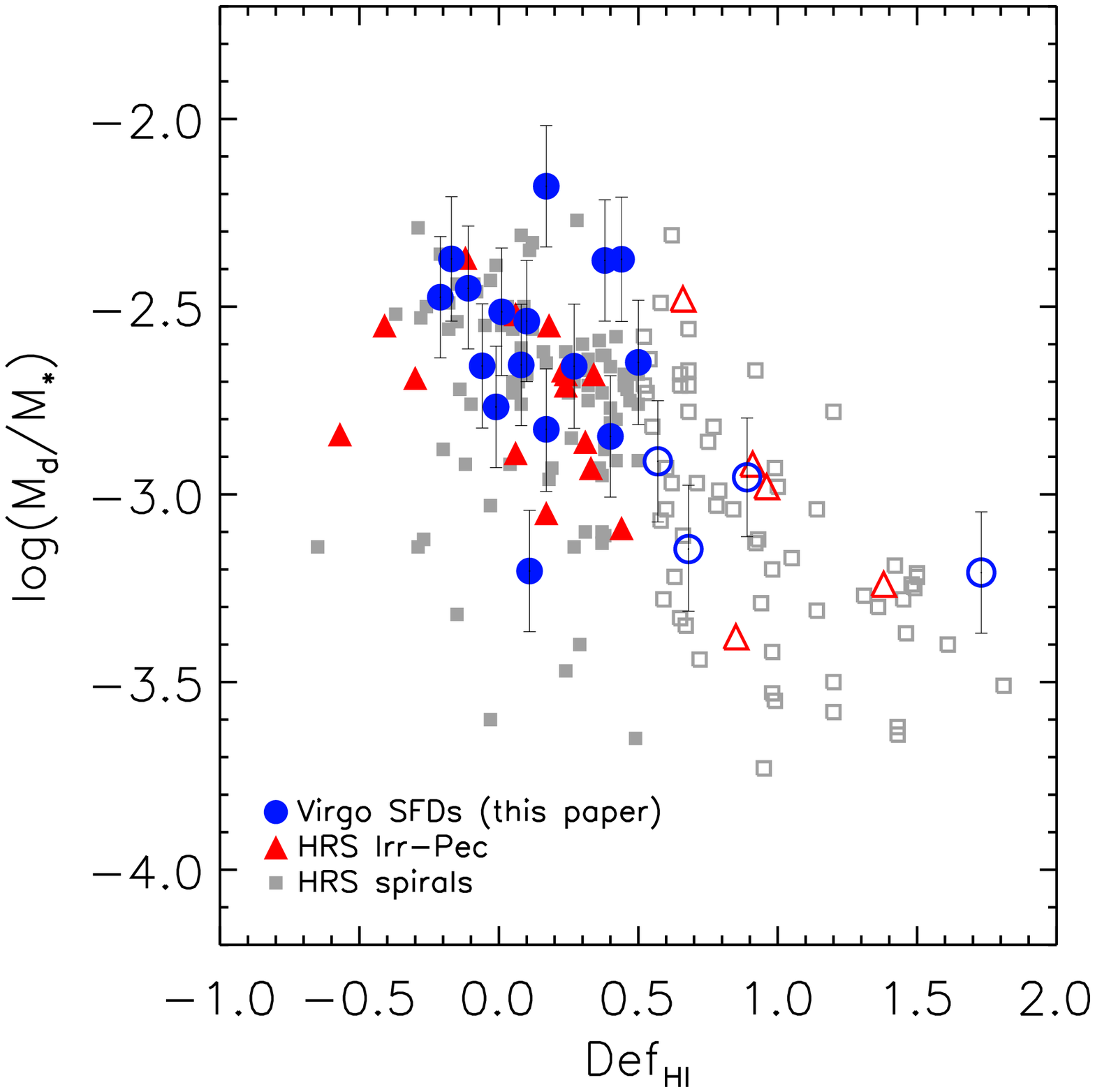}
\includegraphics[bb=50   0 570 550,width=7.0cm]{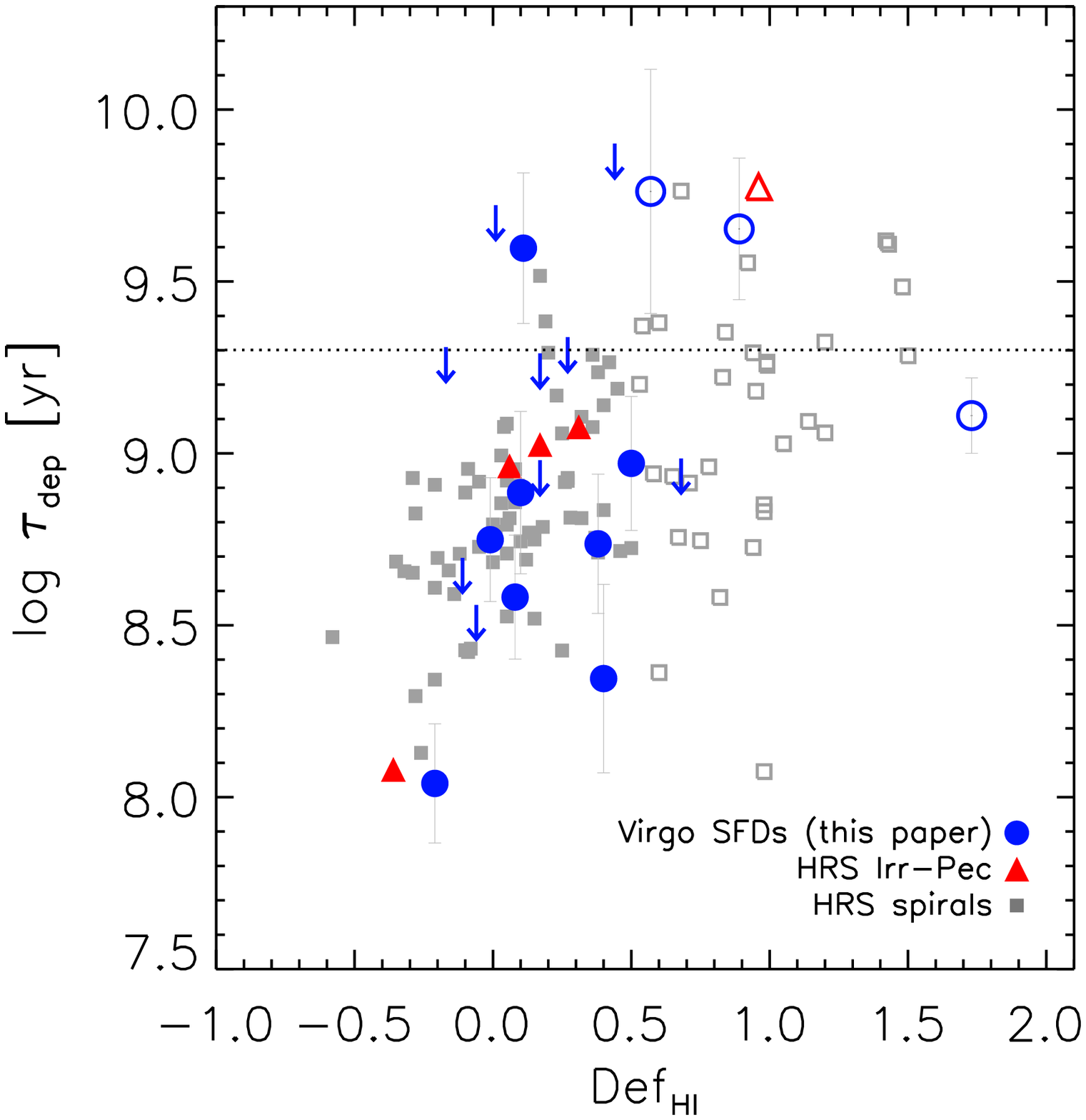}
\caption{Environmental effects on the different ISM components. {\em Top left:} Ratio of \hdue\- to stellar mass against \hi\
 deficiency.
{\em Top right:} Ratio of \hdue\- to \hi\ mass against \hi\  deficiency. {\em Bottom-left:} Ratio of dust to stellar mass against \hi\  deficiency.
{\em Bottom right:} Molecular gas depletion timescale against \hi\  deficiency.
Symbols are the same as in Fig. \ref{fig:H2frac}.
}
   \label{fig:tau_HIdef}%
\end{figure*}

\begin{table}[t]
\centering
\caption{Best-fit coefficients of the integrated star formation laws: $\log$ (SFR) = $a$ x + $b$.}
\begin{tabular}{ccc}
\hline \hline
x & $a$ & $b$ \\
\hline \hline
$\log$ (M$_{H_2}$) & 0.92 $\pm$ 0.07 & -8.32 $\pm$ 0.22 \\
$\log$ (M$_{H_I}$) & 0.88 $\pm$ 0.07 & -8.58 $\pm$ 0.22 \\
$\log$ (M$_{gas}$) & 1.01 $\pm$ 0.08 & -9.74 $\pm$ 0.24 \\
\hline \hline
\end{tabular}
\label{tab:sfr_laws}
\end{table}

The ratio of the molecular mass to the star formation rate, $\tau_{dep} =$ M$_{H_2}$/SFR,
measures the molecular gas depletion timescale, that is, the amount of time for which a galaxy can sustain star formation at the
current rate without accreting gas from the external environment (closed-box model). The inverse of this parameter gives the star formation
rate per unit of molecular gas, or the efficiency at transforming \hdue\ into stars.
Dotted lines show constant
molecular gas depletion timescales of 0.1, 1, and 10 Gyr, with the main locus of the distribution lying at $\tau = $ M$_{H_2}$/SFR =  1 Gyr.
Resolved studies
of the discs of nearby spiral galaxies on kpc scales have shown that $\tau_{dep}$ is roughly constant at $\approx$ 2 Gyr
\citep{2008AJ....136.2846B,2013AJ....146...19L}.
However, systematic variations of the \hdue\ depletion timescale are found when
$\tau_{dep}$ is considered to be averaged over the entire galaxy instead of spatially resolved maps 
\citep{2011MNRAS.415...61S,2013AJ....146...19L,2014A&A...564A..66B,2015A&A...583A.114H}.
Low-mass, low-metallicity, and gas-rich systems are on average more efficient at producing stars than normal spiral discs, and trends are observed
between $\tau_{dep}$ and stellar mass \citep{2013AJ....146...19L}, specific SFR \citep{2015A&A...583A.114H}, and metallicity \citep{2014A&A...564A..66B}.
Molecular gas depletion timescales vary among the Virgo SFDs. 
The dwarfs with a normal \hi\ content are more efficient at forming stars than normal discs, and five out of ten objects have $\tau_{dep}$ between
100 Myr and 1 Gyr,
while all the \hi-deficient objects have $\tau_{dep}$
comparable to or longer than the value found in normal spiral galaxies ($\sim$ 2 - 3 Gyr).

VCC1437 is the dwarf with the longest molecular gas consumption rate among the Virgo sample ($\sim$ 4.5 Gyr). 
It has a low star formation rate ($\sim$ 0.02 \msun\ yr$^{-1}$),
despite its consistent reservoir of both atomic and molecular gas (36\% of the total
baryonic mass), thus it is not clear why star formation appears to be very inefficient in this galaxy.
SDSS images of VCC1437 shows a spherically symmetric stellar component, with a
centrally concentrated region of star formation activity. The galaxy is classified as a nuclear elliptical (nE) BCD \citep{2006AJ....132.2432L,2014A&A...562A..49M} according to the BCD morphological classification of \citet{1986sfdg.conf...73L}, meaning that the nuclear star-forming region is overlaid on an old population of low-mass stars with a spherical or elliptical distribution.
Thus, it is likely that this dwarf galaxy is in a different evolutionary stage than the other systems, which do not show such a spherical symmetric stellar component and host patchy star formation regions throughout the disc. 

\subsection{Environmental effects on the ISM}
\label{subsec:sfe}

From the combined analysis of the dust, atomic, and molecular hydrogen components we can assess how the ISM of Virgo SFDs
is perturbed and depleted as they fall into the cluster substructures.
In this section we compare IRAM data with {\it Herschel} and 21 cm observations
to identify variations in the cold phases of the ISM (and eventually in the star formation activity)
according to the stage of interaction with the cluster environment.
The different levels of \hi\  deficiency can provide information on the degree of perturbation the galaxies are experiencing.

In Fig. \ref{fig:tau_HIdef} we display the molecular gas-to-stellar mass ratio (top-left panel), the \hdue-to-\hi\ ratio (top-right panel),
the dust-to-stellar mass ratio (bottom-left panel), and the molecular gas depletion timescale
against \hi\ deficiency (bottom right panel).
Virgo SFDs are compared to the HRS galaxies.
The M$_{H_2}$/M$_*$ ratio of the \hi-deficient dwarfs does not remarkably differ from that  of the \hi-normal ones (top-left panel),
suggesting that despite the poor statistics, there is no molecular gas deficiency in the \hi-poor dwarfs. A modest variation of M$_{H_2}$/M$_{HI}$ can be seen in the HRS sample.
\citet{2014A&A...564A..67B} claimed that Virgo cluster galaxies have, on average, a lower molecular gas content than similar objects
in the field and that \hdue\ is also removed by the cluster environment through ram-pressure stripping, but less efficiently so than the atomic gas.
This different efficiency is shown in the top right panel, where we display the variation of \hdue/\hi\ mass ratio with $Def_{HI}$.
Virgo SFDs with high values of this ratio are the most \hi-deficient, similarly to the HRS objects, and most of the galaxies in the plot follow a linear relation (dotted line),
which implies a normal \hdue\ content \citep{1986ApJ...301L..13K}.
Both figures suggest that the mechanism removing the low-density atomic gas has left the molecular component mostly intact
at the current stage of evolution of the star-forming dwarfs within the cluster, probably because of their different spatial distribution
\citep[\hdue\ is more centrally concentrated than \hi, with a shorter scale length;][]{2009AJ....137.4670L}.
Highly HI-deficient galaxies are also characterised by lower dust fractions (bottom left panel), which
confirms that dust stripping occurs with \hi\ stripping in both low- and high-mass galaxies in the Virgo cluster \citep{2012A&A...542A..32C,2012A&A...540A..52C,2015A&A...574A.126G}.

The amount of gas available to sustain star formation depends on various factors, including the inflow of gas from the
external environment, the re-accretion of ejected gas, and the availability of large gas reservoirs (extended \hi\ discs).
If all these factors are affected by the interaction with the cluster environment, a reduced star formation activity as well
as longer gas depletion timescales might be expected. In the bottom right panel of Fig. \ref{fig:tau_HIdef} we plot
$\tau_{dep}$ against \hi\ deficiency. The two parameters appear correlated on a
statistical basis (Spearman rank correlation coefficient $r_s = 0.57$, two-sided significance of its deviation from zero $p = 4.8 \times 10^{-11}$), and the most \hi-deficient galaxies show a less efficient star formation
activity.
Both samples show that the molecular gas depletion timescales increase with the \hi\ deficiency.
The 2 Gyr value determined in the HERACLES sample \citep{2008AJ....136.2846B}
is indicated by the dotted line in the figure.
Because the greatest \hi\ depletion so far has occurred mainly in the outer discs, outside the region in which most of the
molecular gas is located, this plot seems to suggest that star formation in the \hi-deficient dwarfs is slowing down
because the gas supply is not replenished from the outer regions of the disc.

\section{Summary and conclusions}
\label{sec:conclusions}

We have presented IRAM 30 m telescope observations at 115 and 230 GHz of a sample of 20 Virgo cluster SFD galaxies.
The dwarfs, selected from the {\it Herschel} Virgo Cluster Survey, have oxygen abundances
within the range 8.1 $\lesssim$ 12 + log(O/H) $\lesssim$ 8.8.
$^{12}$CO(1-0) and $^{12}$CO(2-1) emission is detected in 11 out of 20 objects down to an oxygen abundance of 12 + log(O/H) = 8.1, including a tentative
$\simeq 3\sigma$ detection of VCC1725 at 115 GHz.
The most significant difference between detections and non-detections is the higher stellar mass,
and, consequently, the higher star formation rate.

We corrected CO fluxes taking into account both aperture corrections
(when the CO emission was
more extended than the IRAM 30 m telescope beam size) and the finite source size (if the emitting region is smaller than the beam).
CO fluxes correlate with FIR 250 $\mu$m emission with a slope that is close to linear.
The correlation holds from Virgo spirals to the dwarf sample and covers three orders of magnitude in both CO fluxes and 250 $\mu$m flux densities.
The link between cold dust and molecular gas appears to be stronger than that with warm dust traced by MIR emission.

\hdue\ masses were derived using a $H$-luminosity-dependent factor
\citep{2002A&A...384...33B} and a metallicity-dependent CO-to-\hdue\ conversion factor \citep{2010ApJ...716.1191W}, and we showed that both methods lead to comparable mass estimates.
The ratio of molecular to stellar mass of Virgo SFDs with a normal \hi\  content
remains nearly unchanged below M$_* \sim 10^9$ \msun,
contrary to the \hi\ fraction, M$_{HI}/$M$_*$, which increases inversely with M$_*$ \citep{2015A&A...574A.126G}.
The total gaseous budget is dominated by the atomic hydrogen
component in SFDs, which have a normal \hi\ content.
The mean molecular hydrogen content is $\sim$14\% of the total gas, compared to 25-30\% in spiral galaxies \citep{2014A&A...564A..66B}.
On the other hand, \hdue\ is the main gaseous component in \hi-poor systems.
The molecular-to-atomic ratio is better correlated with stellar surface density than metallicity, confirming that the interstellar gas pressure plays
a key role in determining the balance between the two gaseous components of the ISM.

Virgo dwarfs follow the same linear trend between  molecular, atomic, and total gas mass and star formation rate of more massive spirals,
but the correlation and the scatter around it is tighter for molecular and total gas than for atomic hydrogen.
Molecular gas depletion timescales, $\tau_{dep}$, vary between 100 Myr and 6 Gyr, with one galaxy, VCC1437, showing a remarkably low star formation
efficiency despite its consistent reservoir of both \hi\ and \hdue.

The interaction with the cluster environment is stripping the low-density atomic gas and dust, 
 while the more centrally concentrated molecular gas component appears to be mostly intact at the current stage of evolution
of the dwarfs. We did not find any particular difference in the ratio of molecular to stellar mass of \hi-normal or \hi-deficient galaxies, contrary to the
other components of the ISM.
The correlation between \hi\ deficiency and $\tau_{dep}$ suggests that the lack of gas replenishment from the outer regions of the disc also
lowers the star formation activity.
As the galaxies enters the denser regions of the cluster,  their ISM will be removed by ram-pressure stripping, and star formation will be gradually quenched.
They will evolve into quiescent galaxies, but they might be able to conserve, at
least on short timescales, their angular momentum and rotation \citep{2009ApJ...707L..17T}.
Thus, our sample may give hints about the precursors of rotation-supported early-type dwarfs, with signs of discs and blue nuclei,
which are usually found at larger distances to the core of the Virgo cluster \citep{2006AJ....132.2432L,2009ApJ...707L..17T,2014A&A...562A..49M}.

\section{Acknowledgments}
We thank the anonymous referee for the constructive and timely comments that helped us to improve the manuscript.
M.G. gratefully acknowledges support from CAPES (through grant "A forma\c{c}\~ao de gal\'axias starburst: hist\'orico dos \'ultimos 10 bilh\~oes de anos",
Call n. 001/2010).
I.D.L. gratefully acknowledges the support
of the Flemish Fund for Scientific Research (FWO-Vlaanderen).
L.K.H. acknowledges support from PRIN-INAF 2012/13.
We are grateful to the IRAM 30 m staff for their support during the observations.
We thank Sandra Trevi\~no-Morales for helping us to retrieve observation files from the IRAM archive.
This work has benefited from research funding from the European Community's Seventh Framework Programme.
M.G., L.B., and E.L. acknowledge travel support to Pico Veleta from TNA Radio Net
project funded by the European Commission within the FP7 Programme.
The research leading to these results has received funding from the European Commission Seventh Framework Programme
(FP/2007-2013) under grant agreement N° 283393 (RadioNet3).




\bibliographystyle{aa} 
\bibliography{SFdwarfbib} 

\begin{appendix}

\section{Supplementary tables}
\label{app:tables}

\begin{table}[ht]
\caption{Pointing list, total integration time (ON+OFF), and Herschel/SPIRE photometry at 250 $\mu$m in a
circular aperture of radius $r = 18\farcs6$.}
\centering
\begin{tabular}{lcccr}
\hline \hline
ID & RA     & DEC    & Time  & $S_{250}^{beam}$ \\  
   & [J200] & [J200] & [min] & [mJy] \\  
\hline \hline
    VCC10        &     12:09:24.90  &   13:34:28.0    &   30   &    189.0   $\pm$   8.8    \\
     VCC87        &    12:13:41.30  &   15:27:13.0    &   65   &    105.3   $\pm$   5.2    \\
     VCC135       &    12:15:06.70  &   12:01:00.0    &   15   &    239.3   $\pm$  11.0    \\
     VCC144       &    12:15:18.30  &   05:45:39.0    &   30   &    154.1   $\pm$   7.5    \\
     VCC172       &    12:16:00.40  &   04:39:03.0    &   15   &     66.6   $\pm$   4.6    \\
    VCC213(a)     &    12:16:56.00  &   13:37:31.5    &   20   &    391.3   $\pm$  15.5    \\
    VCC213(b)     &    12:16:56.73  &   13:37:25.8    &   20   &    277.7   $\pm$   9.9    \\
     VCC324       &    12:19:09.90  &   03:51:21.0    &   30   &    220.8   $\pm$   9.2    \\
     VCC334       &    12:19:14.20  &   13:52:56.0    &  130   &     51.1   $\pm$   3.9    \\
     VCC340       &    12:19:22.10  &   05:54:38.0    &   30   &    152.5   $\pm$   6.9    \\
     VCC562       &    12:22:35.90  &   12:09:27.0    &   40   &     53.2   $\pm$   3.5    \\
     VCC693       &    12:24:03.35  &   05:10:50.4    &   30   &     78.7   $\pm$   4.5    \\
    VCC699(a)     &    12:24:07.50  &   06:36:27.8    &   30   &    363.0   $\pm$  12.6    \\
    VCC699(b)     &    12:24:08.14  &   06:36:31.0    &   30   &    343.8   $\pm$  12.1    \\
     VCC737       &    12:24:39.40  &   03:59:44.0    &   55   &     90.8   $\pm$   4.6    \\
     VCC841       &    12:25:47.40  &   14:57:07.5    &   50   &     94.2   $\pm$   6.0    \\
    VCC1437        &    12:32:33.50  &   09:10:25.0   &   20   &    134.2   $\pm$   6.7    \\
   VCC1575(a)      &    12:34:39.96  &   07:09:52.0   &   10   &    467.4   $\pm$  17.2    \\
   VCC1575(b)      &    12:34:39.65  &   07:09:41.8   &   10   &    580.2   $\pm$  20.1    \\
   VCC1575(c)      &    12:34:39.28  &   07:09:32.2   &   10   &    638.0   $\pm$  21.0    \\
   VCC1575(d)      &    12:34:38.80  &   07:09:23.8   &   10   &    527.2   $\pm$  19.0    \\
   VCC1575(e)      &    12:34:40.06  &   07:10:02.8   &   10   &    270.0   $\pm$   8.2    \\
   VCC1686(a)      &    12:36:44.26  &   13:15:27.1   &   30   &    434.8   $\pm$  15.4    \\
   VCC1686(b)      &    12:36:43.73  &   13:15:13.2   &   10   &    226.3   $\pm$   8.1    \\
   VCC1686(c)      &    12:36:43.73  &   13:15:02.2   &   10   &    338.8   $\pm$  11.5    \\
   VCC1699         &    12:37:02.60  &   06:55:31.0   &   70   &    126.3   $\pm$   6.8    \\
   VCC1725         &    12:37:40.57  &   08:33:37.4   &   30   &    139.4   $\pm$   5.9    \\
   VCC1791         &    12:39:25.87  &   07:58:04.6   &   30   &     98.0   $\pm$   6.2    \\
\hline \hline
\end{tabular}
\label{tab:point_list}
\end{table}

\begin{table}[ht]
\caption{Aperture corrections calculated using the methods
of \citet{2011MNRAS.415...32S} and \citet{2011A&A...534A.102L}, source sizes, and correction factors for the finite angular sizes of the source.}
\centering
\begin{tabular}{lcccccc}
\hline \hline
ID & $f_{ap}^{S11}$ & $f_{ap}^{L11}$ & $2R_u$  & $\theta_g$  & $f_{s}^u$ & $f_{s}^g$  \\
   &                &                & [$\arcsec$] & [$\arcsec$] &         &            \\
\hline \hline
  VCC10  &   1.48  &   1.46$\pm$0.44  &   $<$ 20             &  $<$ 19              & $1.07^{0.24}_{0.07}$ &  $1.19^{0.56}_{0.19}$  \\
  VCC87  &   1.82  &   2.04$\pm$0.61  &   --                 &  --                  & --                   &   --                   \\
 VCC135  &   1.58  &   1.72$\pm$0.52  &   $3.7^{6.1}_{3.7}$  &  $2.6^{4.9}_{2.6}$   & $1.01^{0.06}_{0.01}$ &  $1.01^{0.10}_{0.01}$  \\
 VCC144  &   1.22  &   1.26$\pm$0.38  &   $<$ 13             &  $<$ 10.4            & $1.03^{0.10}_{0.03}$ &  $1.06^{0.17}_{0.06}$  \\
 VCC172  &   1.66  &   1.78$\pm$0.53  &   --                 &  --                  & --                   &   --                   \\
 VCC213  &   1.41  &   1.65$\pm$0.49  &   ext                &  ext                 & ext                  &   ext                  \\
 VCC324  &   1.73  &   2.32$\pm$0.70  &   $18.9^{7.4}_{6.2}$ &  $18.0^{14.4}_{7.7}$ & $1.28^{0.30}_{0.16}$ &  $1.67^{0.33}_{0.45}$  \\
 VCC334  &   1.18  &   1.30$\pm$0.39  &   --                 &  --                  & --                   &   --                   \\
 VCC340  &   1.53  &   1.59$\pm$0.48  &   $17.7^{6.9}_{6.4}$ &  $16.3^{12.1}_{7.4}$ & $1.24^{0.25}_{0.15}$ &  $1.55^{0.45}_{0.38}$  \\
 VCC562  &   1.22  &   1.33$\pm$0.40  &   --                 &  --                  & --                   &   --                   \\
 VCC693  &   1.58  &   2.03$\pm$0.61  &   --                 &  --                  & --                   &   --                   \\
 VCC699  &   3.12  &   3.11$\pm$0.93  &   ext                &  ext                 & ext                  &   ext                  \\
 VCC737  &   1.51  &   1.53$\pm$0.46  &   --                 & --                   & --                   &   --                   \\
 VCC841  &   1.35  &   1.37$\pm$0.41  &   --                 & --                   & --                   &   --                   \\
VCC1437  &   1.20  &   1.29$\pm$0.39  &   $12.5^{2.5}_{2.8}$ &  $10.1^{2.8}_{2.7}$  & $1.12^{0.05}_{0.05}$ &  $1.21^{0.13}_{0.10}$  \\
VCC1575  &   3.54  &   3.20$\pm$0.96  &   ext                &  ext                 & ext                  &   ext                  \\
VCC1686  &   3.17  &   4.45$\pm$1.33  &   ext                &  ext                 & ext                  &   ext                  \\
VCC1699  &   1.91  &   2.21$\pm$0.66  &   --                 & --                   & --                   &   --                   \\
VCC1725  &   1.91  &   2.33$\pm$0.70  &   ext                & ext                  & ext                  &   ext                  \\
VCC1791  &   1.68  &   1.86$\pm$0.56  &   --                 & --                   & --                   &   --                   \\
\hline \hline
\end{tabular}
\label{tab:apcorr}
\end{table}

\begin{table}[hb]
\caption{Stellar masses used to compare Virgo SFDs to the HRS sample, and $H$-band photometry.}
\centering
\begin{tabular}{lcc}
\hline \hline
ID & $\log$($M_*$)   &  $m_H$ \\
   &  [M$_{\odot}$]  & [mag] \\
   \hline \hline
VCC10     &   8.67    &  12.81  \\
VCC87     &   8.09    &  13.16    \\
VCC135    &   9.40    &  11.48  \\
VCC144    &   8.17    &  13.04  \\
VCC172    &   8.55    &  12.62  \\
VCC213    &   8.75    &  11.48  \\
VCC324    &   8.35    &  11.98  \\
VCC334    &   7.77    &  13.28  \\
VCC340    &   8.82    &  12.29  \\
VCC562    &   7.37    &  15.70  \\
VCC693    &   8.21    &   --    \\
VCC699    &   8.91    &  11.28  \\
VCC737    &   8.10    &  12.73  \\
VCC841    &   8.35    &  13.16  \\
VCC1437   &   8.43    &  12.32  \\
VCC1575   &   9.19    &  10.76  \\
VCC1686   &   8.82    &  10.99  \\
VCC1699   &   8.12    &  12.37  \\
VCC1725   &   8.42    &  12.46  \\
VCC1791   &   8.15    &  12.81  \\
\hline \hline
\end{tabular}
\label{tab:mstar_mdust_HRS_like}
\end{table}

\section{CO(1-0) and CO(2-1) spectra of Virgo SFDs}
\label{app:spectra}

\begin{figure*}
  \centering
\includegraphics[width=4.5cm]{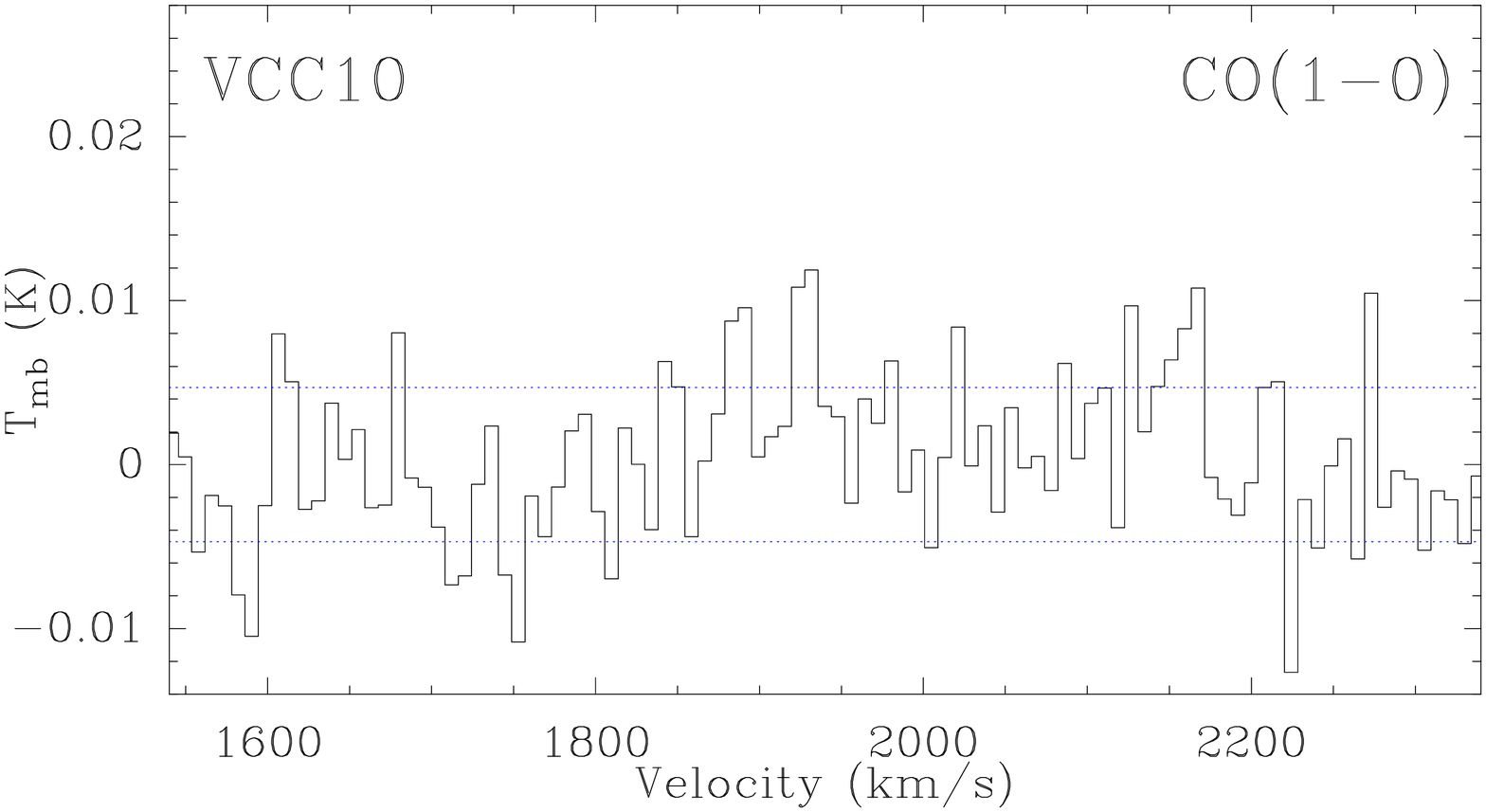}
\includegraphics[width=4.5cm]{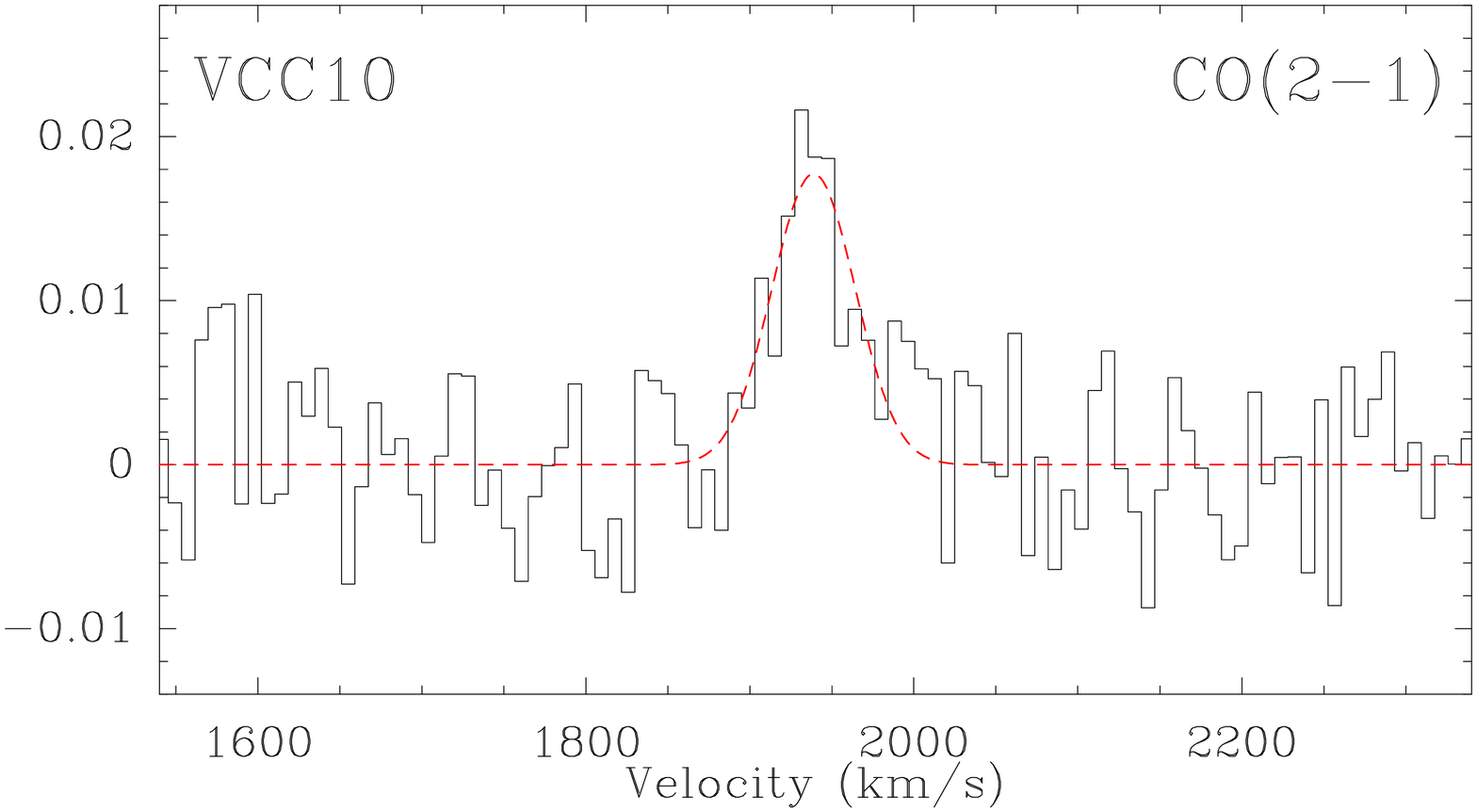}
\includegraphics[width=4.5cm]{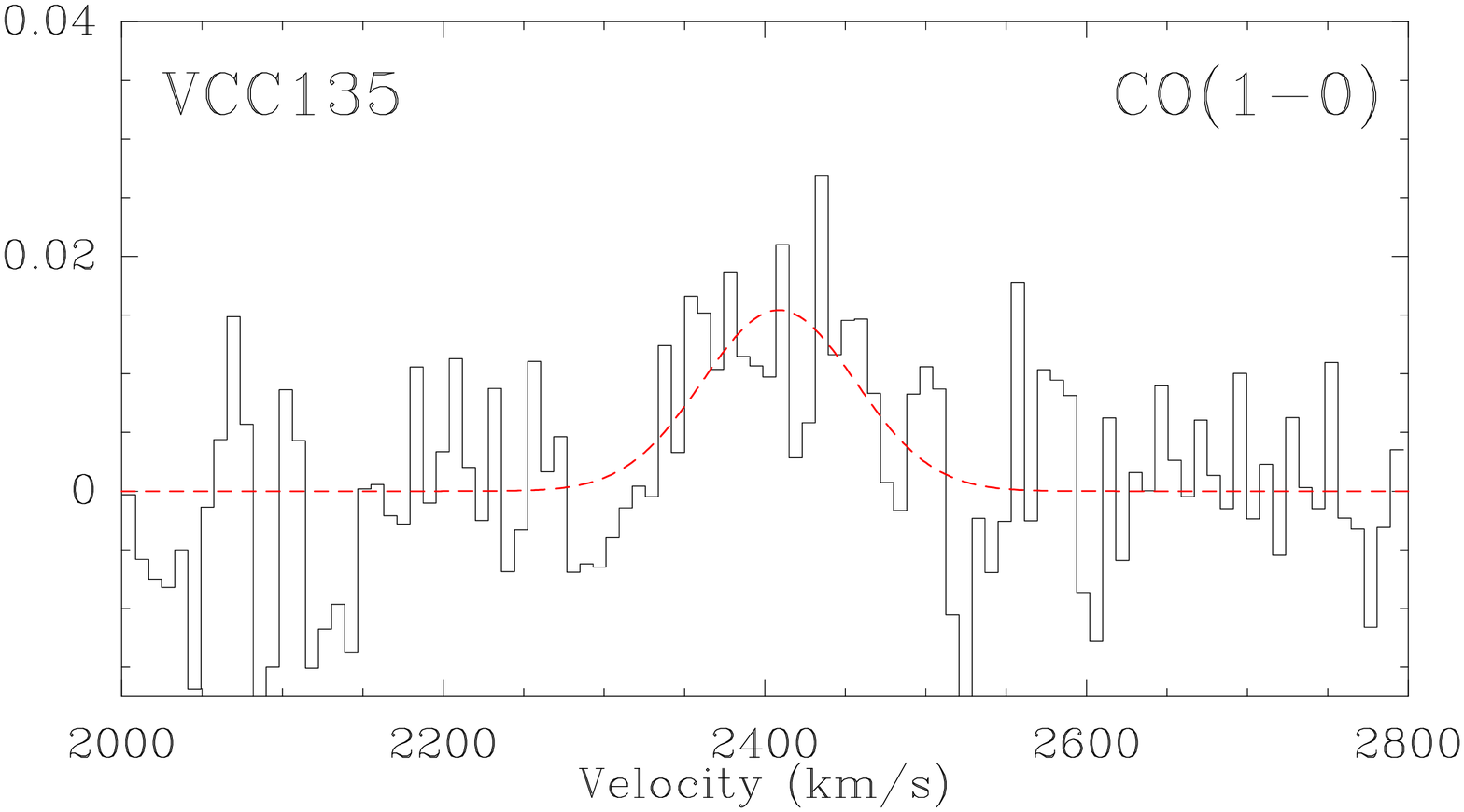}
\includegraphics[width=4.5cm]{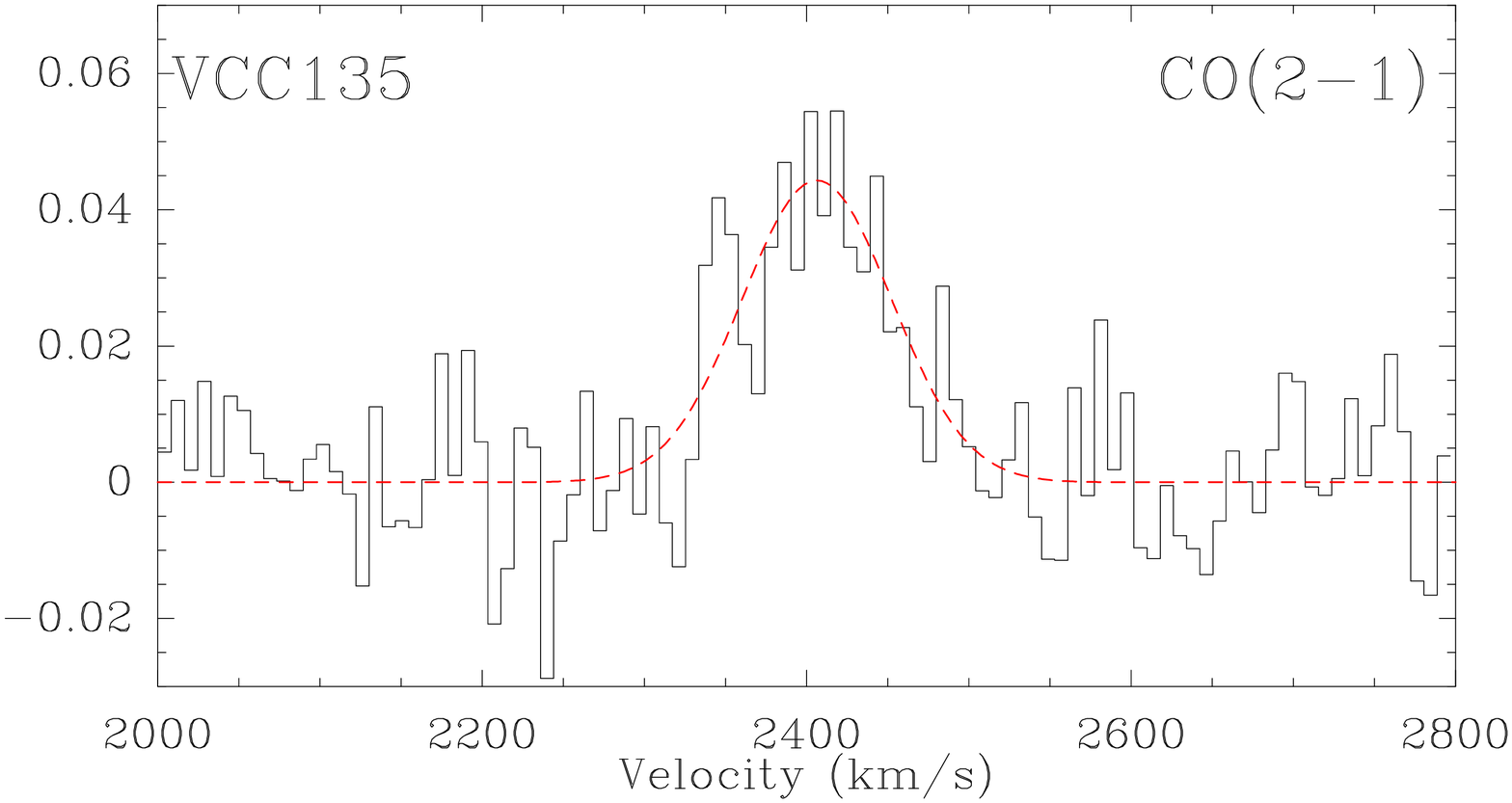}\\
\includegraphics[width=4.5cm]{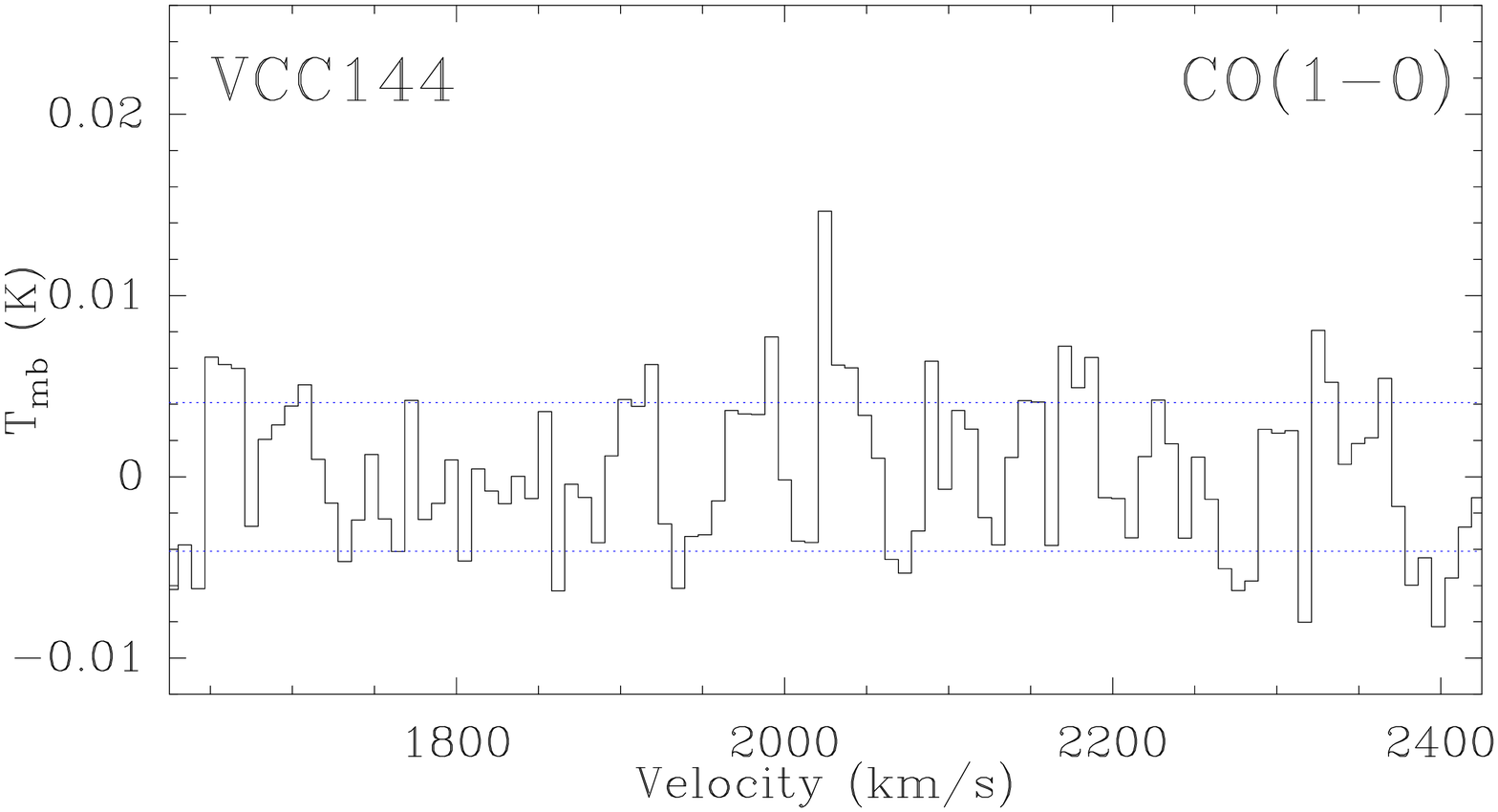}
\includegraphics[width=4.5cm]{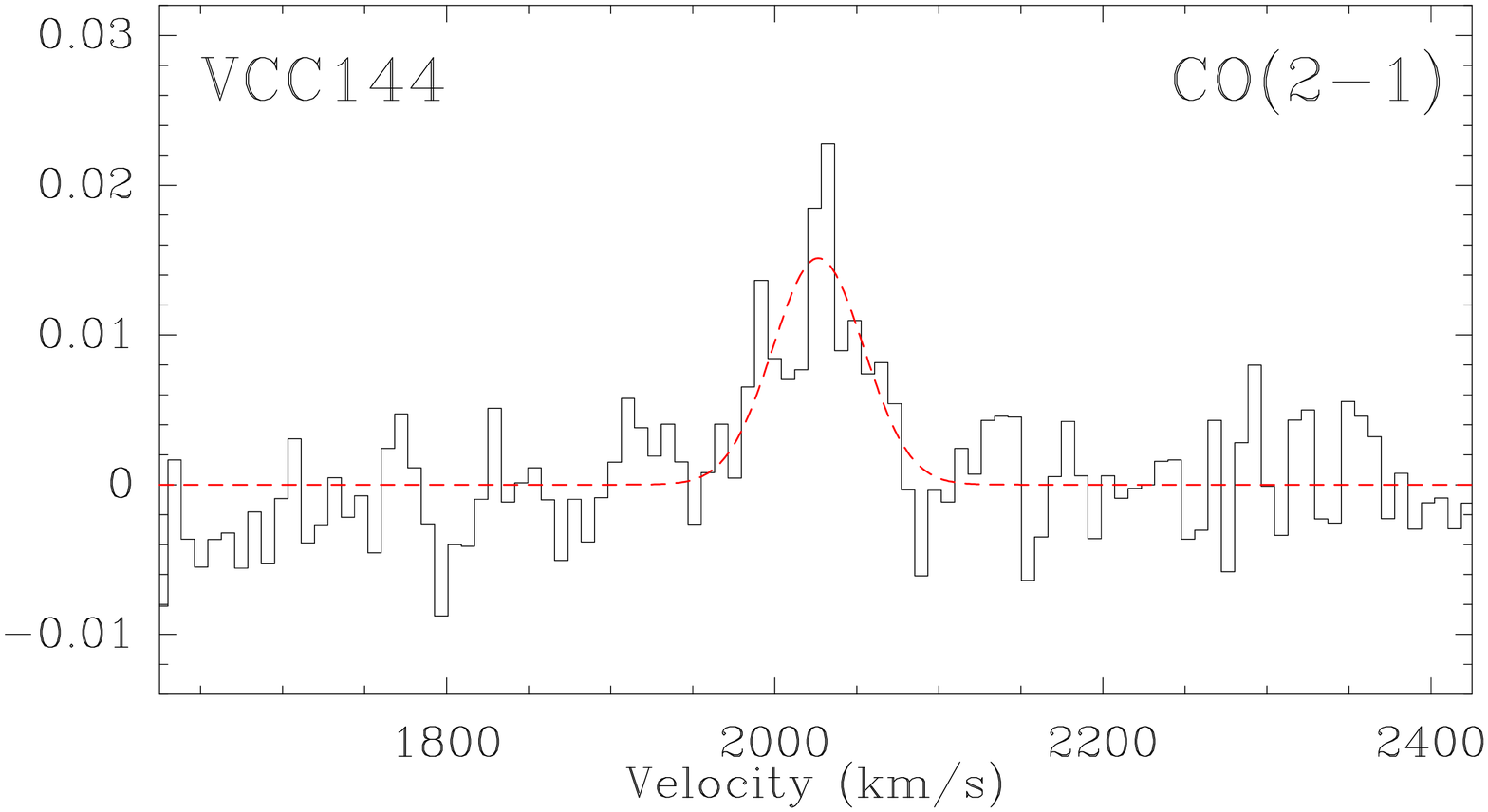}
\includegraphics[width=4.5cm]{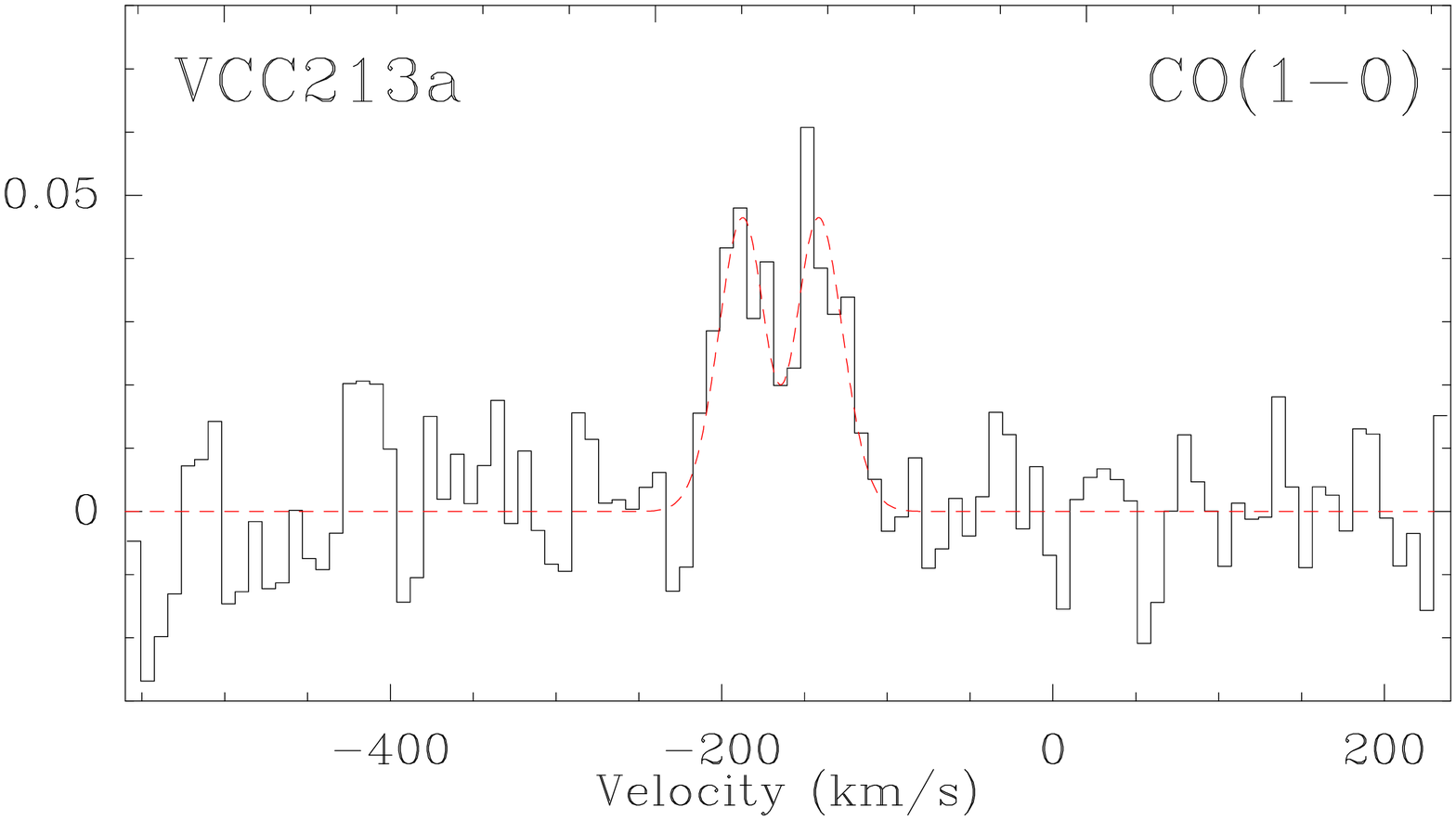}
\includegraphics[width=4.5cm]{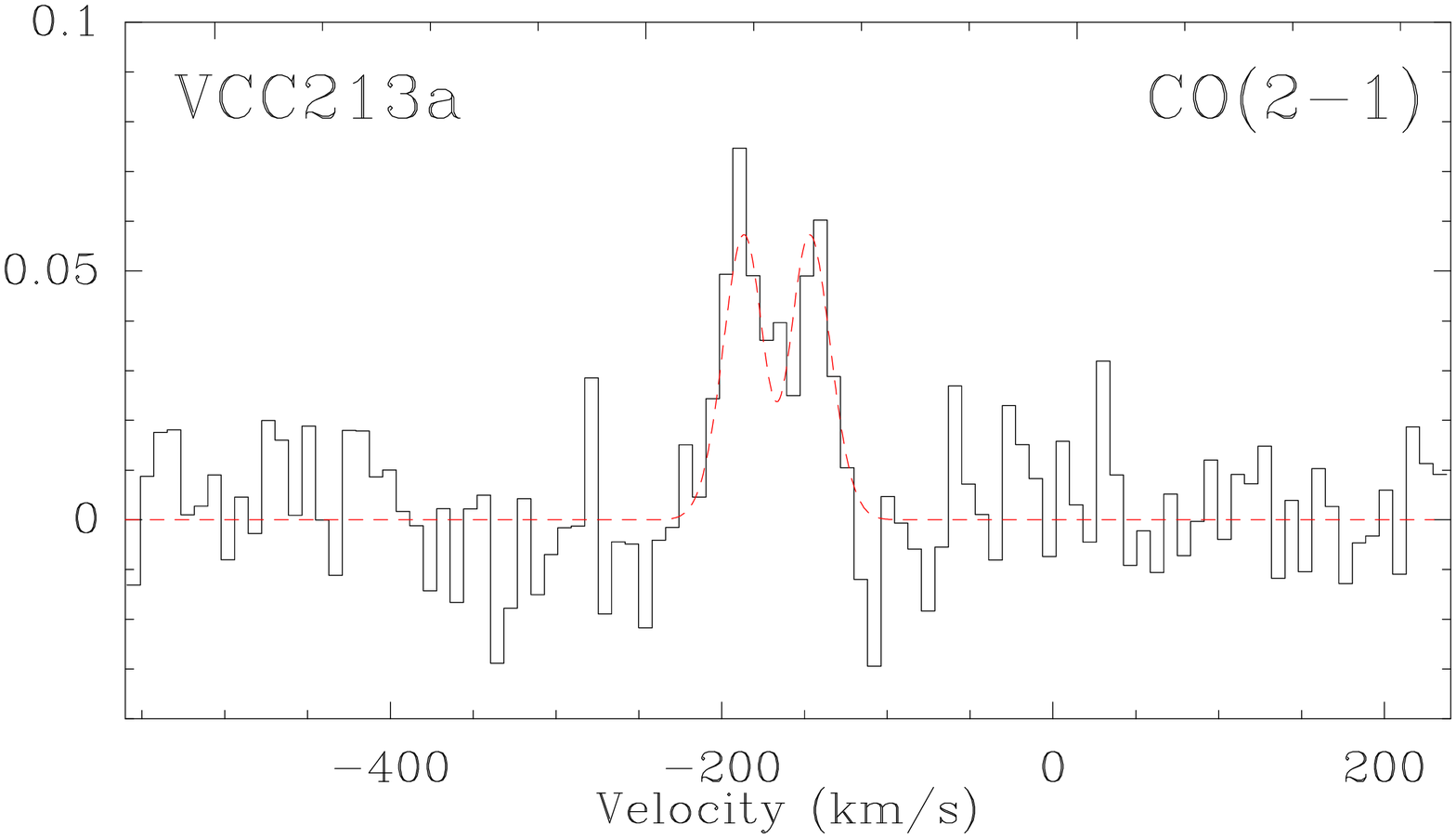}\\
\includegraphics[width=4.5cm]{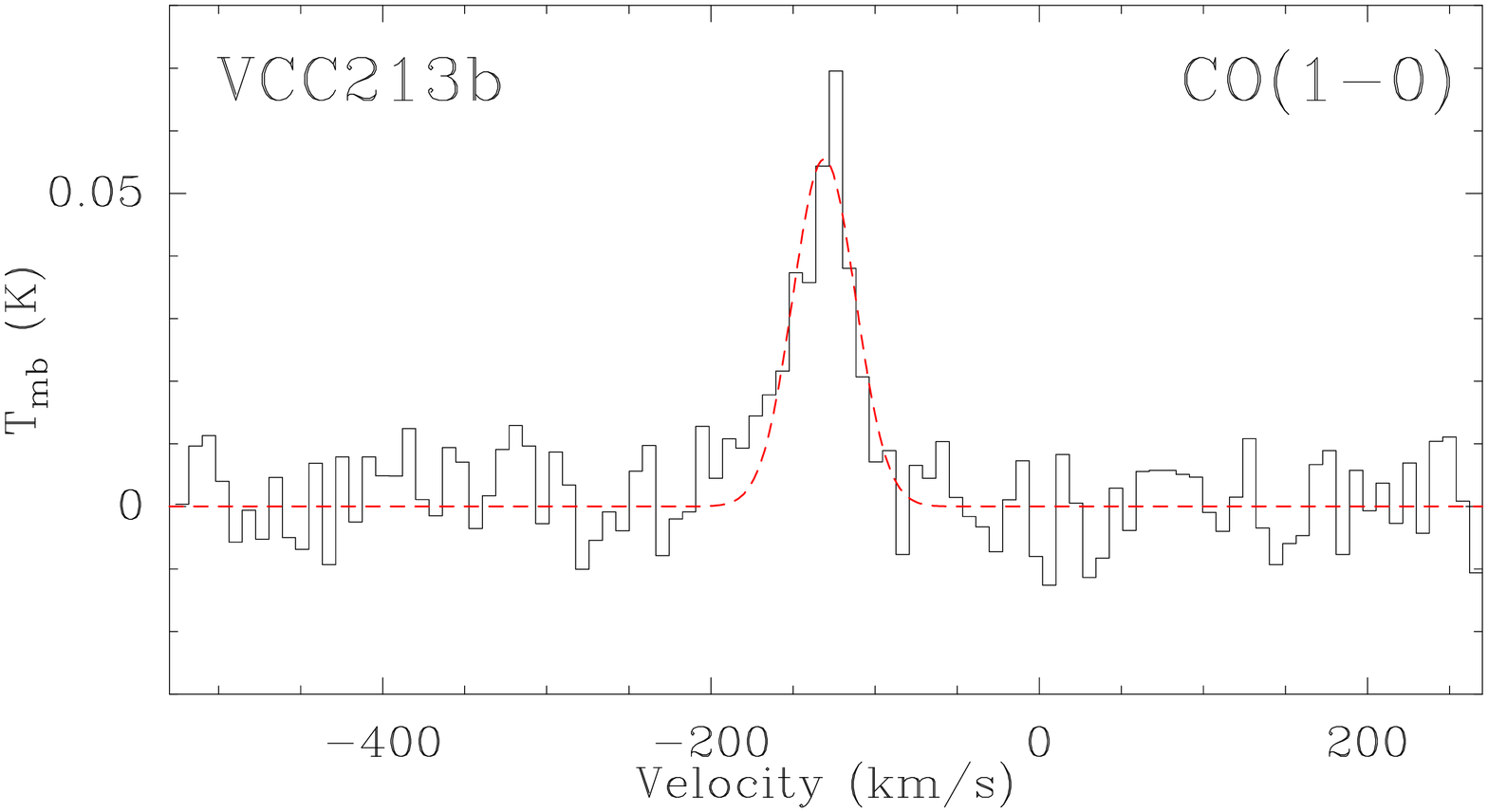}
\includegraphics[width=4.5cm]{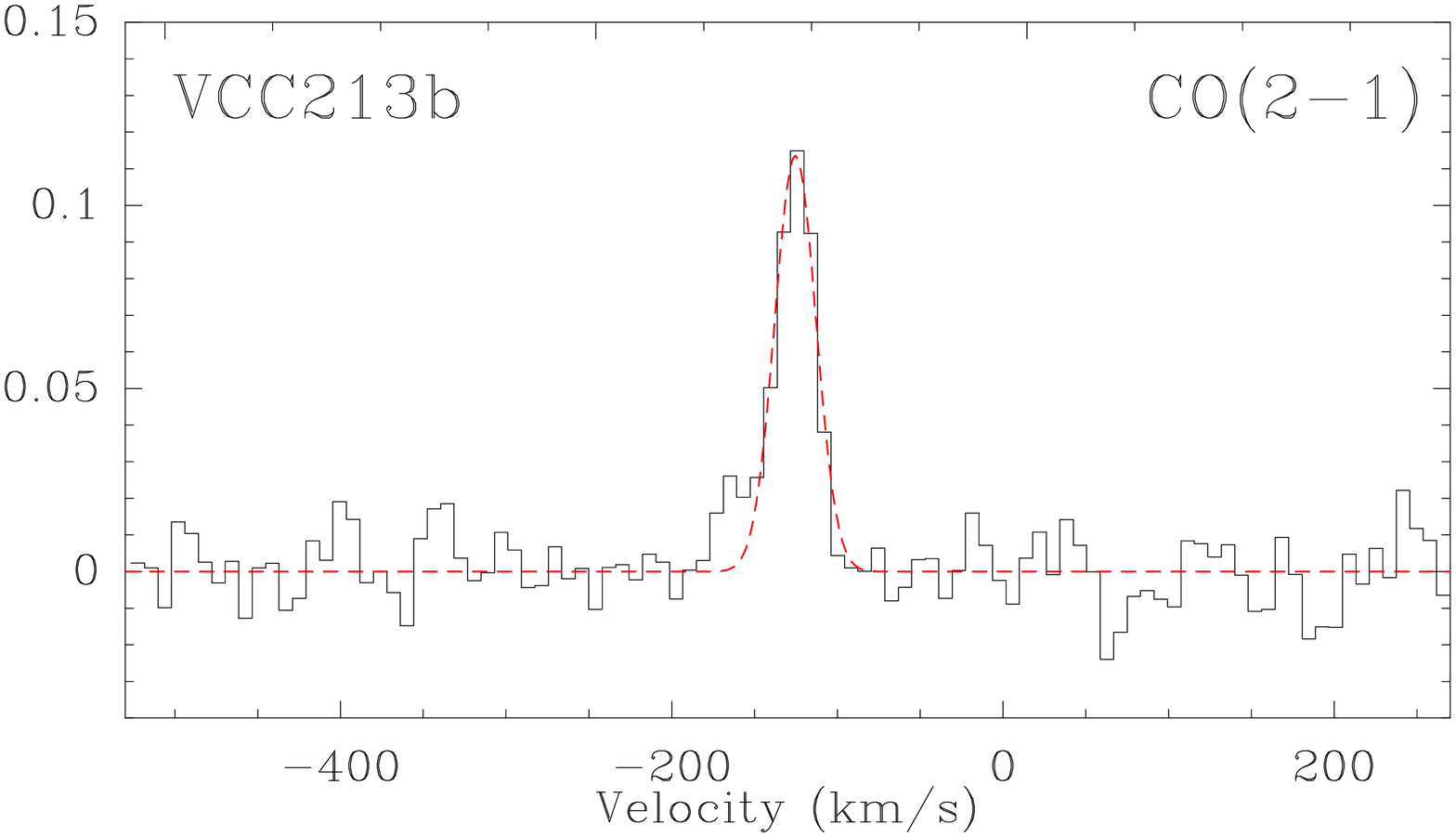}
\includegraphics[width=4.5cm]{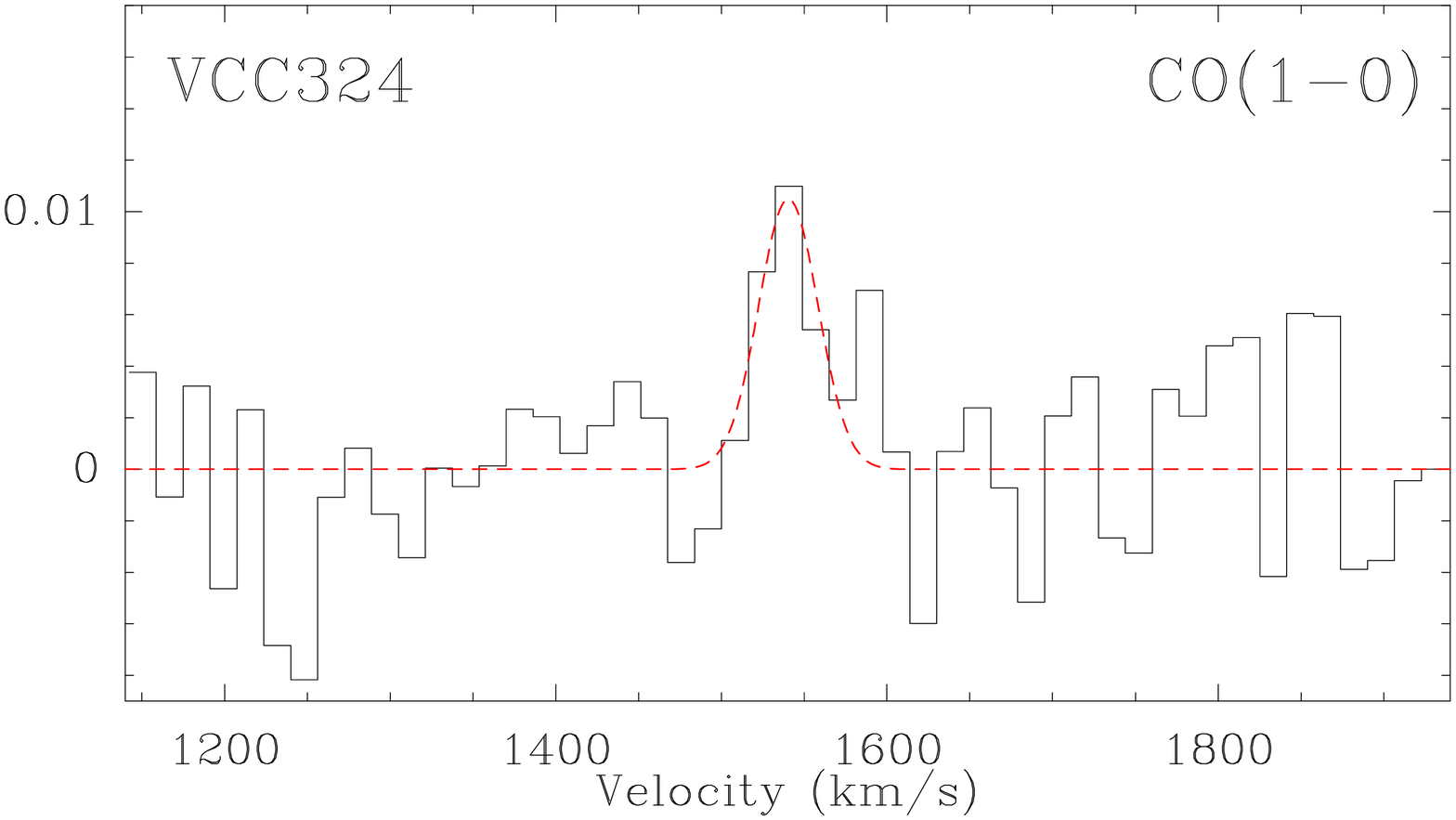}
\includegraphics[width=4.5cm]{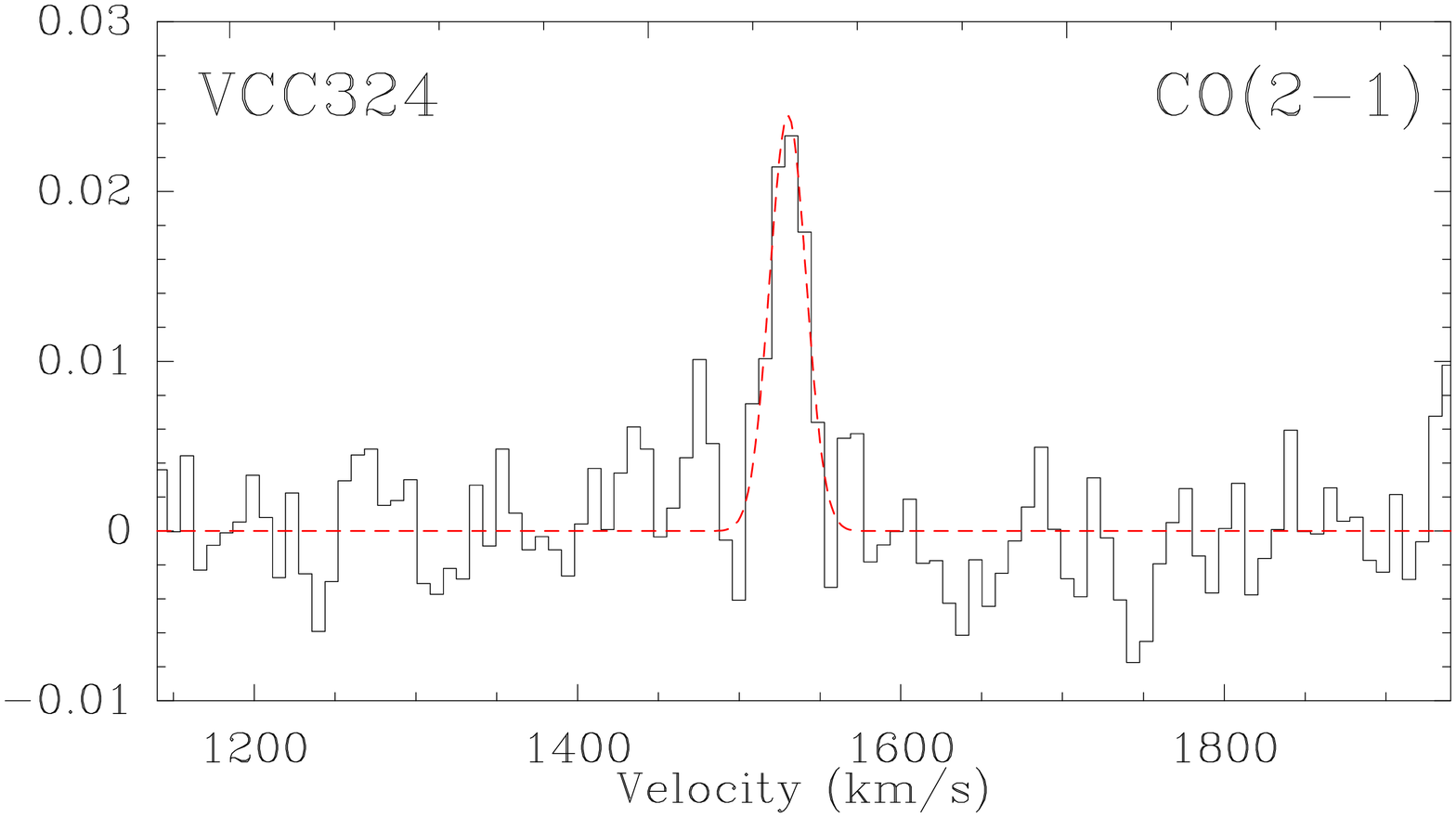}\\
 \includegraphics[width=4.5cm]{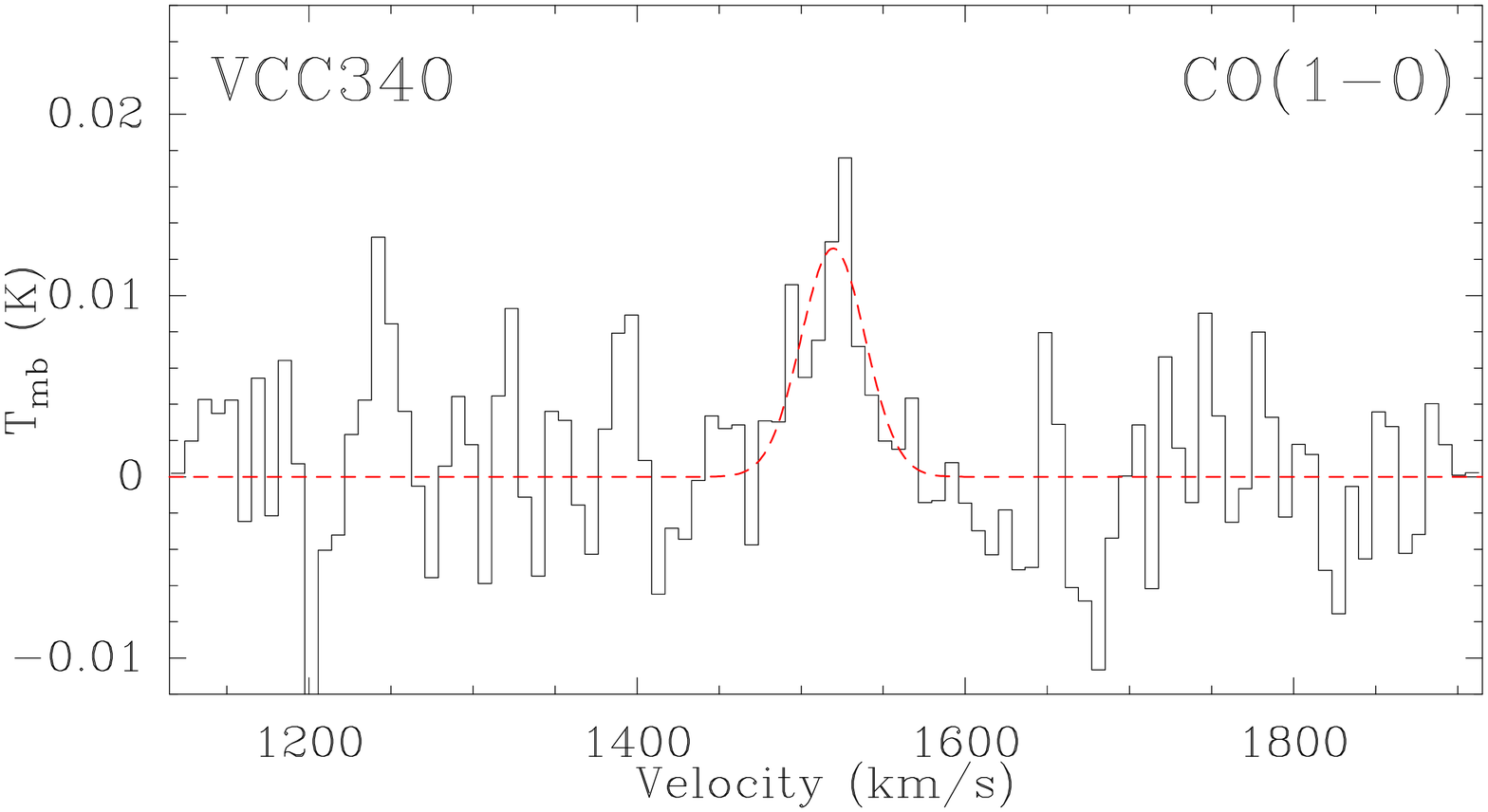}
 \includegraphics[width=4.5cm]{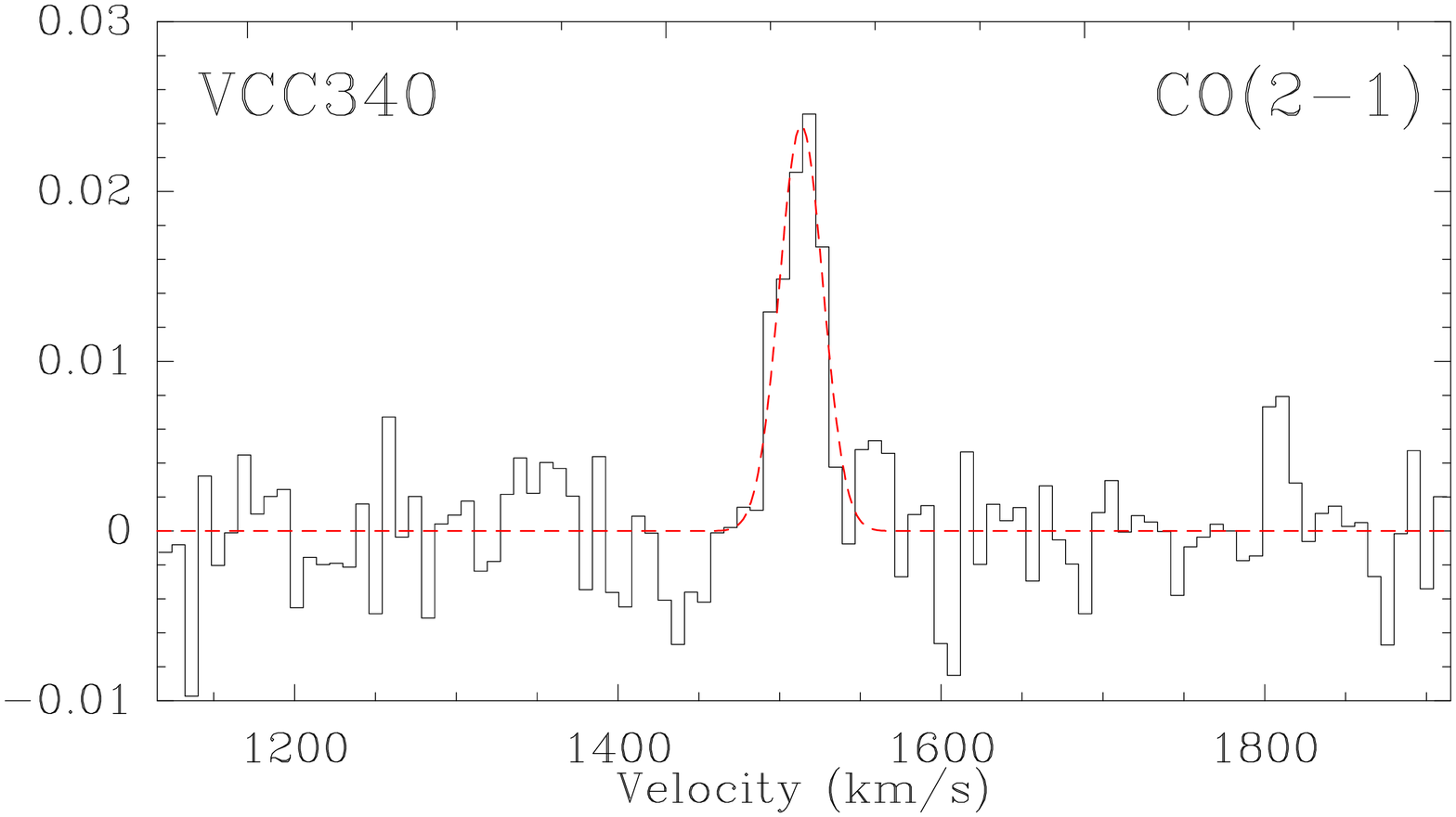}
 \includegraphics[width=4.5cm]{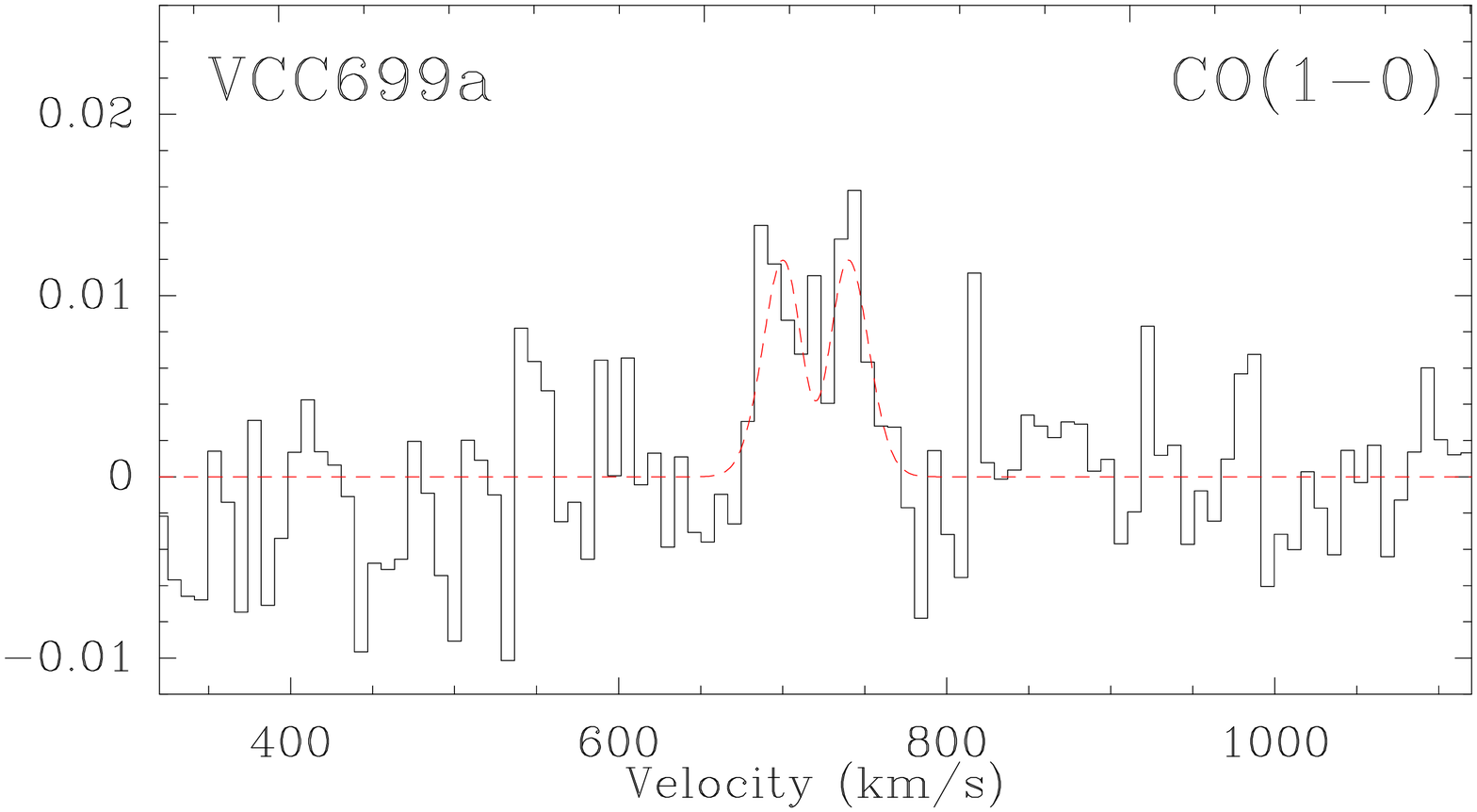}
 \includegraphics[width=4.5cm]{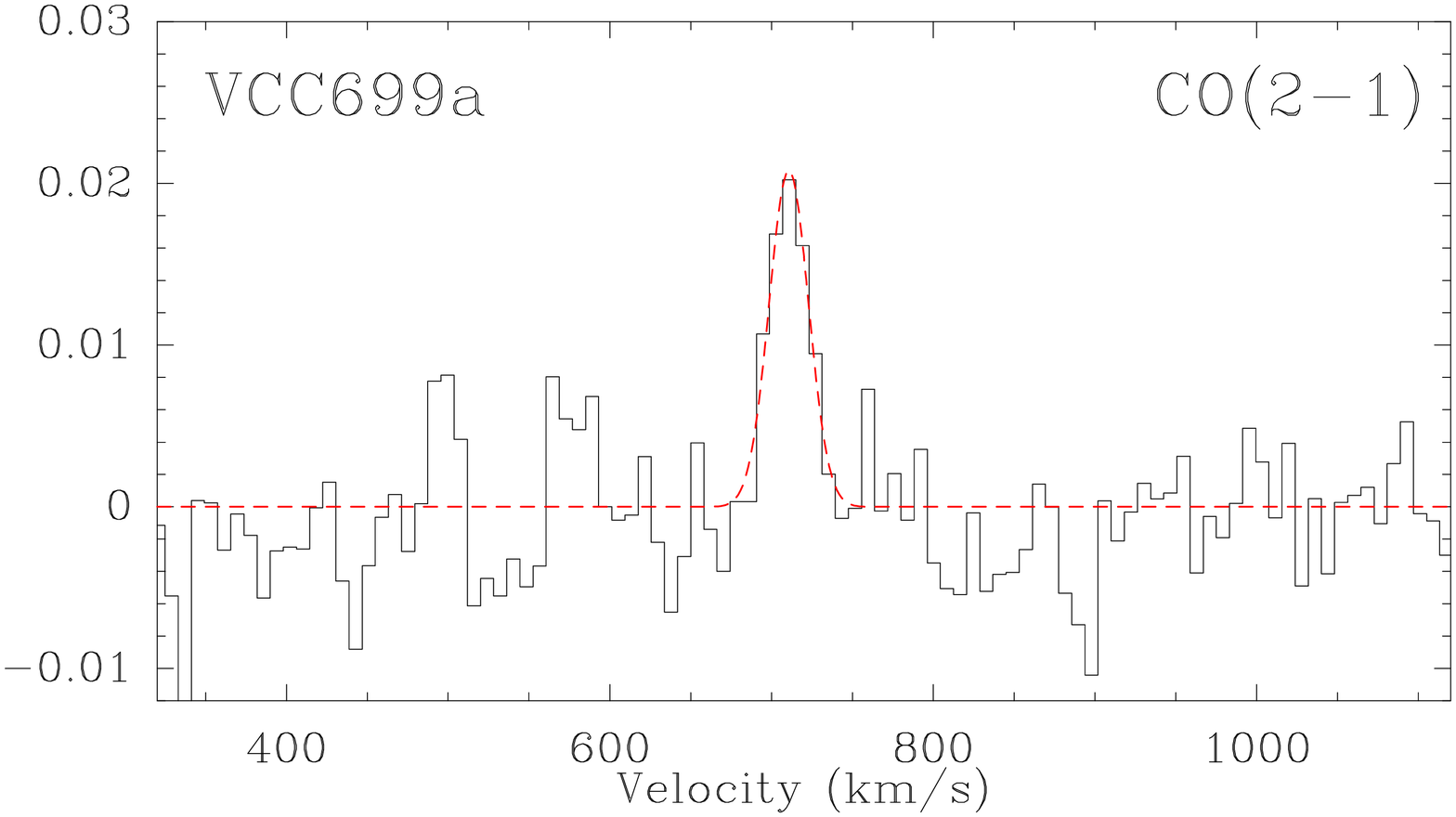}\\
 \includegraphics[width=4.5cm]{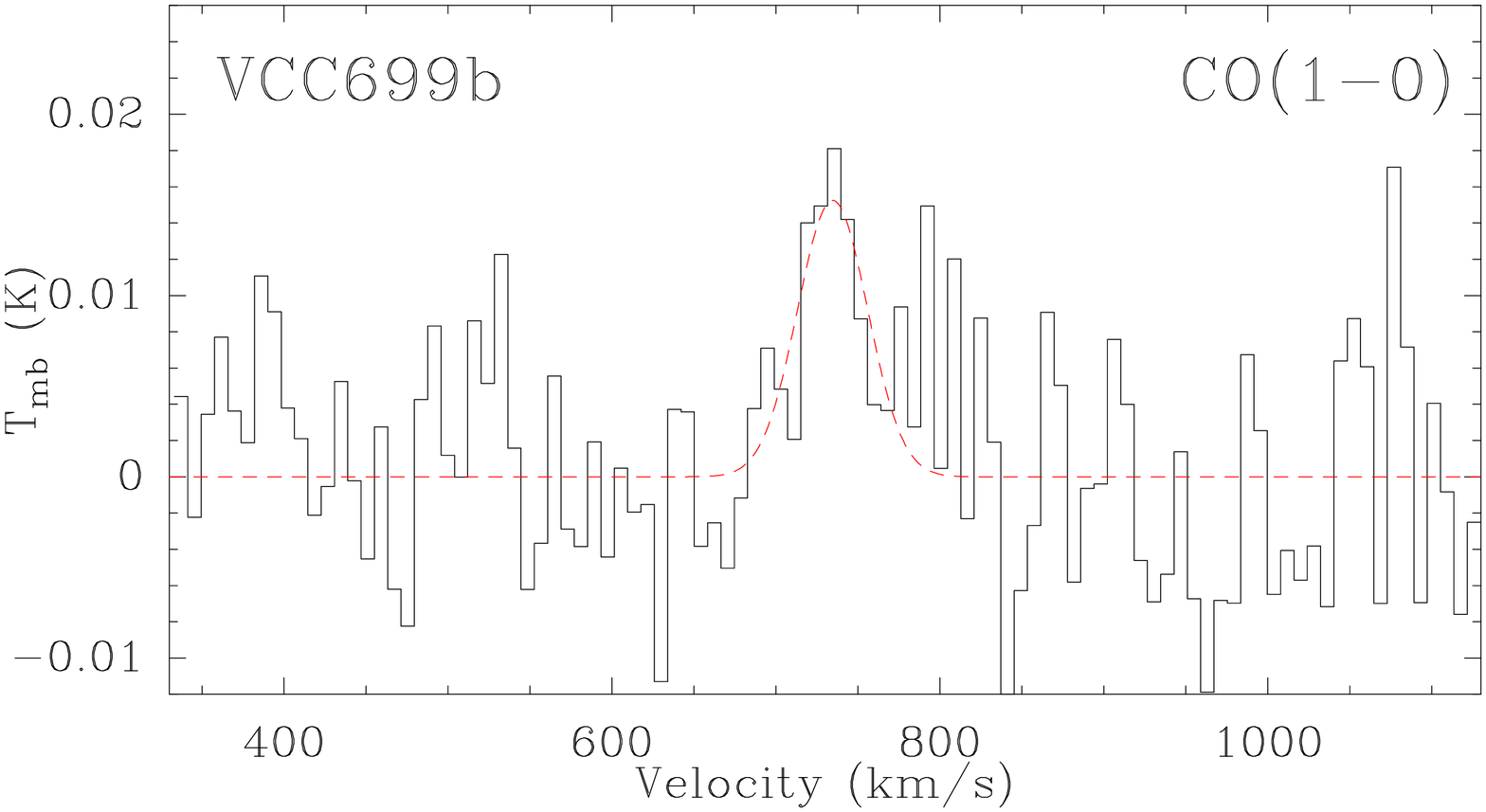}
 \includegraphics[width=4.5cm]{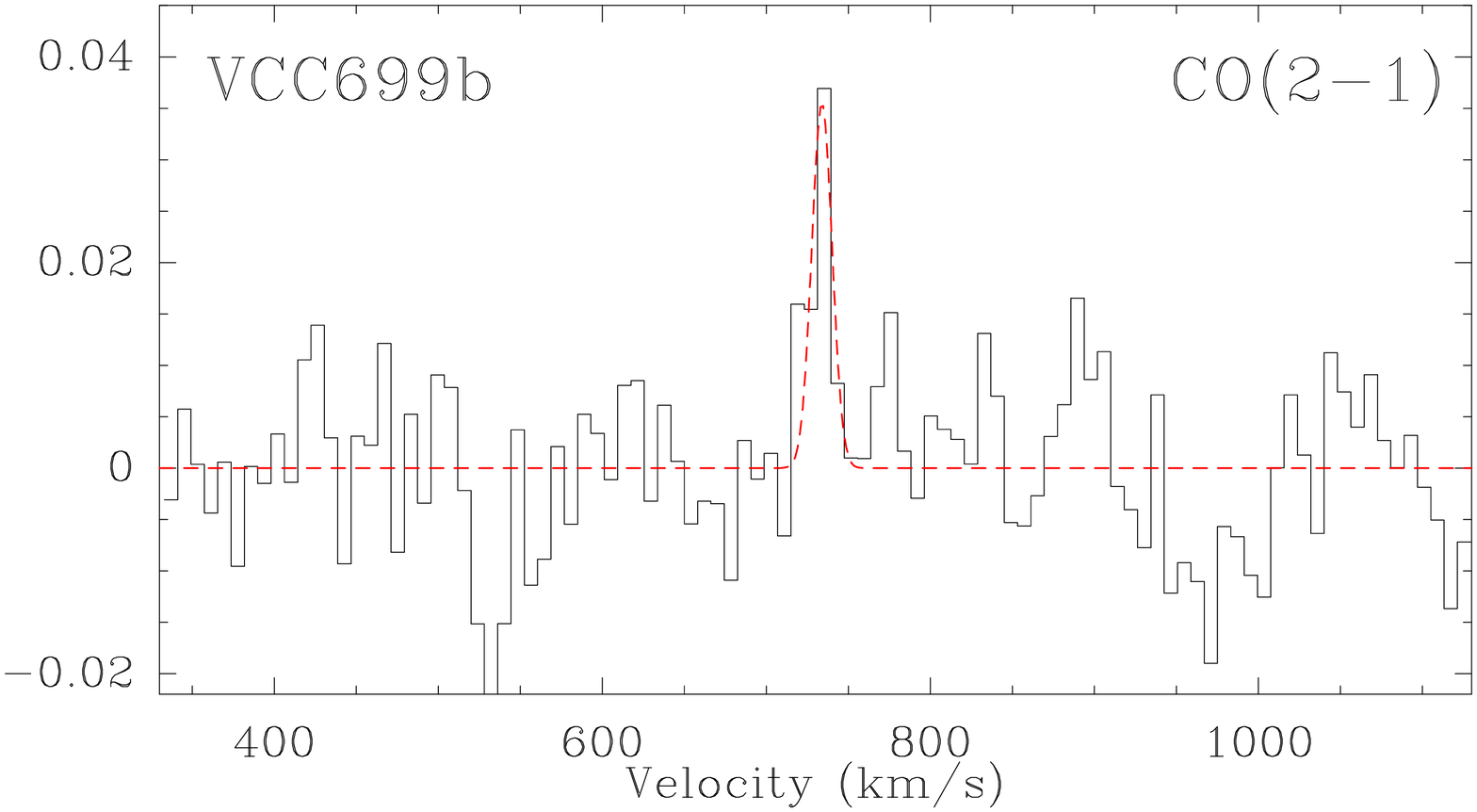}
 \includegraphics[width=4.5cm]{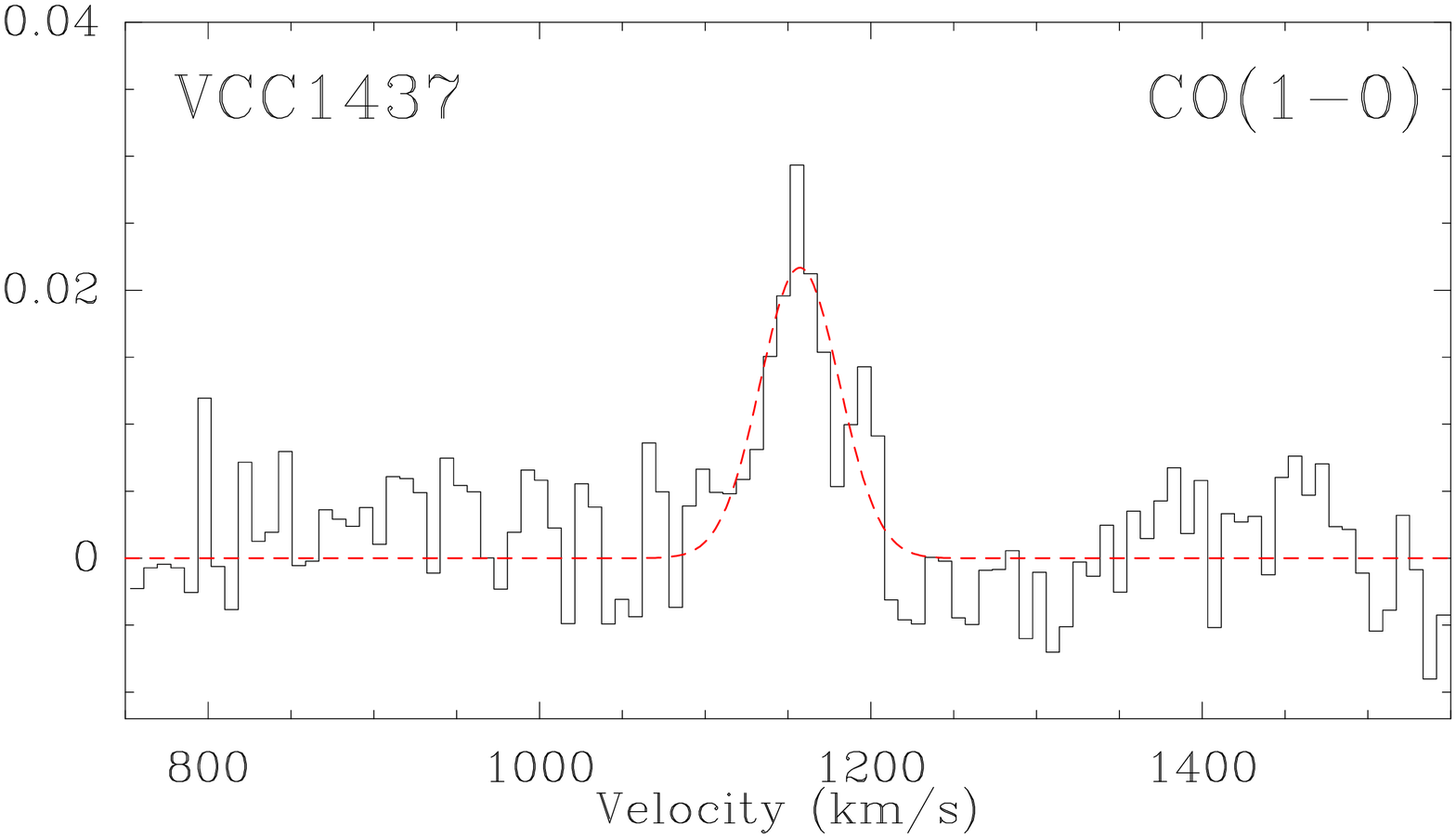}
 \includegraphics[width=4.5cm]{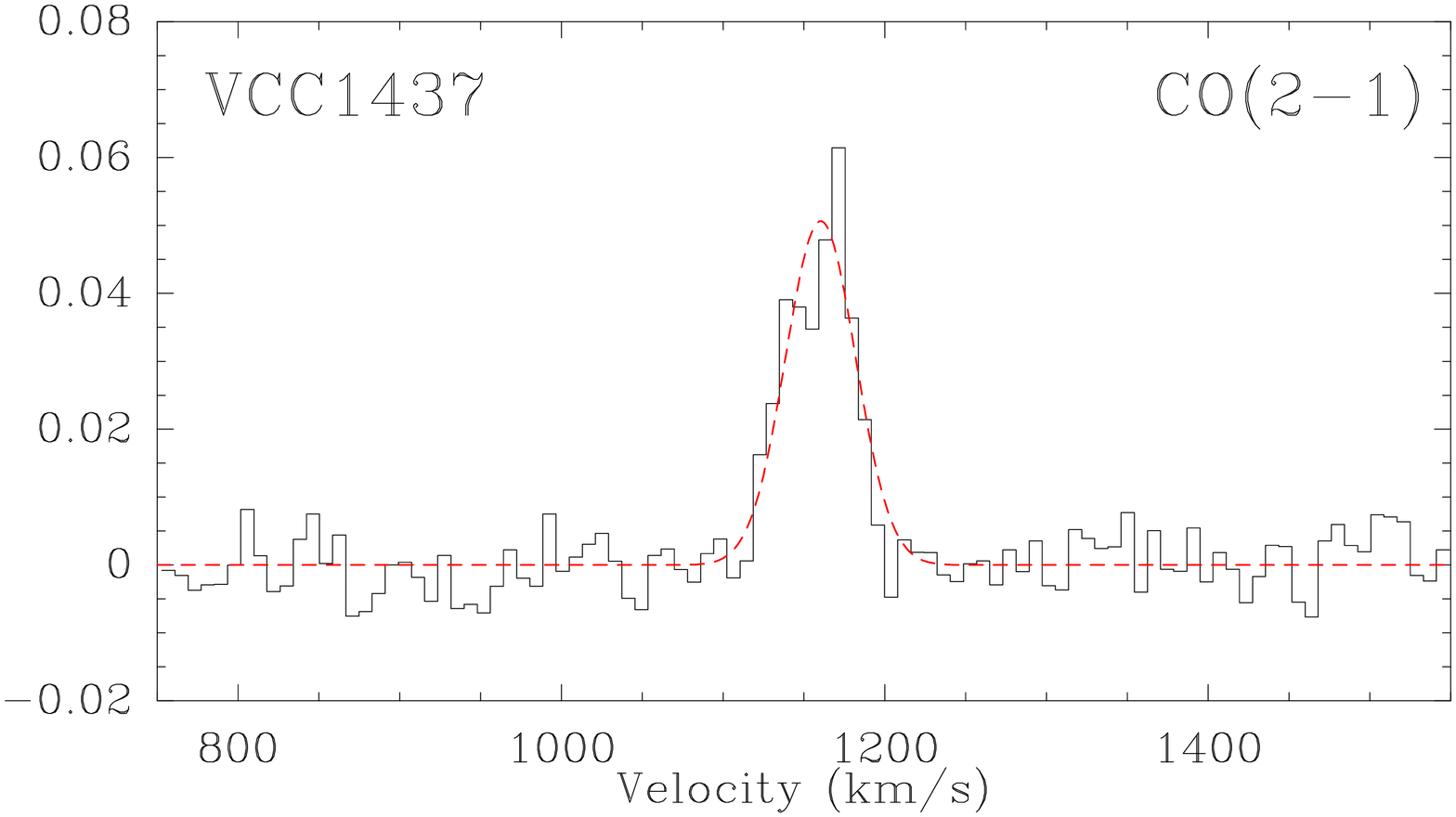}\\
 \includegraphics[width=4.5cm]{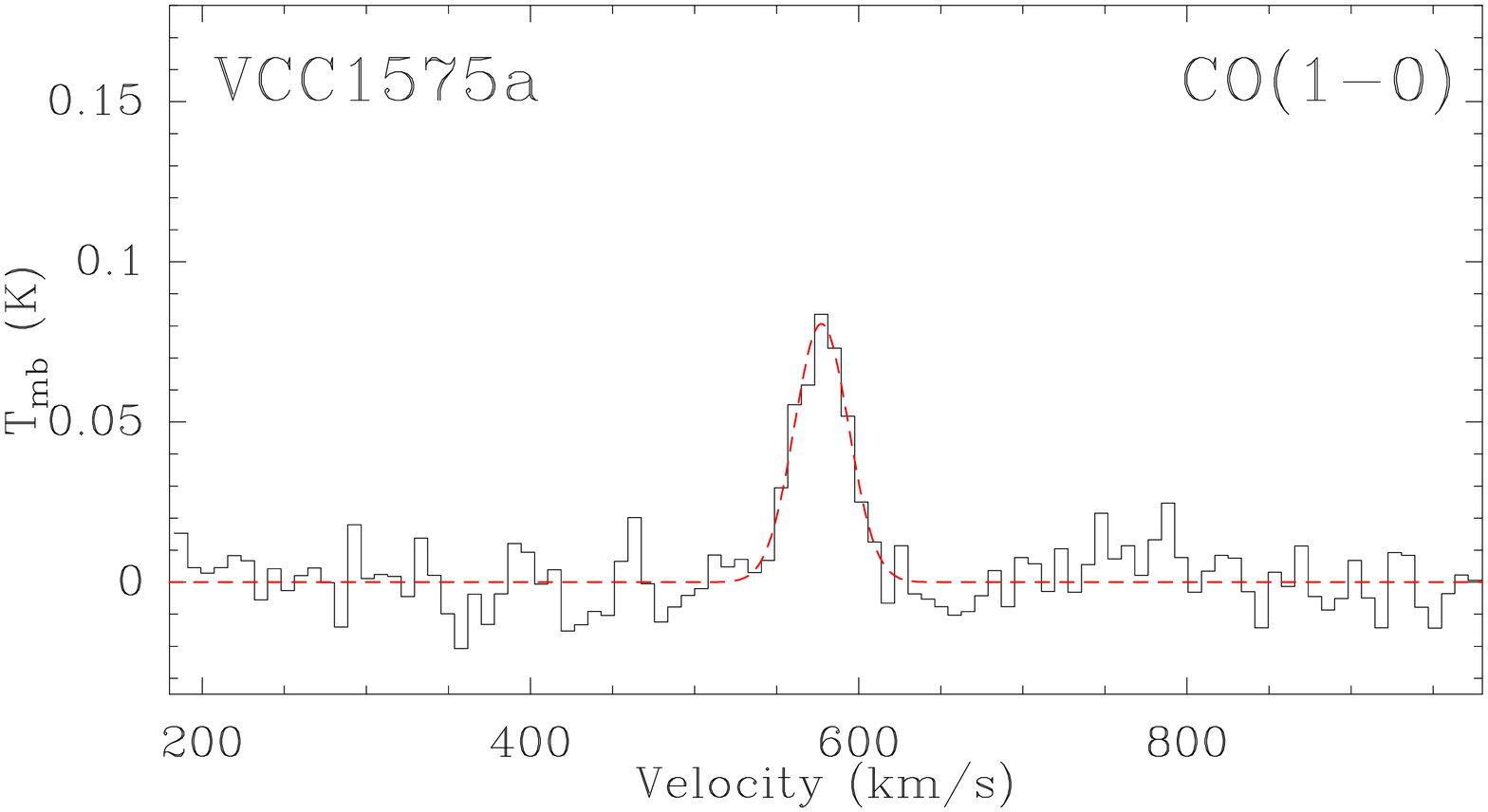}
 \includegraphics[width=4.5cm]{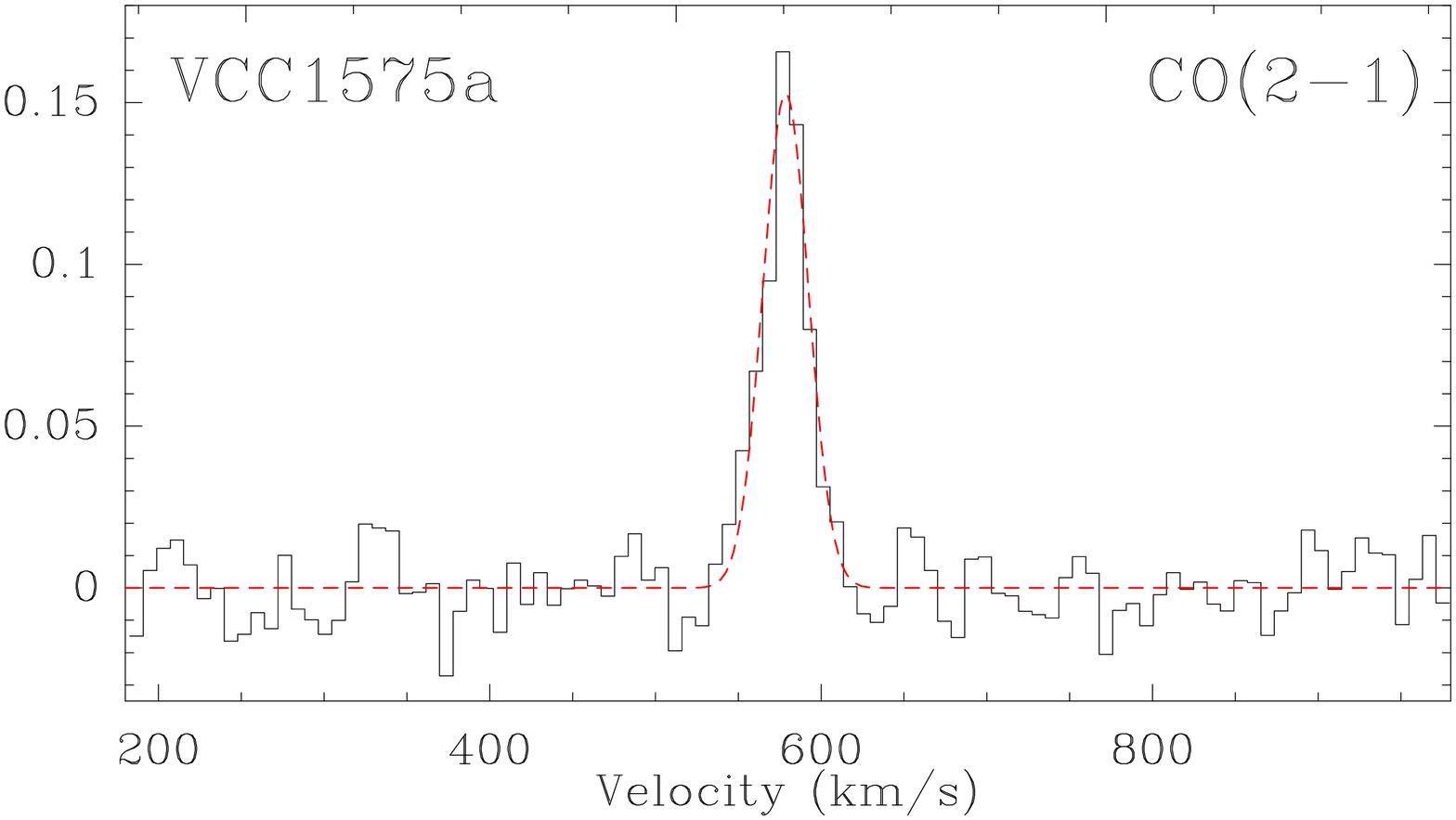}
 \includegraphics[width=4.5cm]{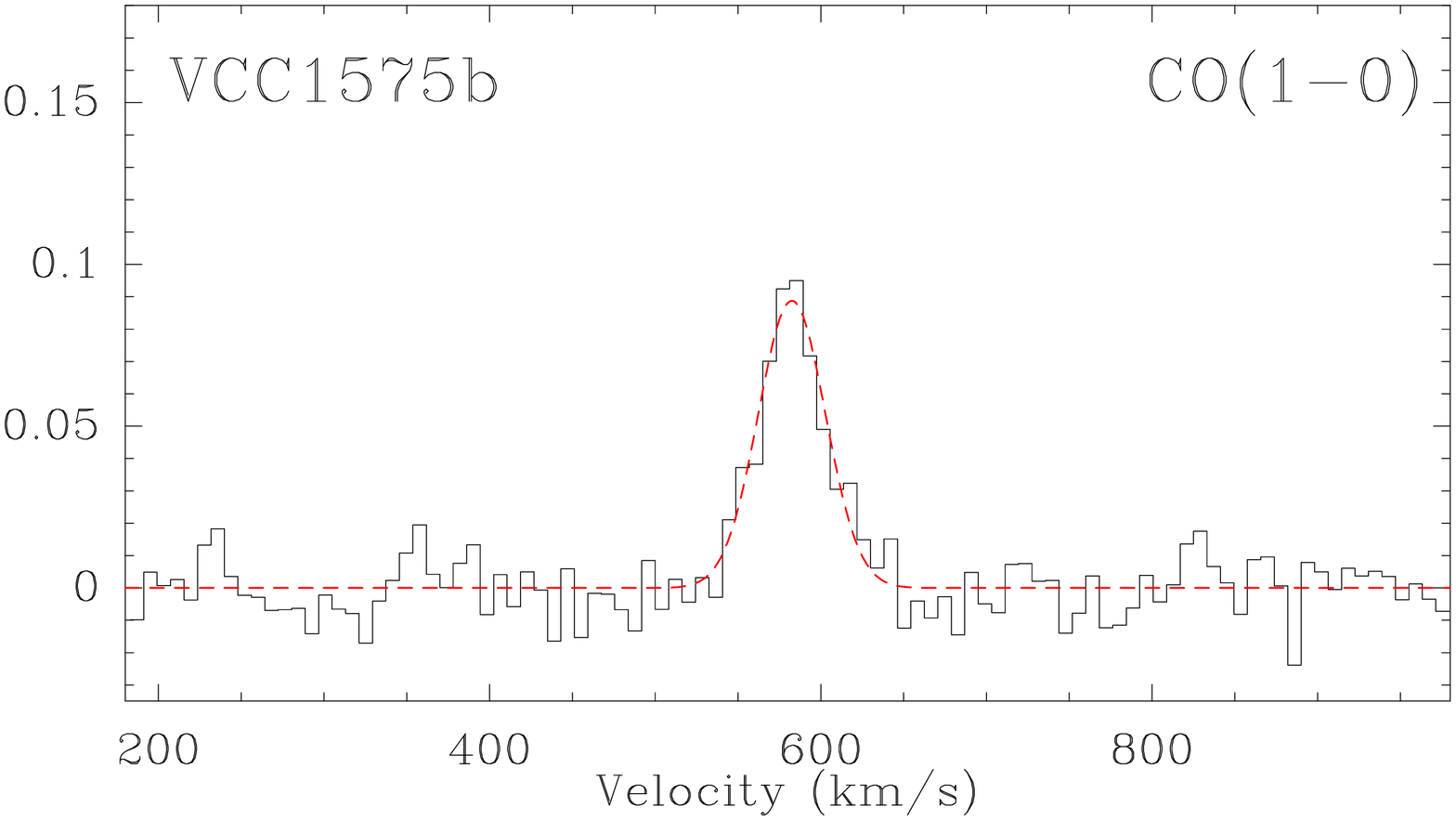}
 \includegraphics[width=4.5cm]{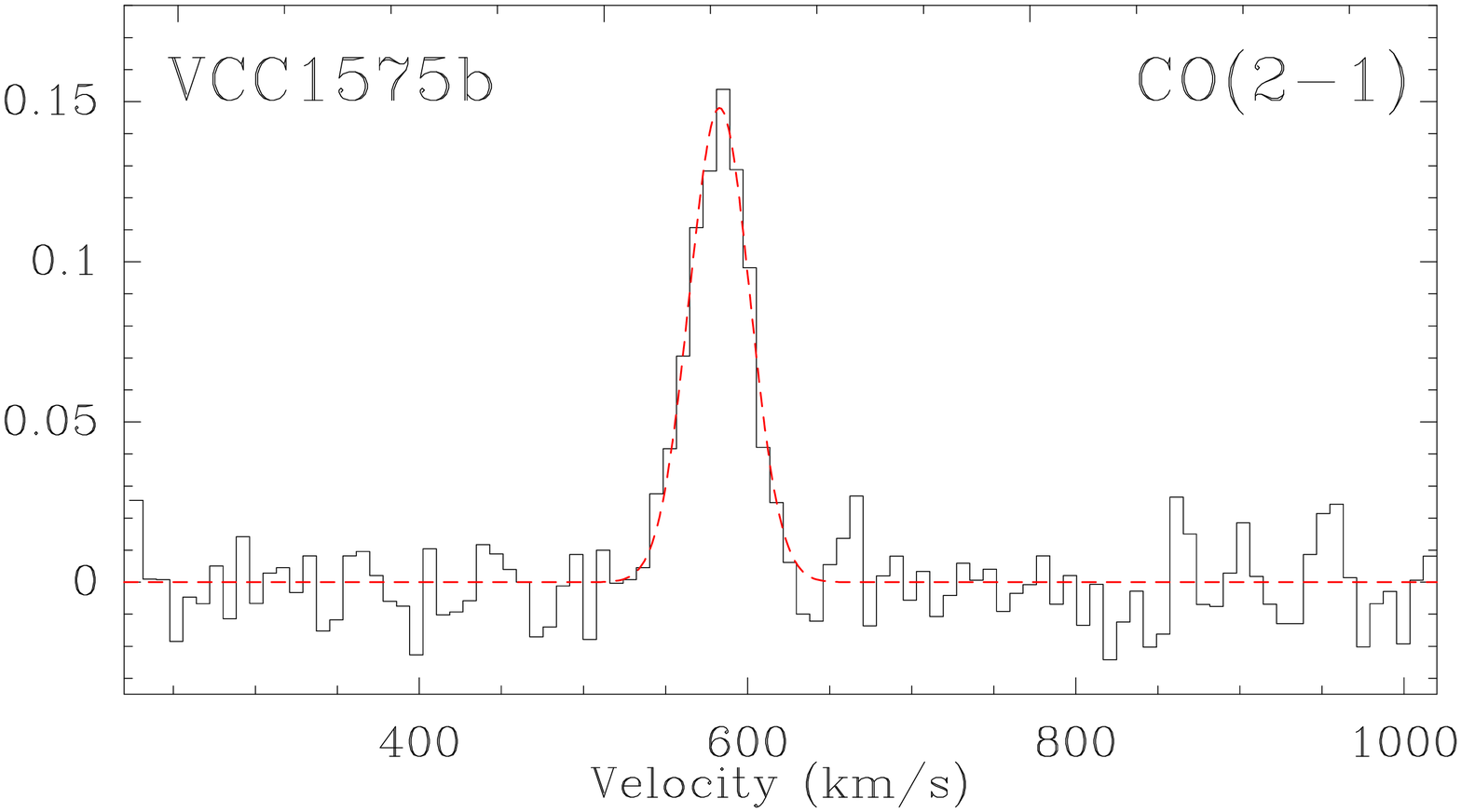}\\
 \includegraphics[width=4.5cm]{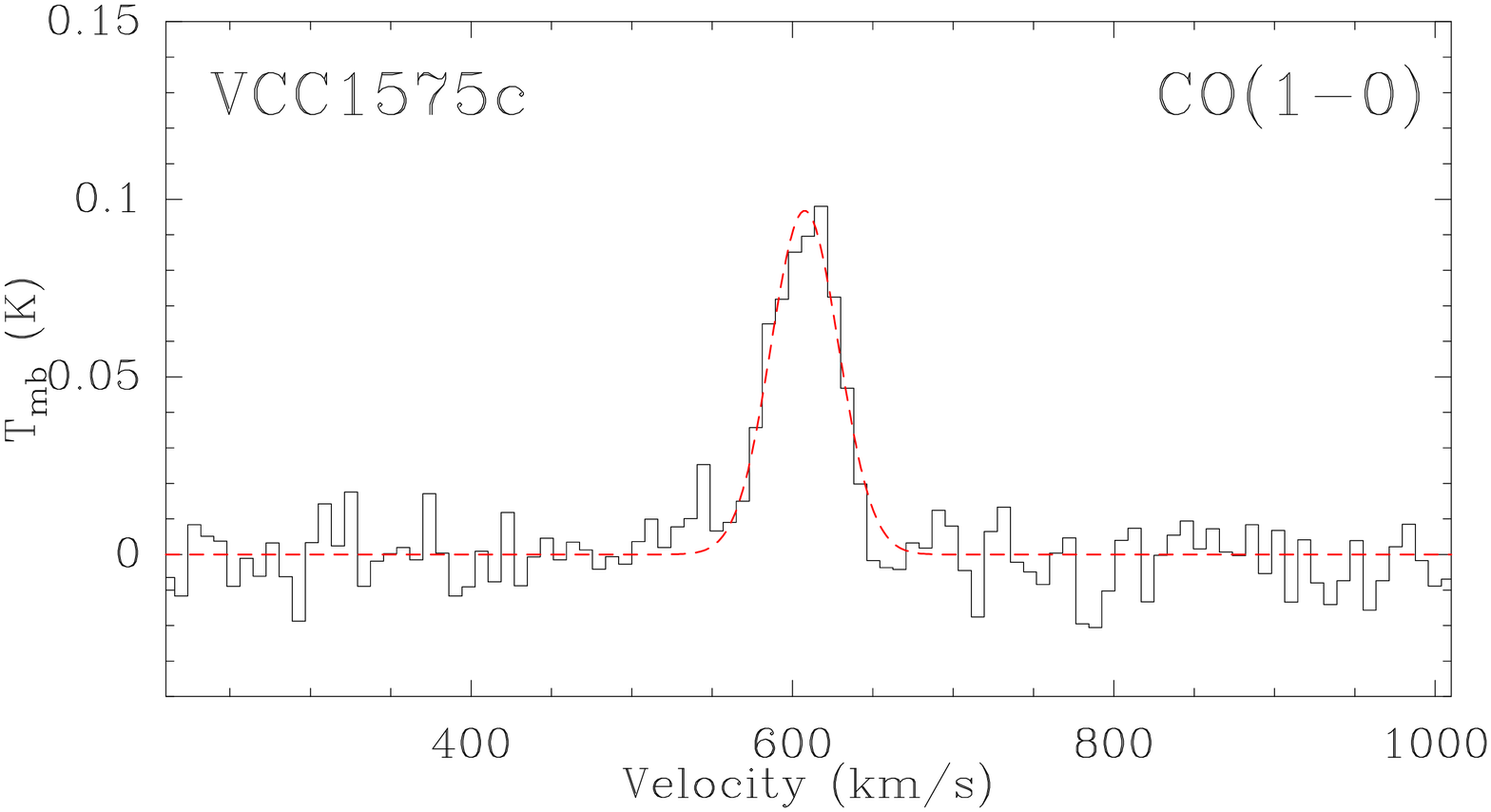}
 \includegraphics[width=4.5cm]{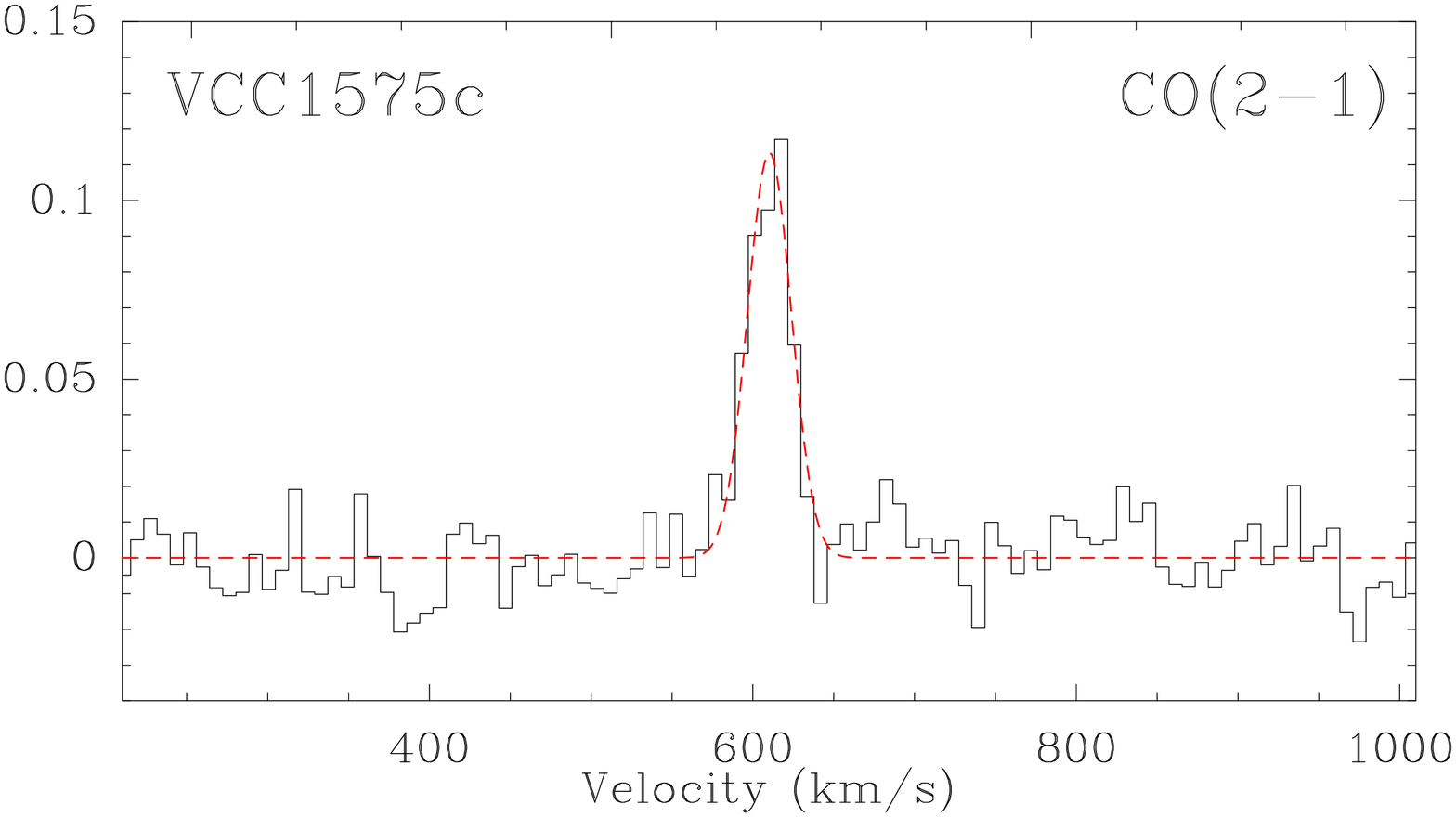}
 \includegraphics[width=4.5cm]{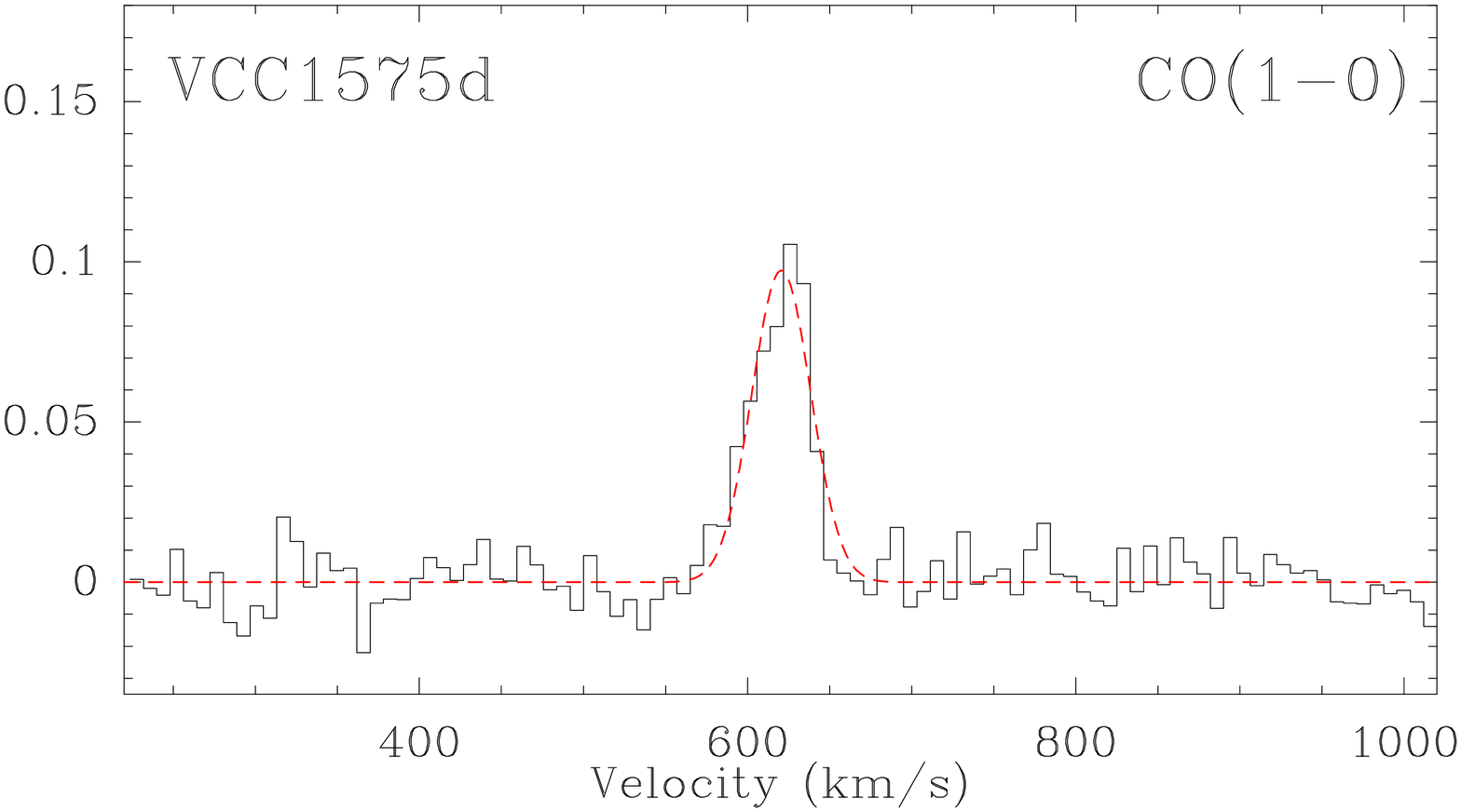}
 \includegraphics[width=4.5cm]{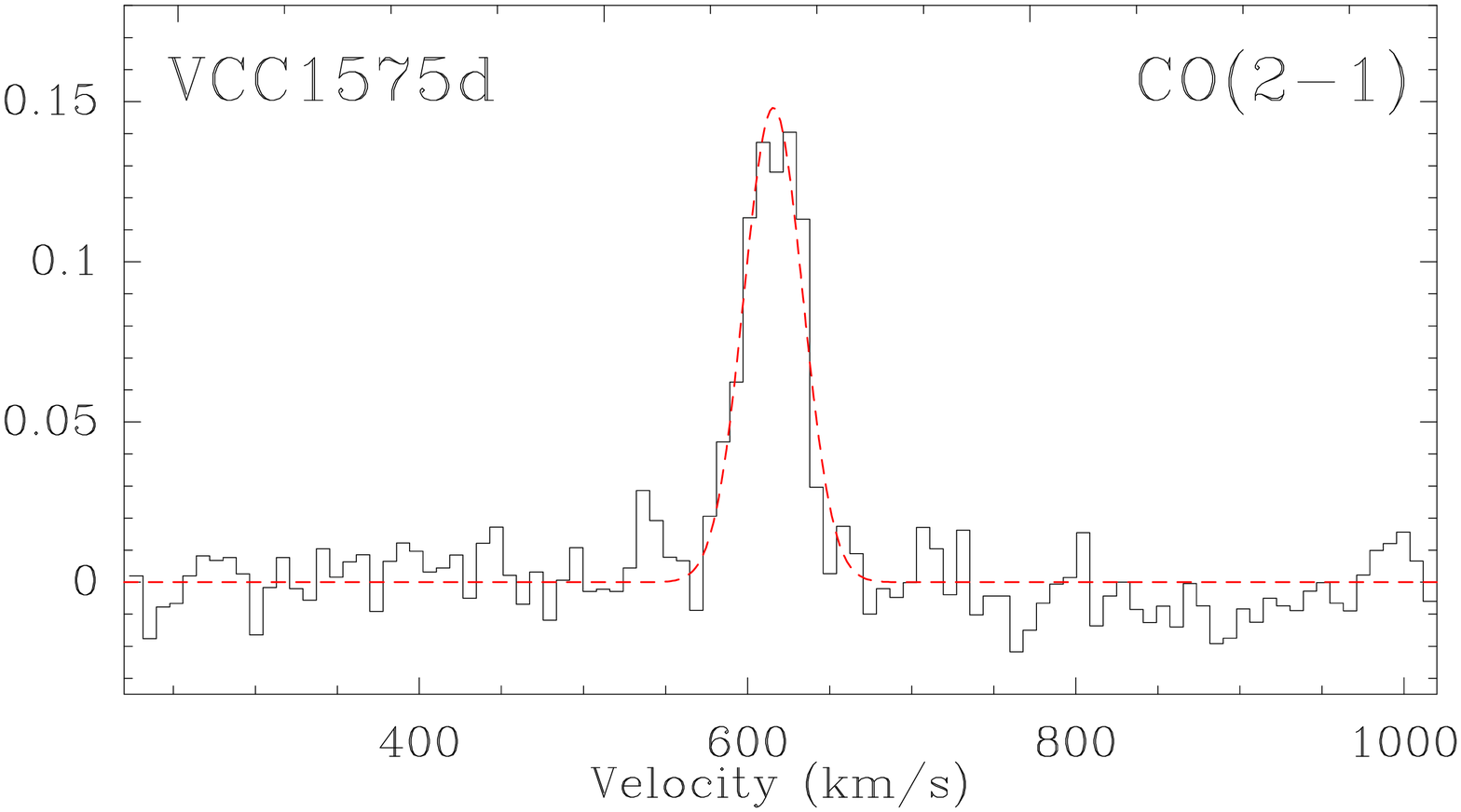}\\
 \includegraphics[width=4.5cm]{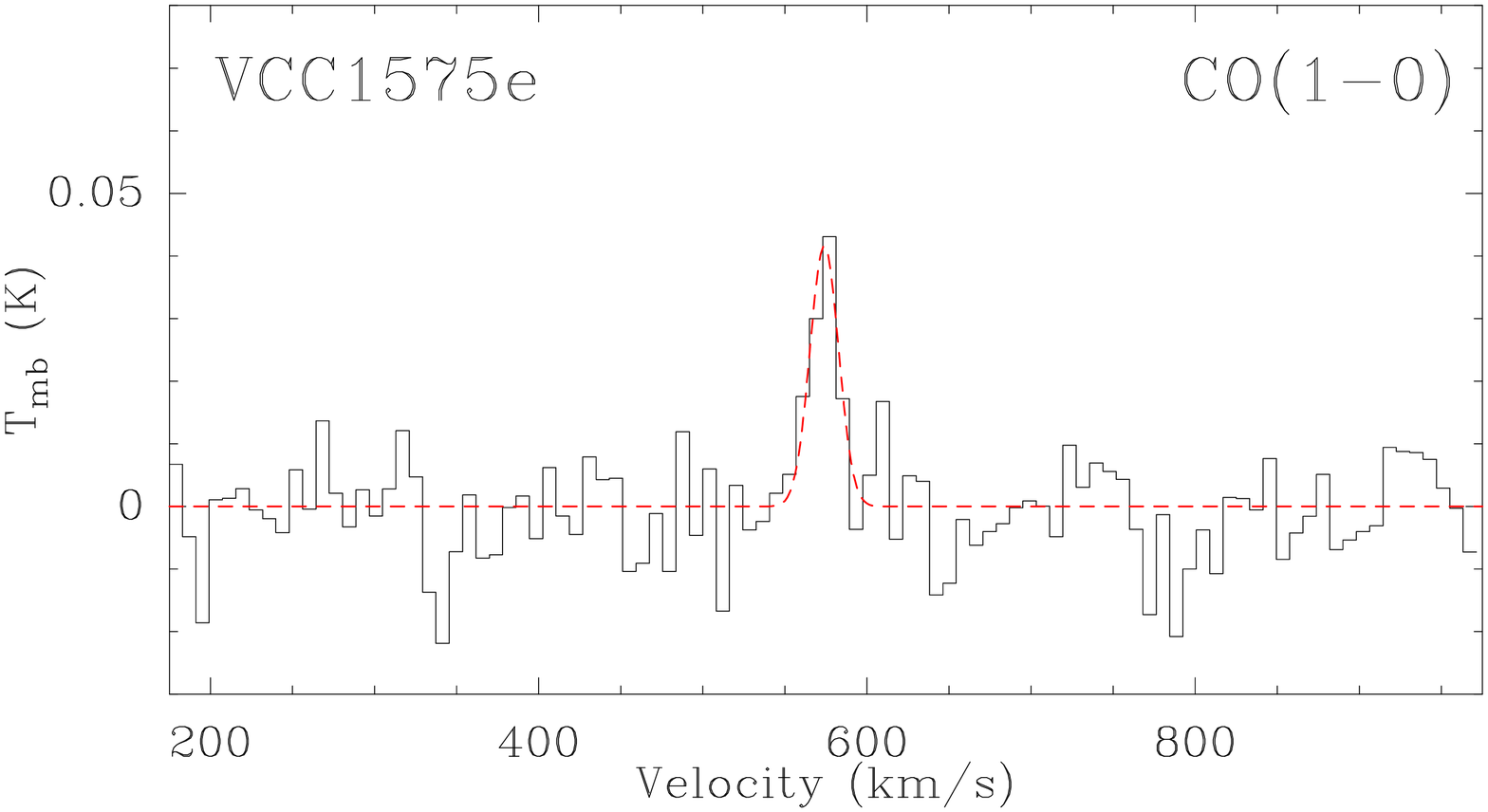}
 \includegraphics[width=4.5cm]{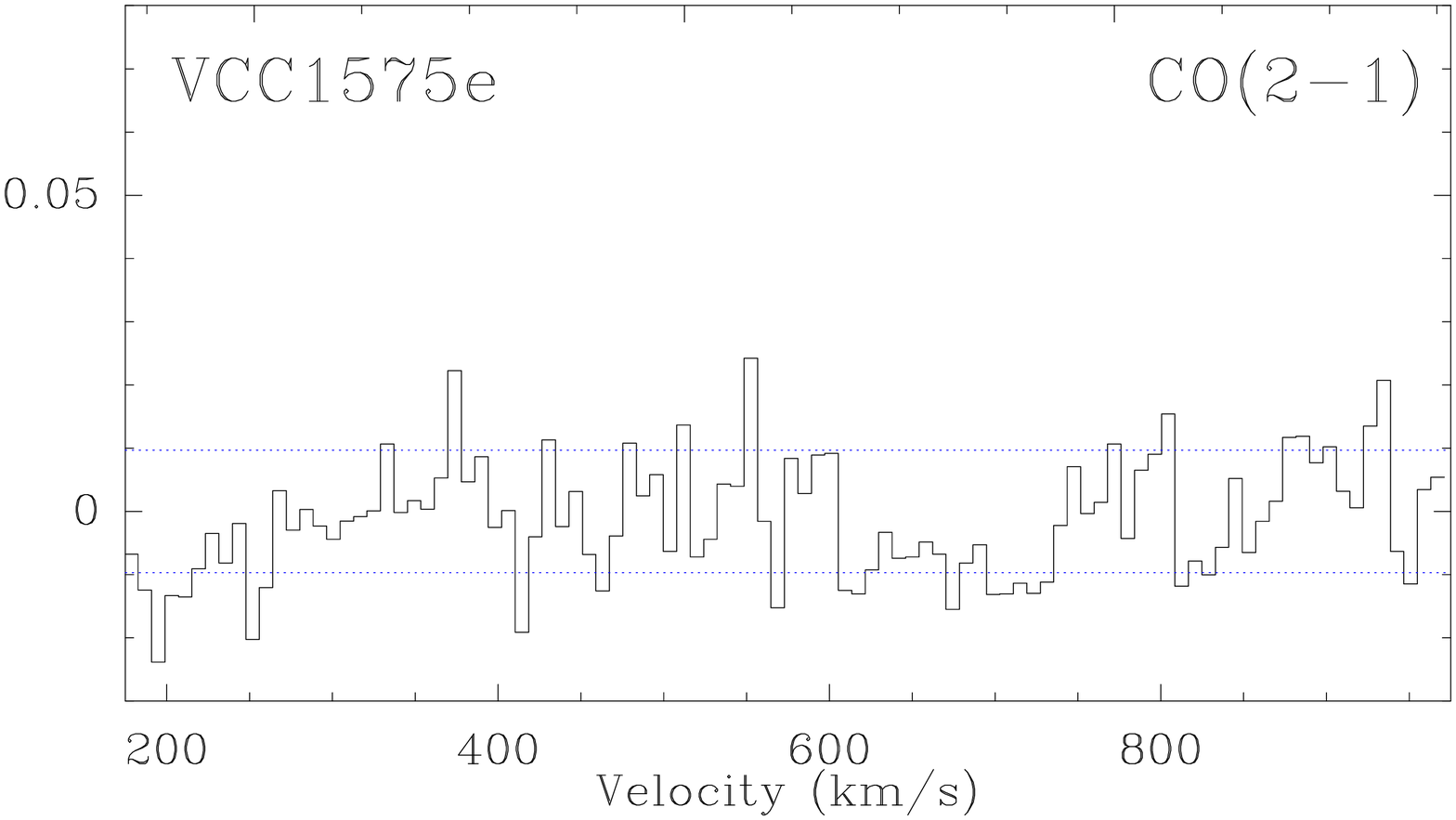}
 \includegraphics[width=4.5cm]{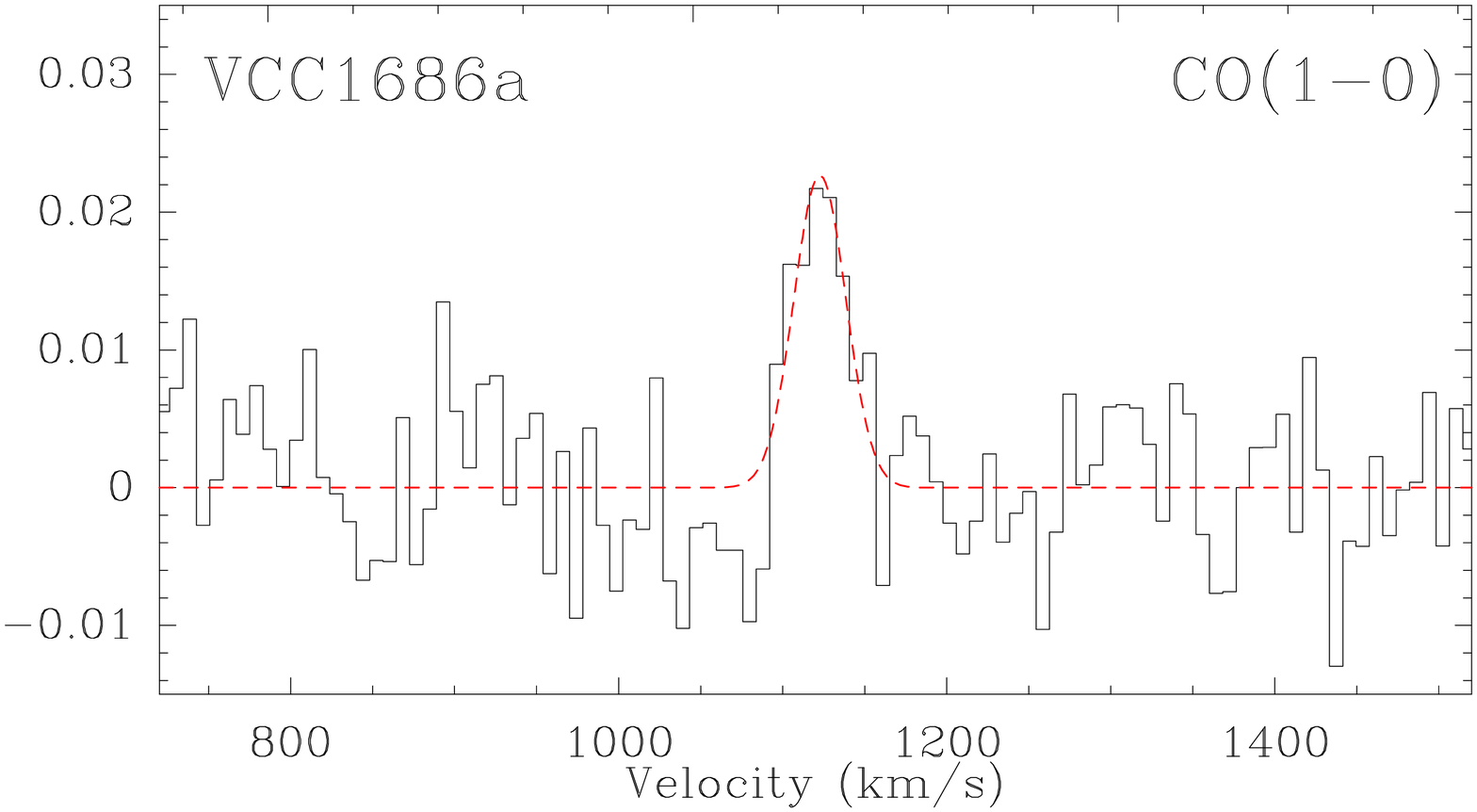}
 \includegraphics[width=4.5cm]{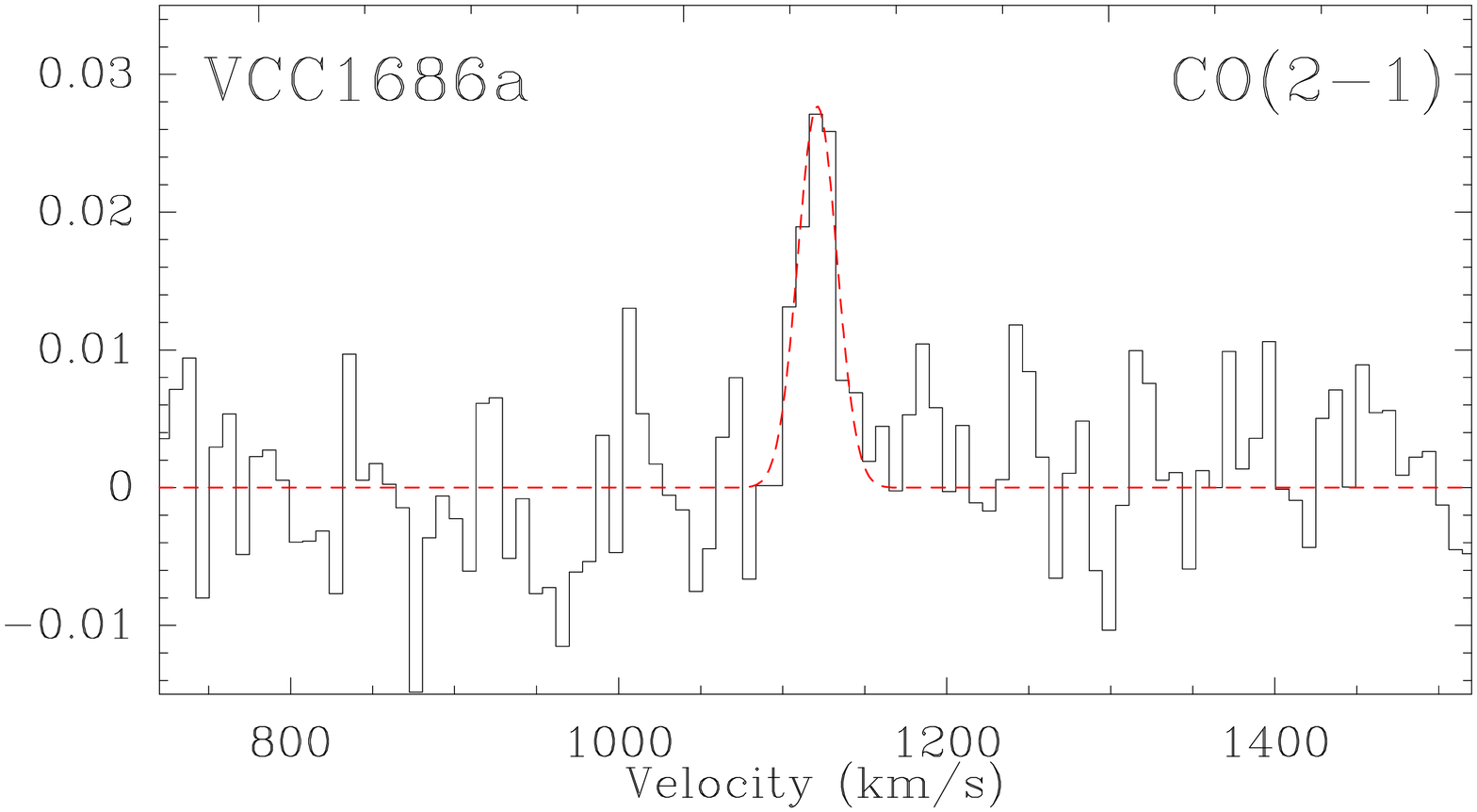}\\
 \includegraphics[width=4.5cm]{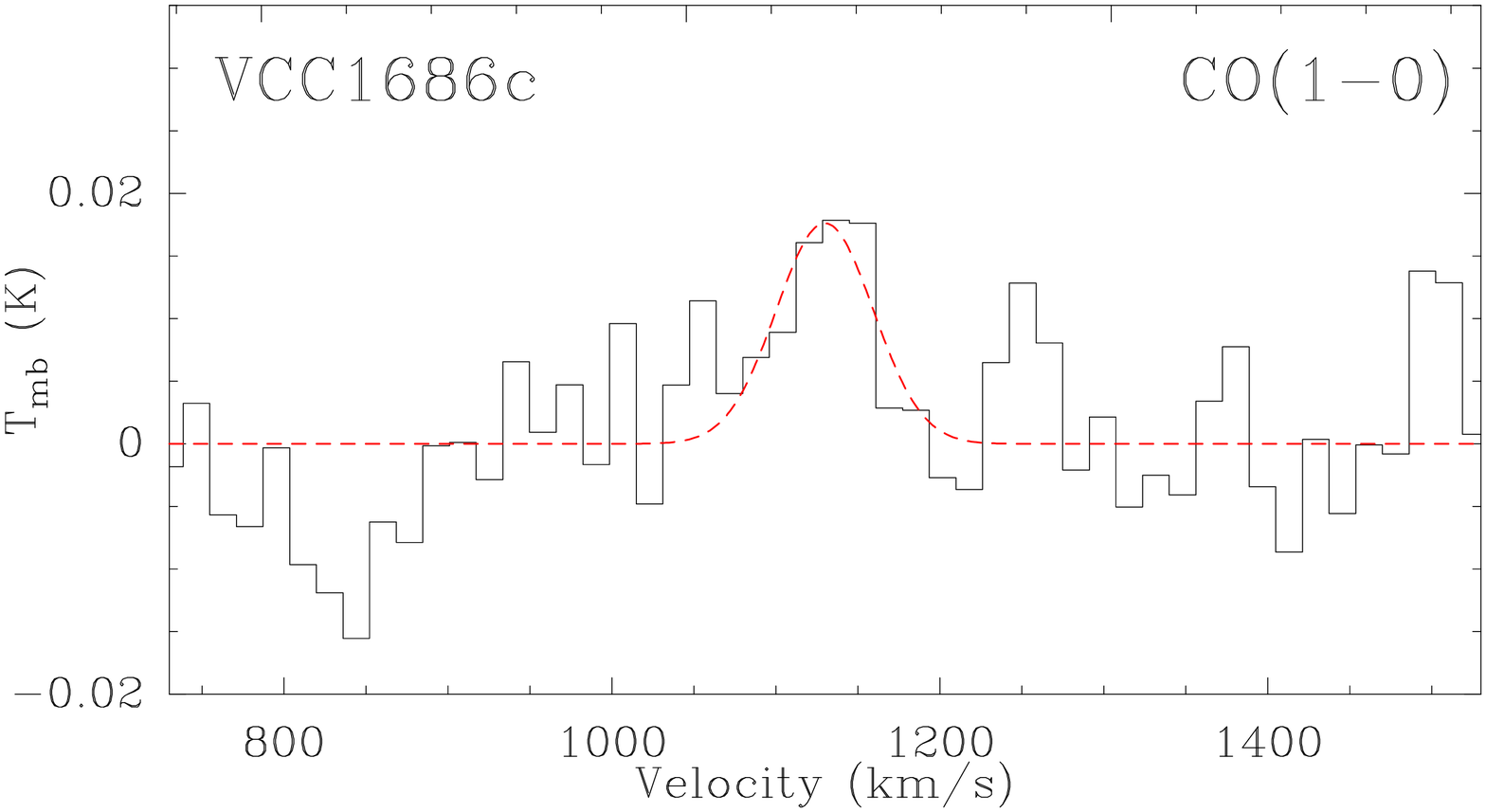}
 \includegraphics[width=4.5cm]{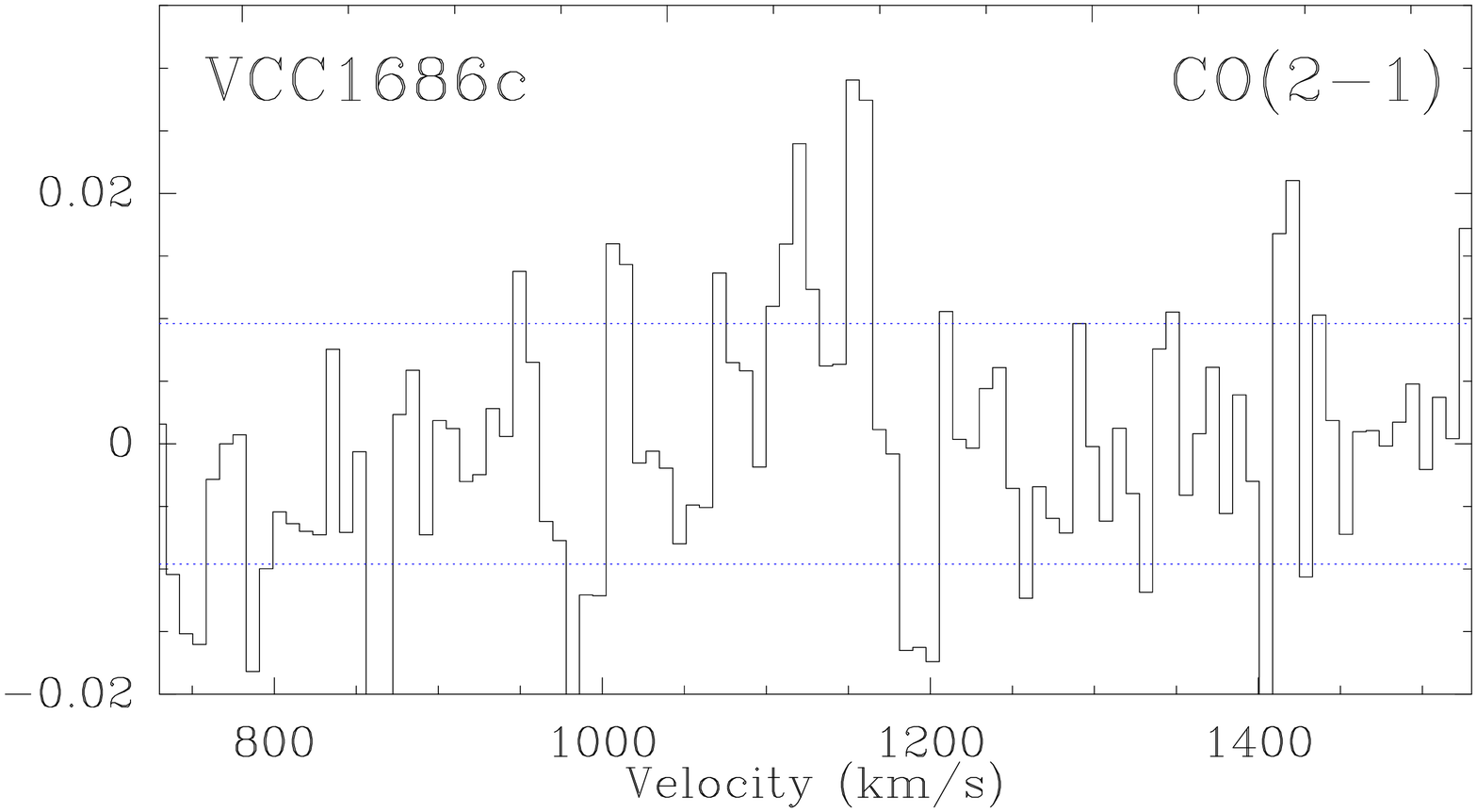}
 \includegraphics[width=4.5cm]{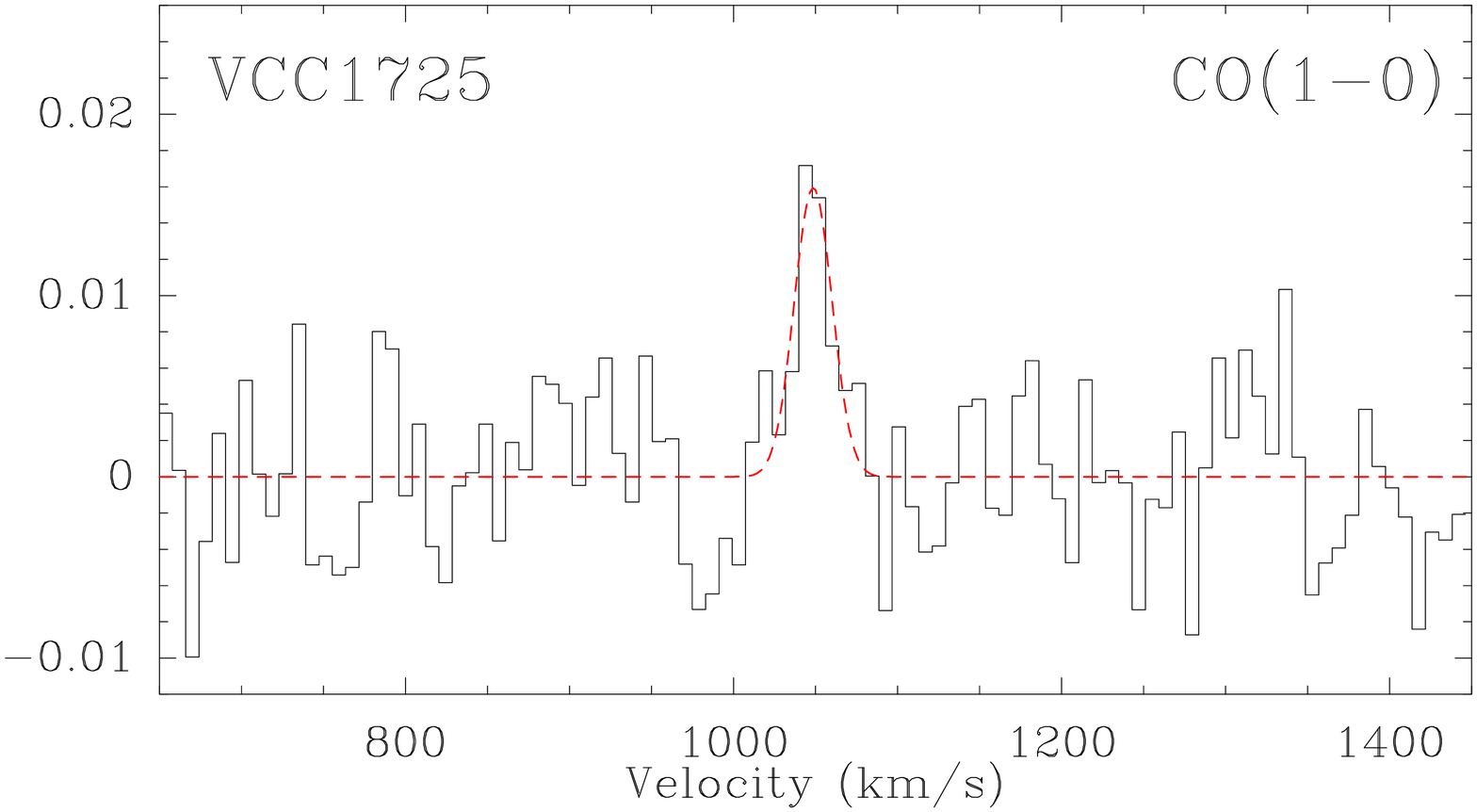}
\includegraphics[width=4.5cm]{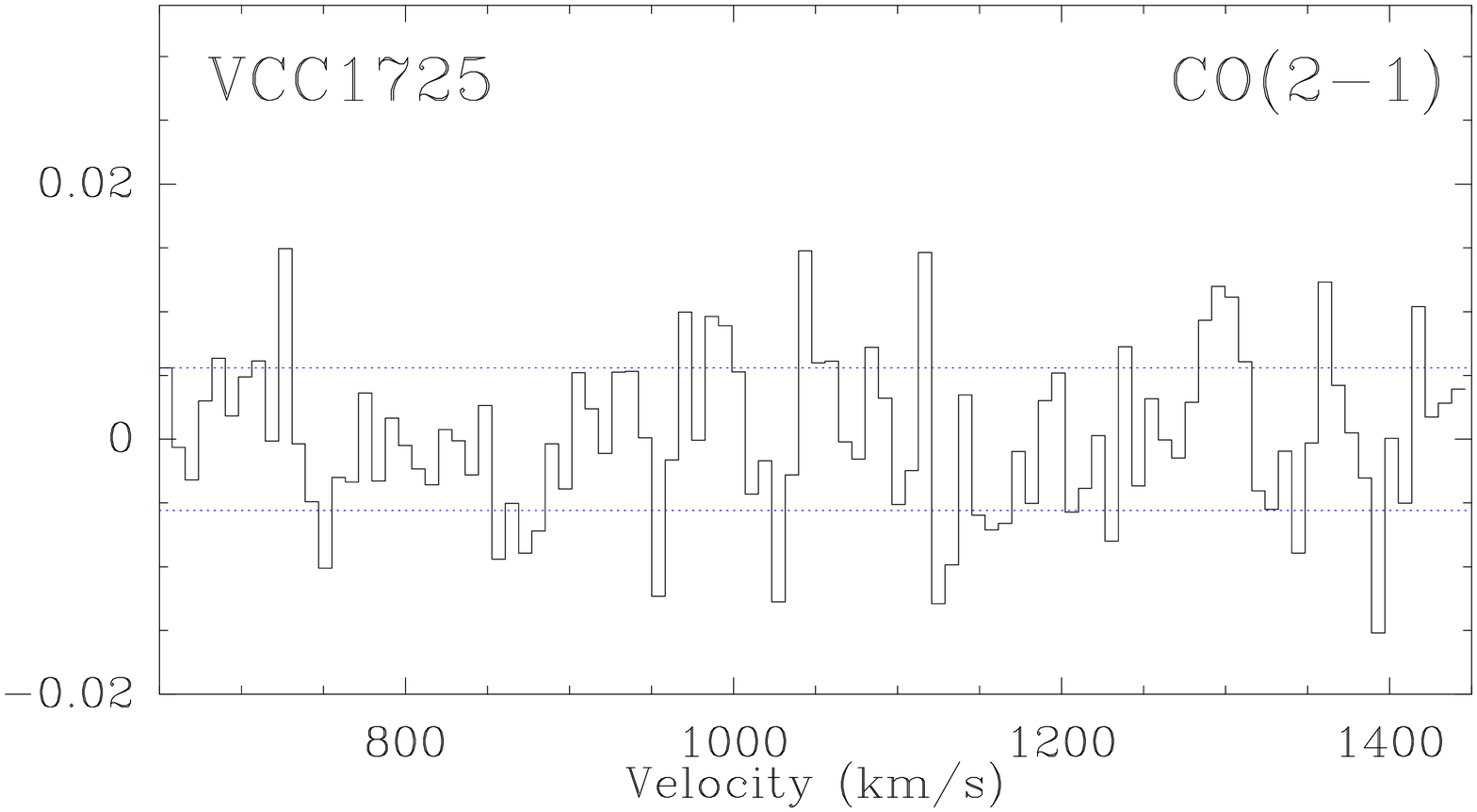}\\
\caption{CO(1-0) and CO(2-0) spectra of galaxies detected in at least one band. The dashed curves show the best fit to the lines.
The dotted horizontal lines indicate the $\pm$1$\sigma$ rms level in case of a non-detection.}
   \label{fig:CO_spec}%
\end{figure*}

\begin{figure*}
  \centering
\includegraphics[width=4.5cm]{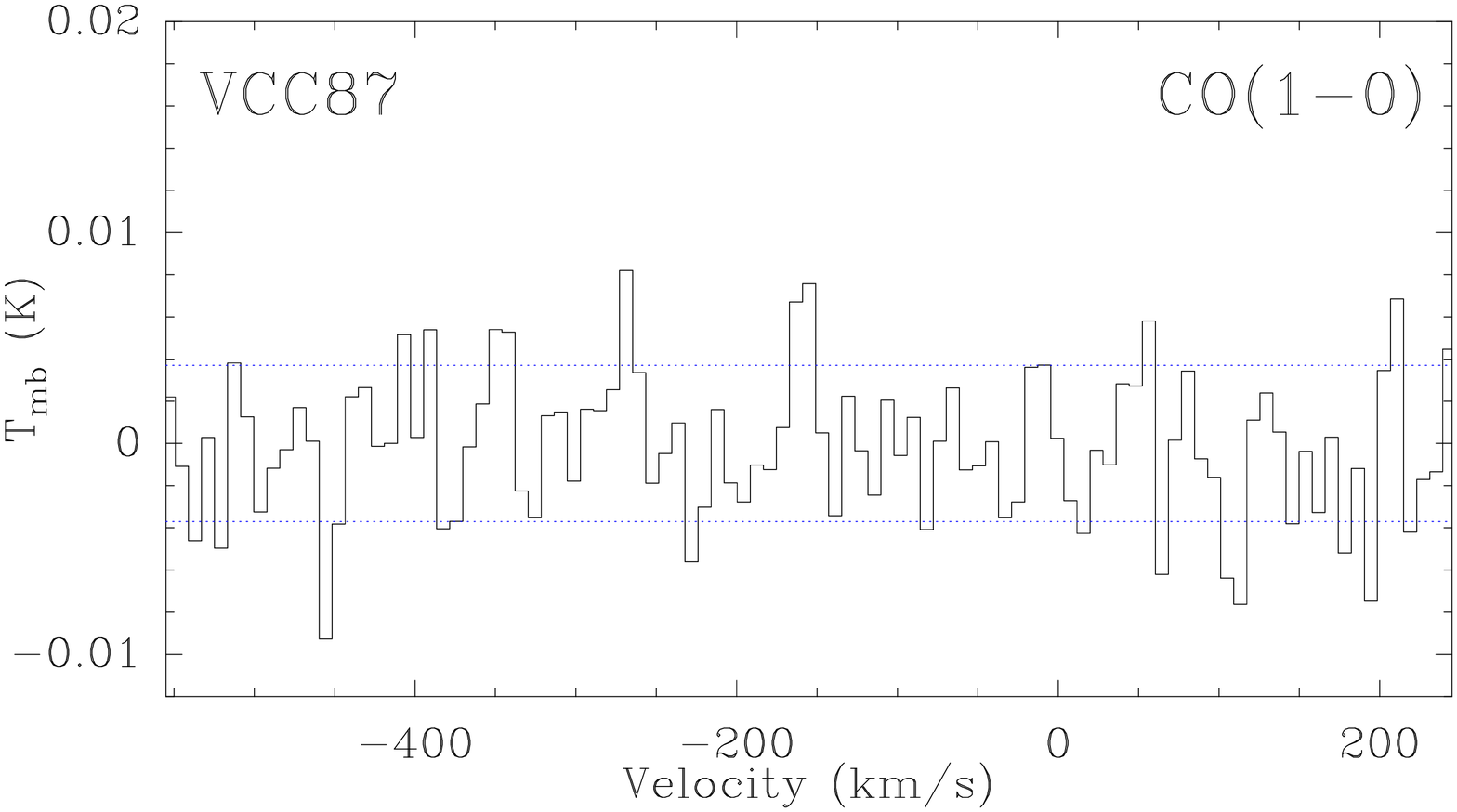}
\includegraphics[width=4.5cm]{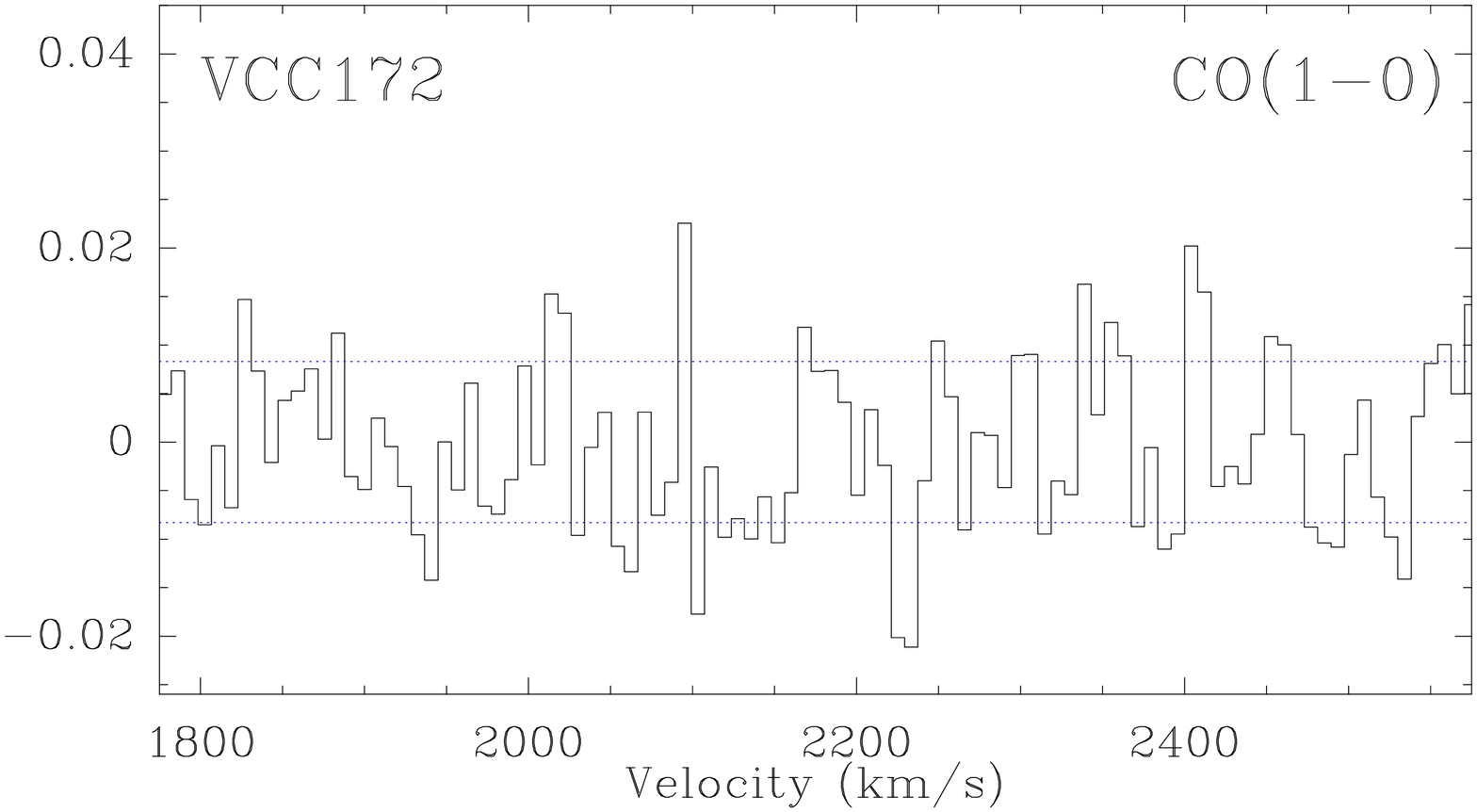}
\includegraphics[width=4.5cm]{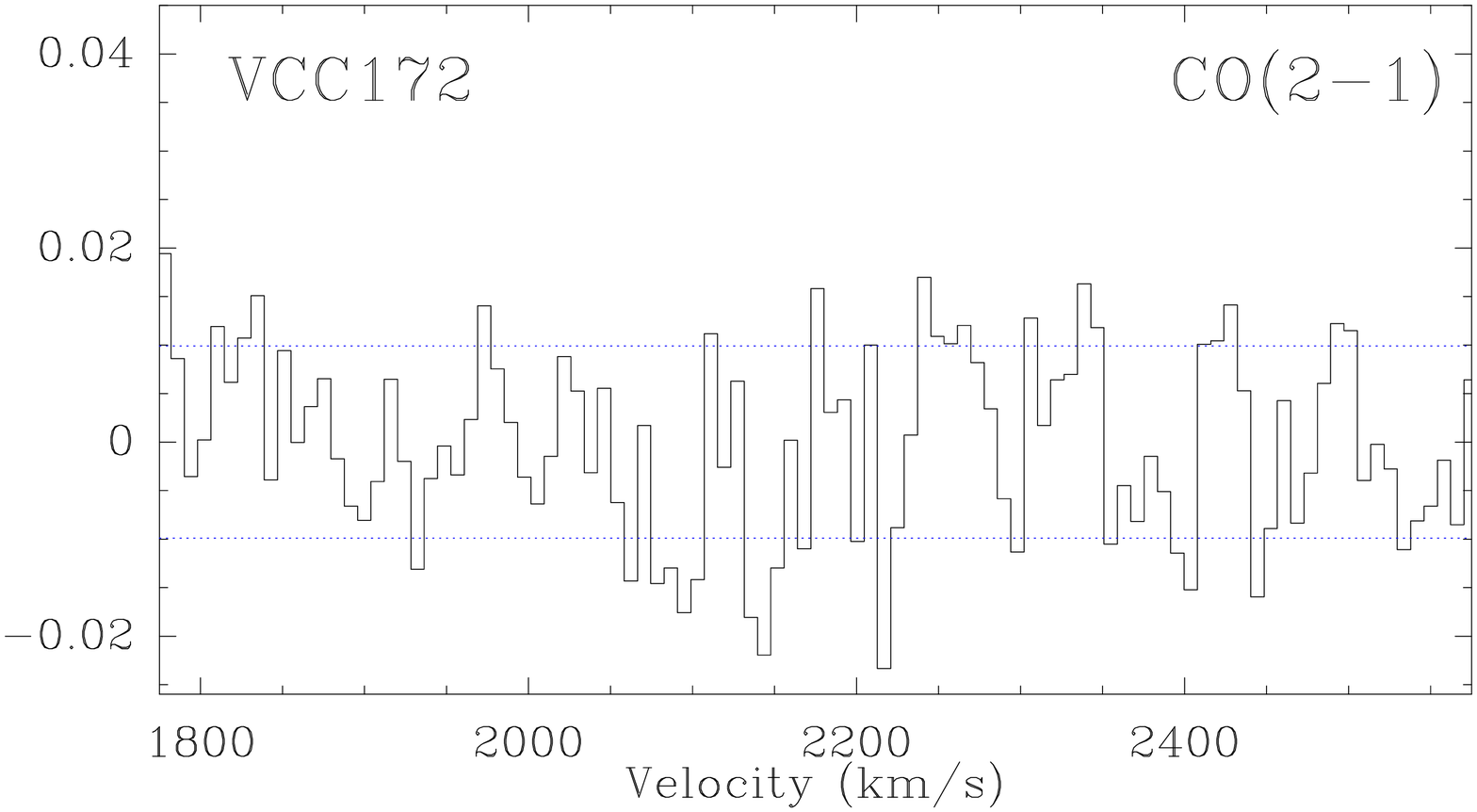}
\includegraphics[width=4.5cm]{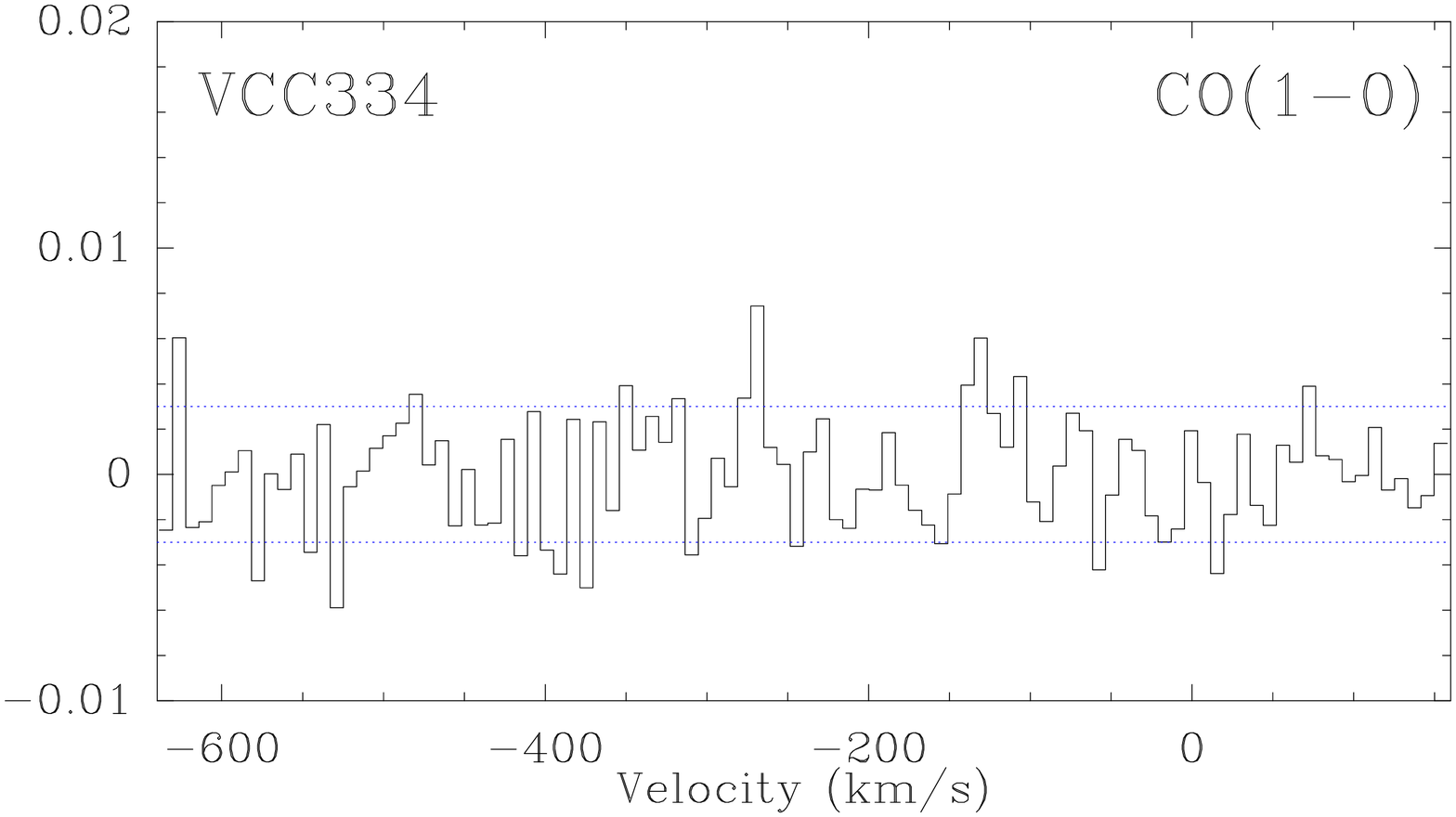}\\
\includegraphics[width=4.5cm]{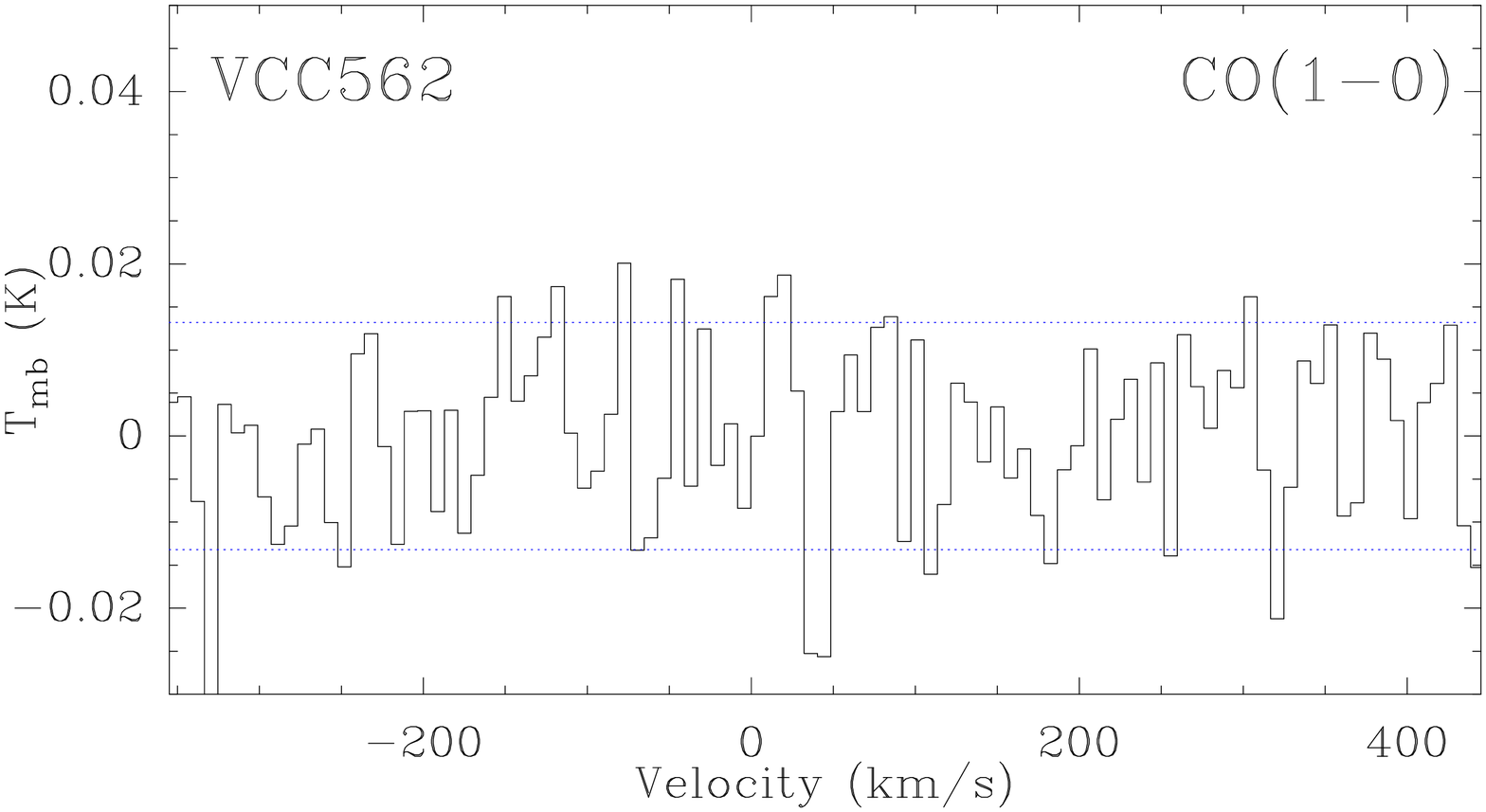}
\includegraphics[width=4.5cm]{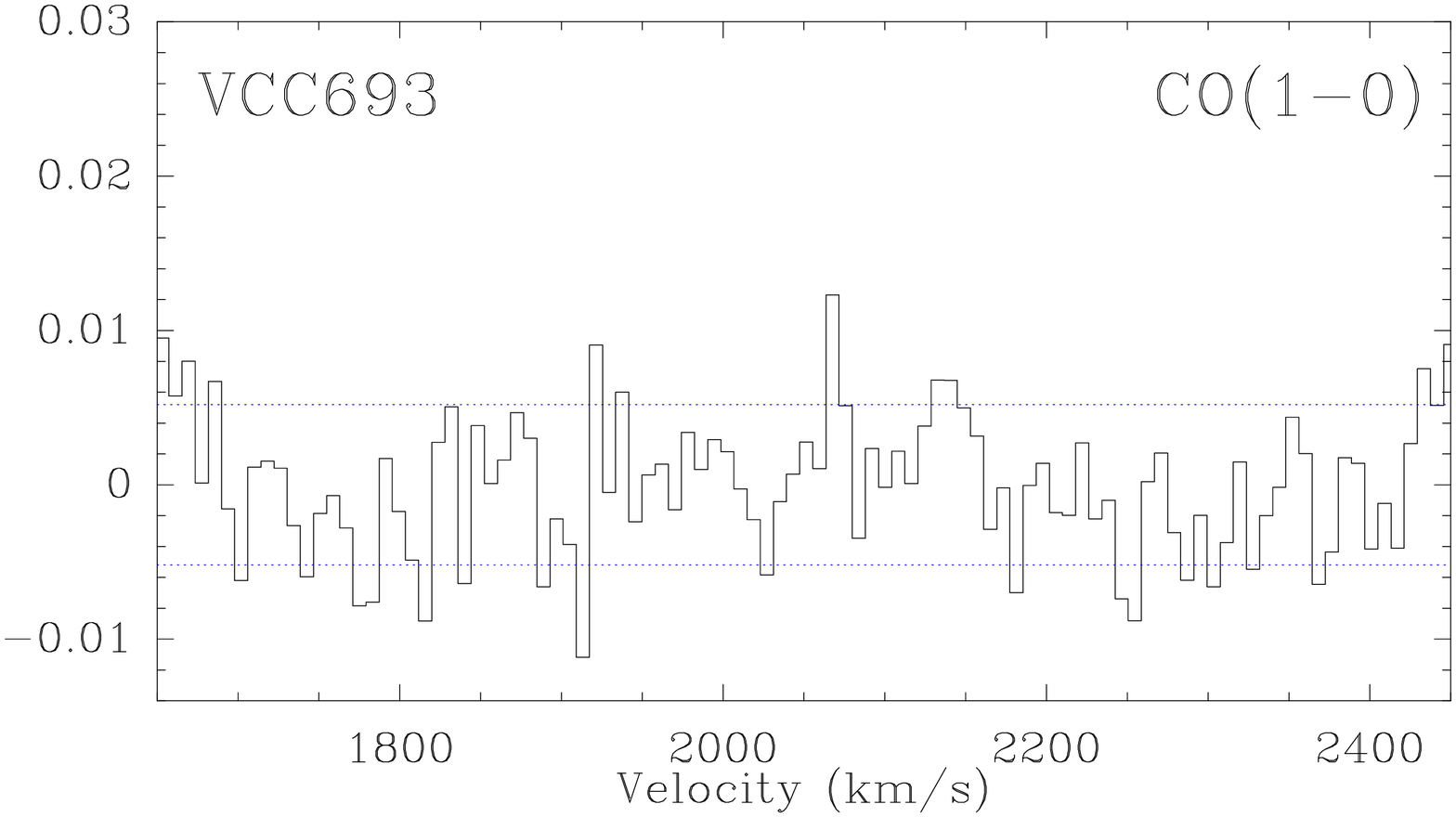}
\includegraphics[width=4.5cm]{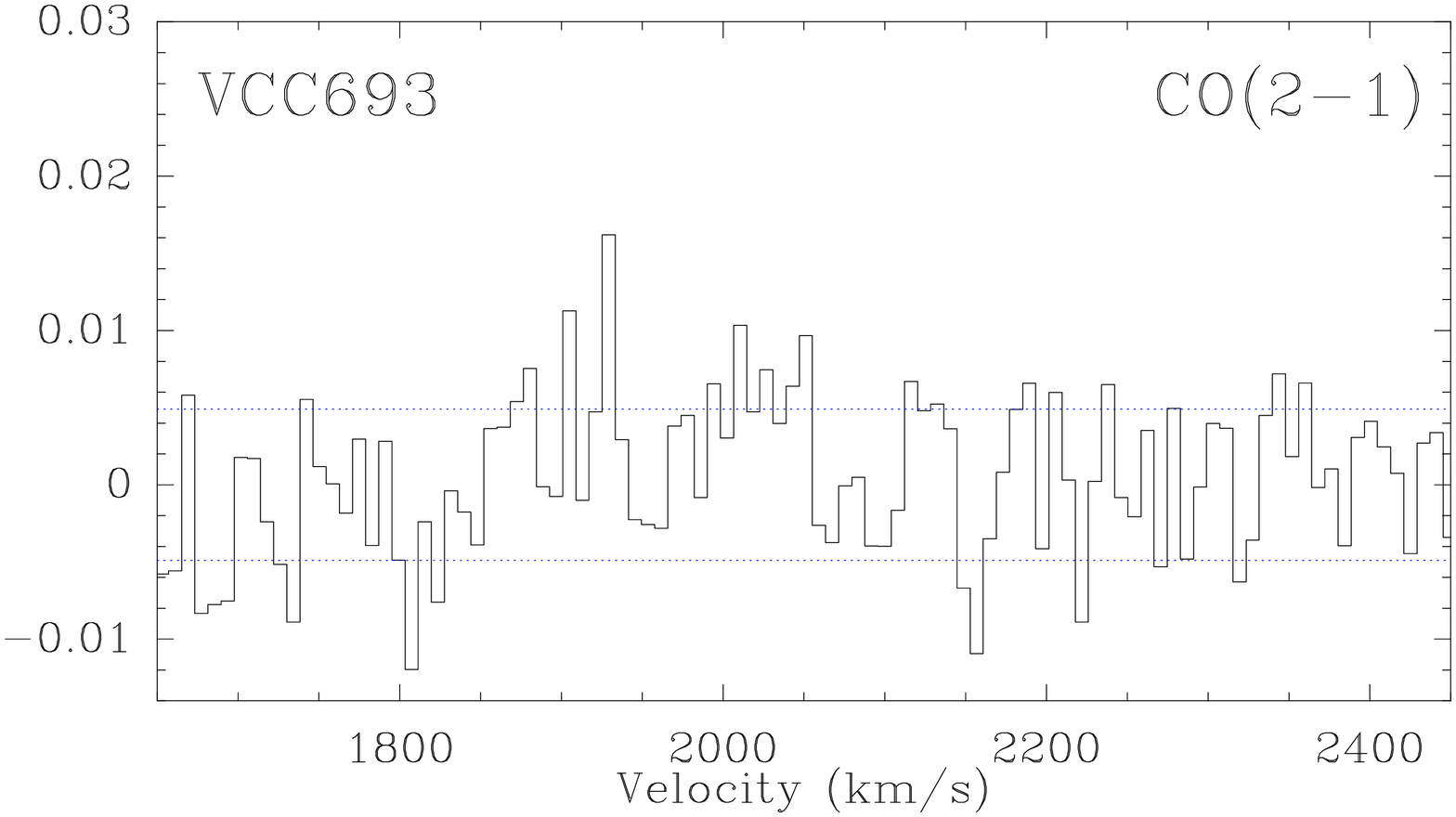}
\includegraphics[width=4.5cm]{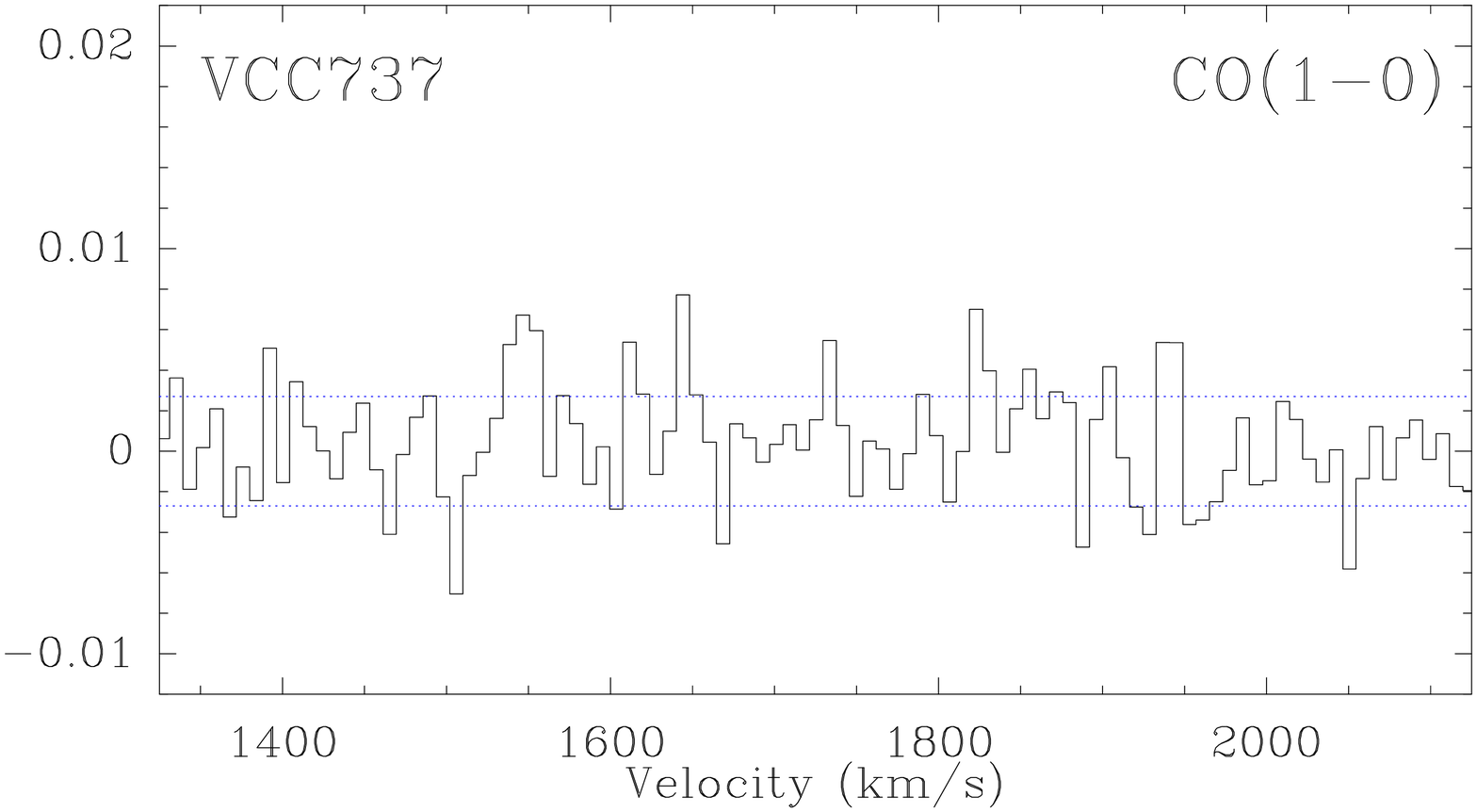}\\
\includegraphics[width=4.5cm]{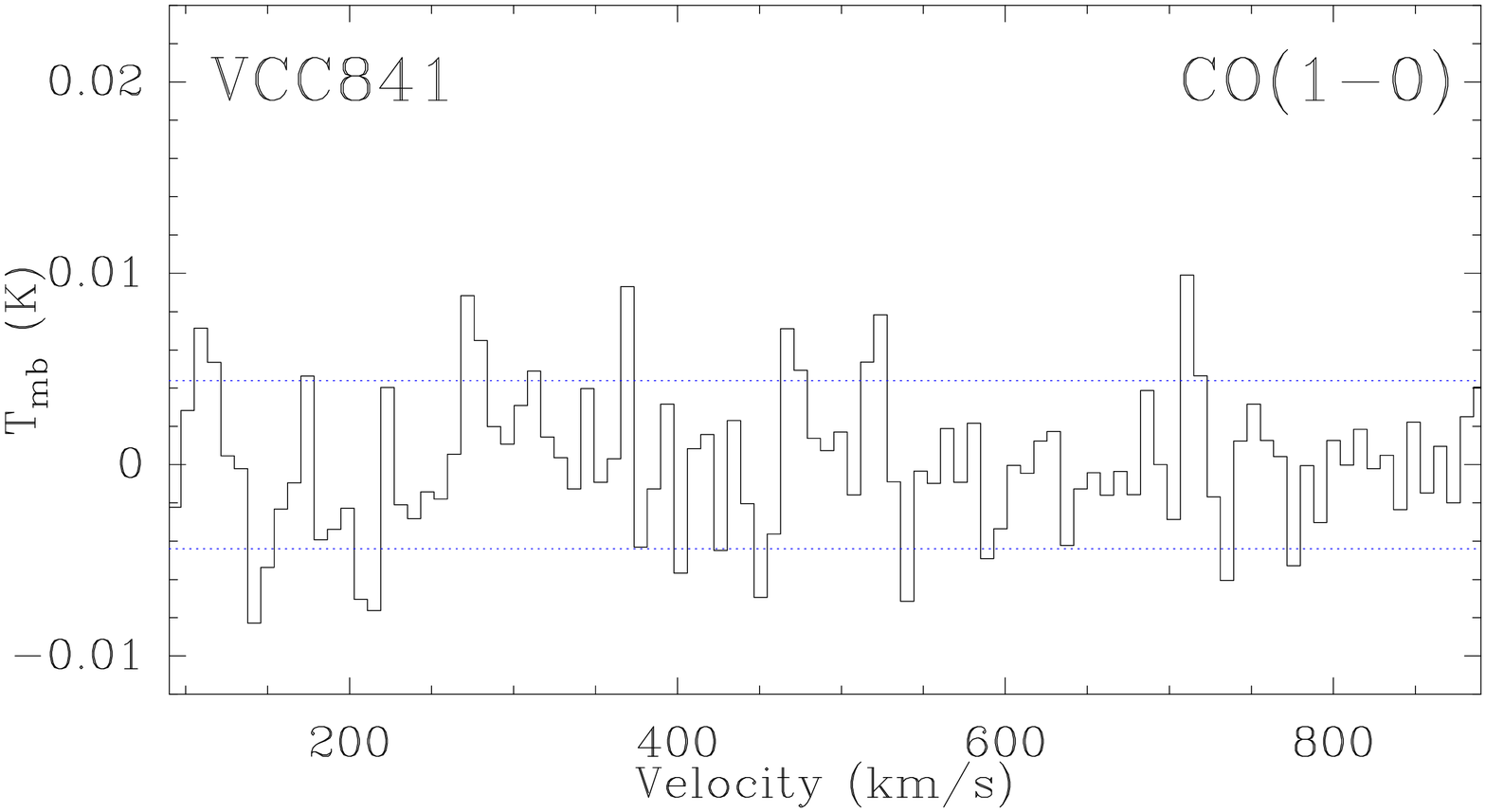}
\includegraphics[width=4.5cm]{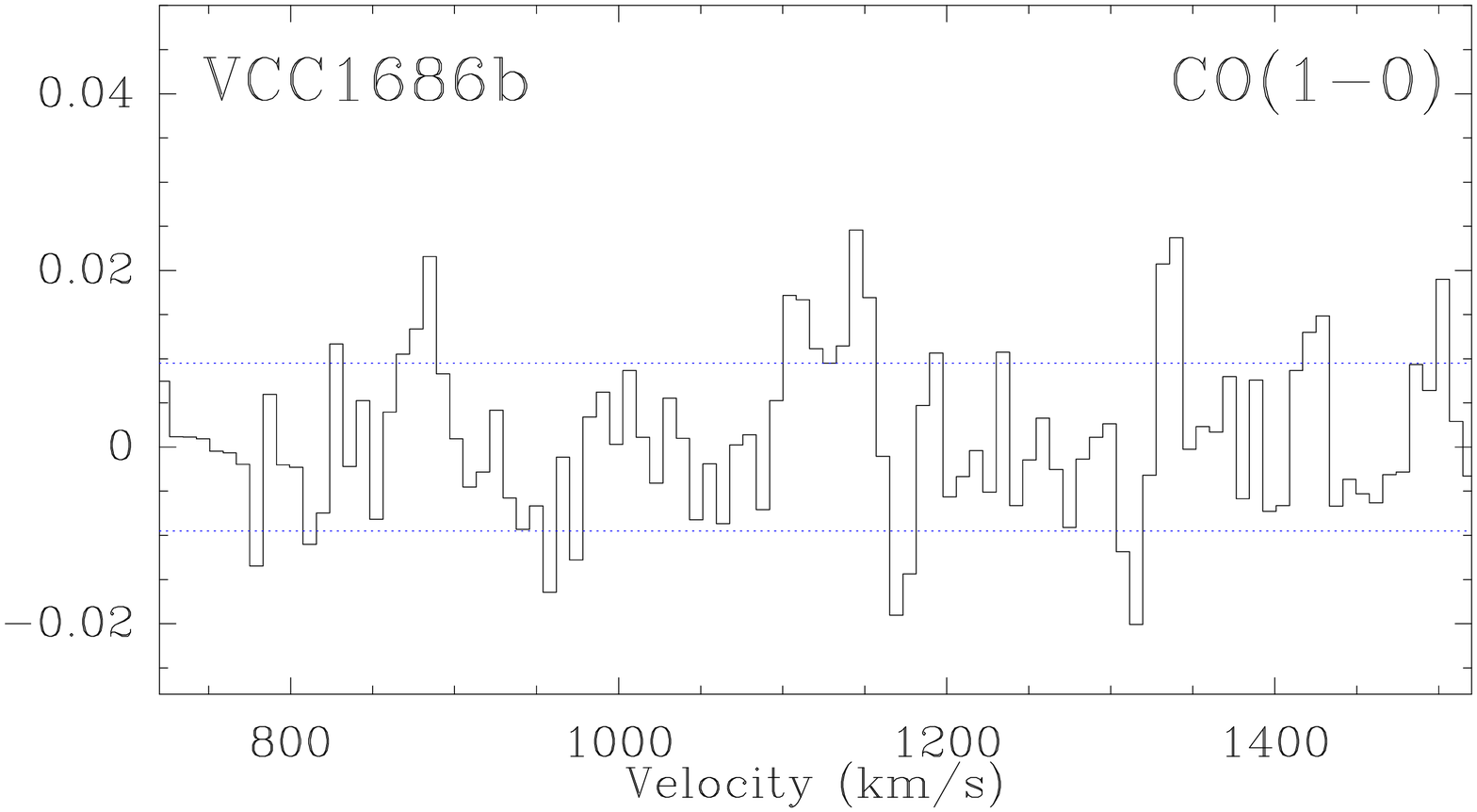}
\includegraphics[width=4.5cm]{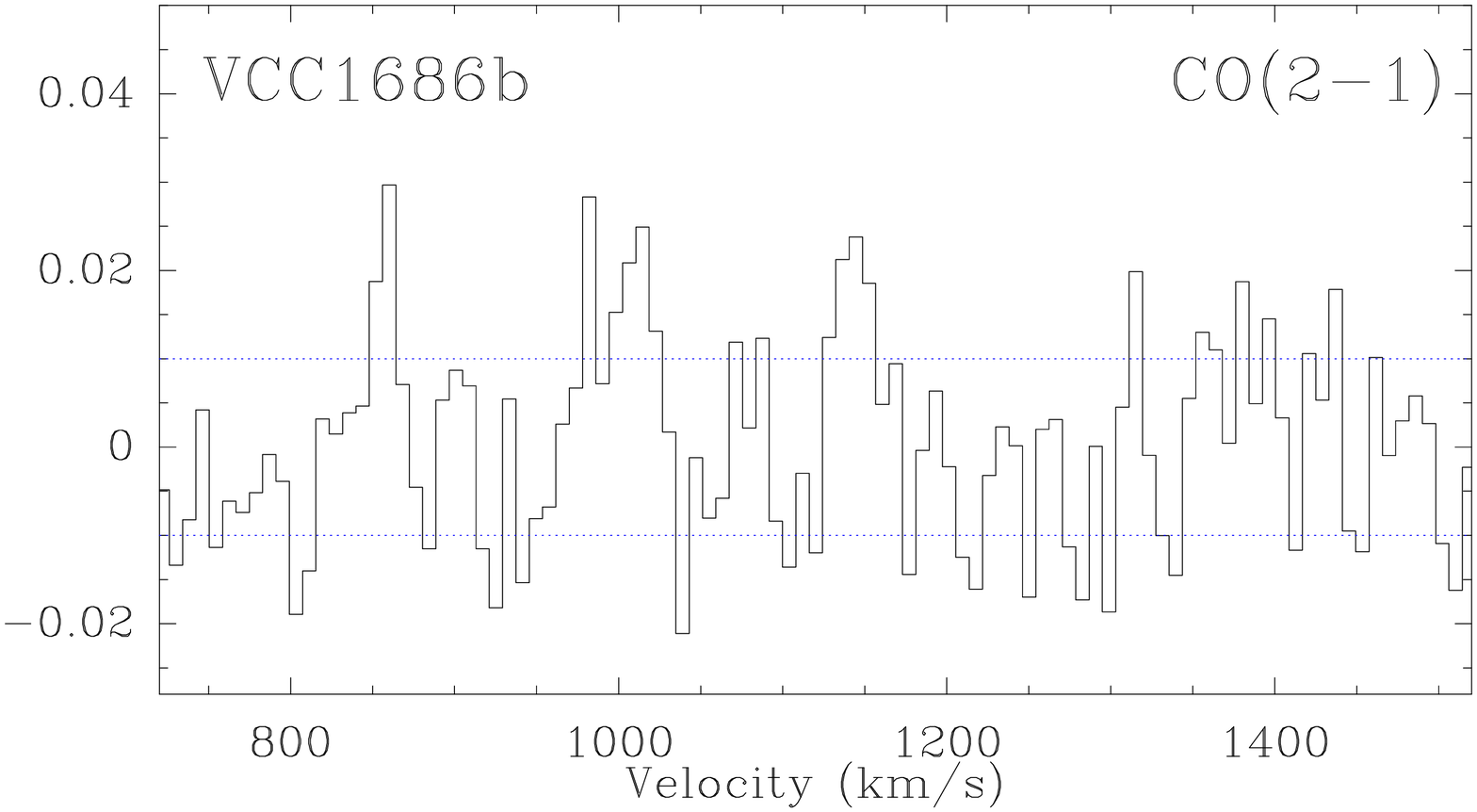}
\includegraphics[width=4.5cm]{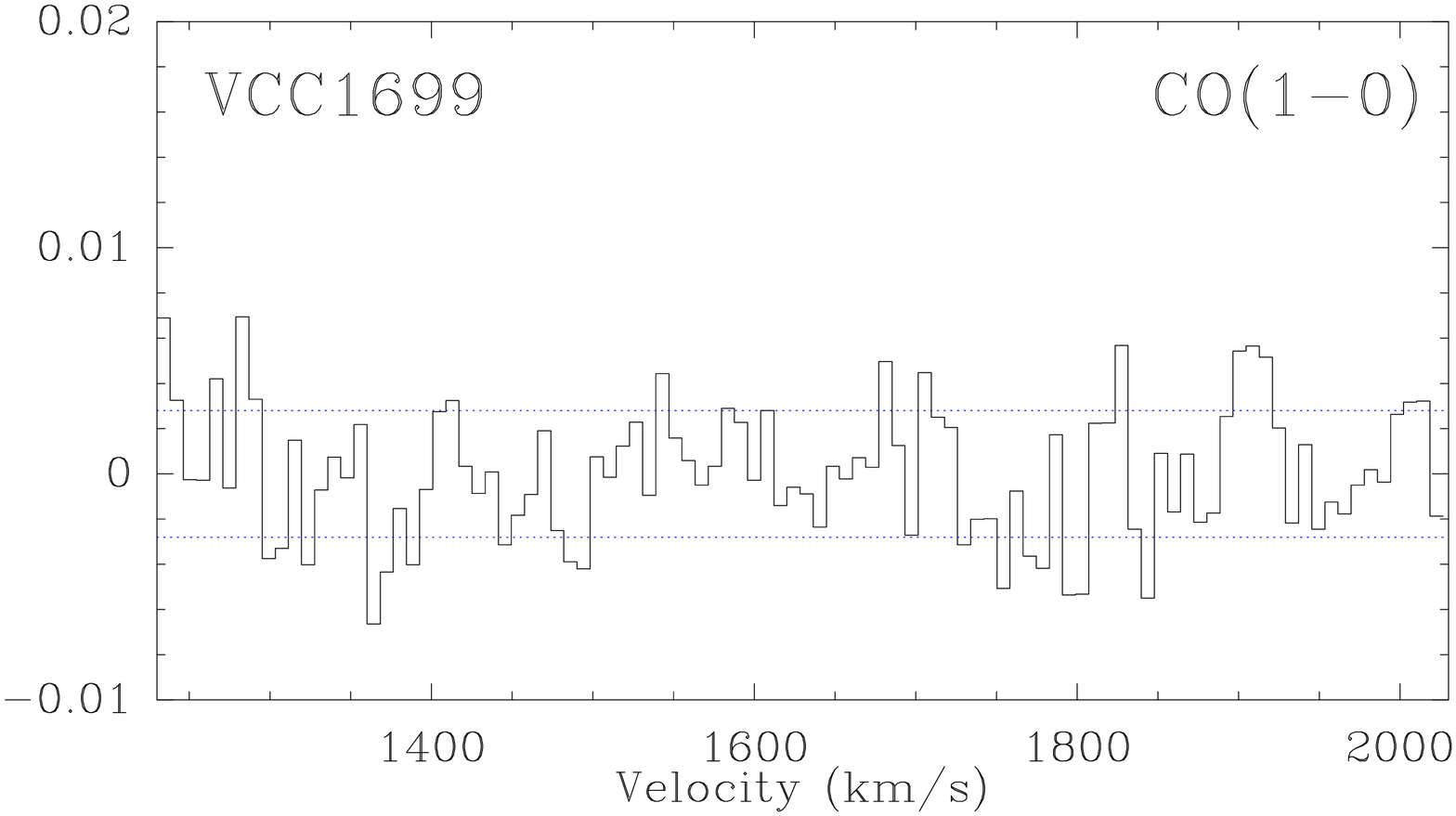}\\
\includegraphics[width=4.5cm]{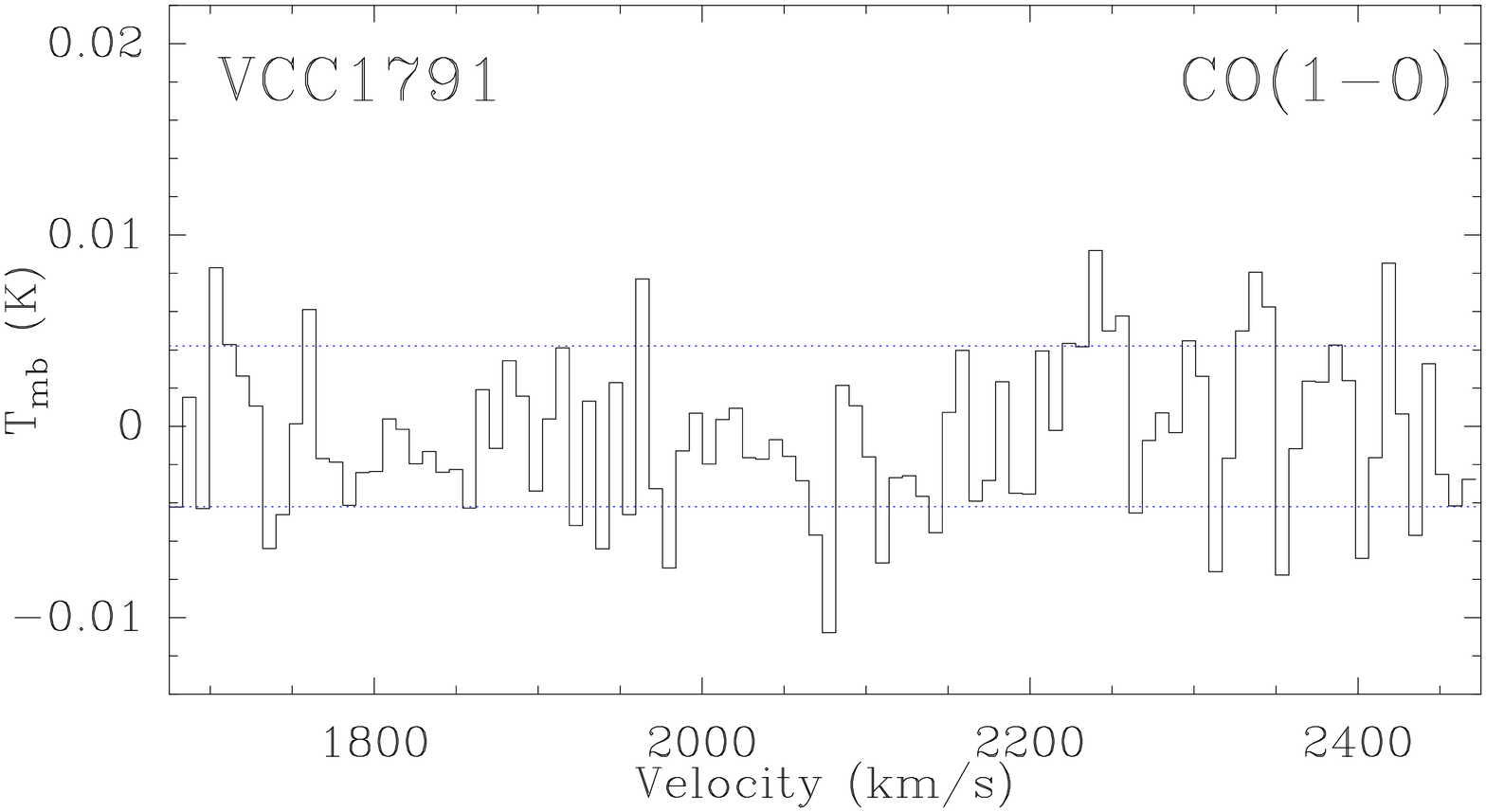}
\caption{CO(1-0) and CO(2-1) spectra of non-detections. The dotted horizontal lines show the $\pm$1$\sigma$ rms level.}
   \label{fig:CO21_spec}%
\end{figure*}

\end{appendix}

\end{document}